\documentclass[a4paper,11pt]{article}
\pdfoutput=1 

\usepackage{jheppub} 

\usepackage[T1]{fontenc} 
\usepackage{mathrsfs}
\usepackage[export]{adjustbox} 
\usepackage{microtype}

\usepackage[mathscr]{euscript}
\DeclareMathAlphabet{\mathpzc}{OT1}{pzc}{m}{it}

\def\cW{{\cal W}}
\def\cM{{\cal M}}
\def\cN{{\cal N}}

\def\cQ{{\cal Q}}

\newcommand{\bea}{\begin{eqnarray}}
\newcommand{\eea}{\end{eqnarray}}
\newcommand{\be}{\begin{eqnarray}}
\newcommand{\ee}{\end{eqnarray}}
\newcommand{\nn}{\nonumber}
\newcommand{\Tr}{\textrm{Tr}}
\newcommand{\Res}{\textrm{Res}}

\def\Mon			{\mathpzc{M}}
\def\Non			{\mathpzc{N}}

\def \checkPhi 		{\Omega}
\def \FFPhi		{\Theta}
\def \Flip 			{ {F}}

\def\m	 		{\mathpzc{m}}
\def\n	 		{\mathpzc{n}}

\def\Num 			{N+1}

\usepackage{tikz}
\usetikzlibrary{arrows}

\usepackage[T1]{fontenc}
\usepackage{graphicx}
\def\chk#1{#1^{\smash{\scalebox{.9}[1.5]{\rotatebox{90}{\guilsinglleft}}}}}

\usetikzlibrary{arrows}
\tikzset{>=stealth}

\title{
Flipping the head of $T[SU(N)]$: \\
mirror symmetry, spectral duality and monopoles
}

\author[1]{Francesco Aprile}
\author[1]{Sara Pasquetti}
\author[1,2]{Yegor Zenkevich}

\affiliation[1]{Dipartimento di Fisica, Universit\`a di Milano-Bicocca \& INFN, Sezione di Milano-Bicocca, \\
I-20126 Milano, Italy
}
\affiliation[2]{ITEP, Moscow 117218, Russia\\
}
\emailAdd{}
\emailAdd{sara.pasquetti@gmail.com}
\emailAdd{yegor.zenkevich@gmail.com}
\emailAdd{francesco.aprile1@unimib.it}

\abstract{ We consider $T[SU(N)]$ and its mirror, and we argue that
  there are two more dual frames, which are obtained by adding
  flipping fields for the moment maps on the Higgs and Coulomb branch.
  Turning on a monopole deformation in $T[SU(N)]$, and following its
  effect on each dual frame, we obtain four new daughter theories dual
  to each other.  We are then able to construct pairs of $3d$ spectral
  dual theories by performing simple operations on the four dual
  frames of $T[SU(N)]$.  Engineering these $3d$ spectral pairs as
  codimension-two defect theories coupled to a trivial $5d$ theory,
  via Higgsing, we show that our $3d$ spectral dual theories descends
  from the $5d$ spectral duality, or fiber base duality in topological
  string.  We provide further consistency checks about the web of
  dualities we constructed by matching partition functions on
  $\mathcal{S}^3_b$, and in the case of spectral duality, matching
  exactly topological string computations with holomorphic blocks.}

\begin{document} 
\maketitle
\flushbottom

\section{Introduction and Summary}

The study of $3d$ supersymmetric gauge theories and their dualities
has received a great deal of attention in the last decade. 
Thanks to important achievements in study of supersymmetric localisation (for a review see \cite{Pestun:2016zxk}),  
it has been possible to calculate exactly quantities such as partition functions of $\mathcal{N}\geq 2$ theories on various 3-manifolds, 
and test a plethora of old and new IR dualities.

Much progress stemmed from the idea of
obtaining $3d$ dualities from $4d$ Seiberg-like dualities, which was
revamped in~\cite{Aharony:2013dha}. $3d$ dualities obtained from
$4d$ have been in turn observed to generate old and new dual pairs
when subject to various types of real mass deformations. 
Identifying a common  ancestor of various
apparently unrelated $3d$ dualities seems a useful organizing principle to attempt charting the vast 
landscape of  $3d$ dualities. For recent
results in this direction see~\cite{Benvenuti:2018bav},
\cite{Amariti:2018wht} and references therein.

In this paper we are interested in a different type of $3d$ dualities:
We will consider mirror and {\it spectral} dualities for which we can identify a  
 $5d$ ancestor, and not a $4d$ one. 

Our starting point is the so-called $T[SU(N+1)]$ quiver theory of  \cite{Gaiotto:2008ak}, depicted in the Figure \eqref{inidiagram}, 
which can be realised  on a set of D3 branes stretched between NS5 and
D5 branes. The action of $S$-duality on Type IIB three- and
five-branes can then be used to show that the $T[SU(N+1)]$ theory
has a mirror dual, which we call $\chk{T[SU(\Num)]}$, that is described by the same theory with Higgs and Coulomb
branch swapped.

Our first observation is that  $T[SU(N+1)]$  has \emph{two
  more} dual descriptions, which we denote by $FFT[SU(N+1)]$ and
$\chk{FFT[SU(\Num)]}$. In more detail,  
$FFT[SU(N+1)]$ is obtained from the very same quiver tail of 
$T[SU(N+1)]$, by adding two extra sets of gauge singlets which
couple linearly to the moment maps of the Coulomb and Higgs
branches. Pictorially, $FFT[SU(N+1)]$ is the $T[SU(N+1)]$ theory
with \emph{flipped} Coulomb and Higgs branches.  Similarly, $\chk{FFT[SU(\Num)]}$ is the $\chk{T[SU(\Num)]}$ theory with \emph{flipped} Coulomb and Higgs branches.
From our construction follows the diagram of dualities shown here below. 

\be
\label{inidiagram}
             \begin{tikzpicture}
             \def\ptAx{-2 }
             \def\ptAy{.2}
             \def\lato {5}
             \def\latoA{3.5}
          
             \def\spazio{.3}

             \draw (\ptAx-.5,\ptAy+.075) --  (\ptAx+.6,\ptAy+.075) ;
             \draw (\ptAx-.5,\ptAy-.075) --  (\ptAx+.6,\ptAy-.075) ;
	    \draw (\ptAx-2.5,\ptAy+.075) --  (\ptAx-1,\ptAy+.075) ;
             \draw (\ptAx-2.5,\ptAy-.075) --  (\ptAx-1,\ptAy-.075) ;
 	    
	    \filldraw[fill=white,draw=black] (\ptAx,\ptAy+.3) circle (.25cm);
             \filldraw[fill=white,draw=black] (\ptAx,\ptAy) circle (.35cm);

             \filldraw[fill=white,draw=black] (\ptAx-1.5,\ptAy+.3) circle (.25cm);
             \filldraw[fill=white,draw=black] (\ptAx-1.5,\ptAy) circle (.35cm);
                   
             \draw (\ptAx-.72,\ptAy) node[font=\footnotesize] {\ldots};
            
             \filldraw[fill=white,draw=black] (\ptAx-2.5,\ptAy+.3) circle (.25cm);
             \filldraw[fill=white,draw=black] (\ptAx-2.5,\ptAy) circle (.35cm);

             \filldraw[fill=white,draw=black] (\ptAx+.6,\ptAy-.3) rectangle (\ptAx+1.4,\ptAy+.3);
             
             \draw (\ptAx-2.5,\ptAy) node[font=\footnotesize] {$1$};
             \draw (\ptAx-1.5,\ptAy) node[font=\footnotesize] {$2$};
             \draw (\ptAx,\ptAy) node[font=\footnotesize] {$N$};
             \draw (\ptAx+1,\ptAy) node[font=\footnotesize] {$N$+1};

             \draw (\ptAx+\lato-.8,\ptAy+.075) --  (\ptAx+\lato+1.1,\ptAy+.075) ;
             \draw (\ptAx+\lato-.8,\ptAy-.075) --  (\ptAx+\lato+1.1,\ptAy-.075) ;            
             \draw (\ptAx+\lato+1.6,\ptAy+.075) --  (\ptAx+\lato+.6+2.5,\ptAy+.075) ;
             \draw (\ptAx+\lato+1.6,\ptAy-.075) --  (\ptAx+\lato+.6+2.5,\ptAy-.075) ;
             \filldraw[fill=white,draw=black] (\ptAx+\lato+.6,\ptAy+.3) circle (.25cm);
             \filldraw[fill=white,draw=black] (\ptAx+\lato+.6,\ptAy) circle (.35cm);
                       
              \filldraw[fill=white,draw=black] (\ptAx+\lato+.6+1.5,\ptAy+.3) circle (.25cm);
              \filldraw[fill=white,draw=black] (\ptAx+\lato+.6+1.5,\ptAy) circle (.35cm);
                 
              \draw (\ptAx+\lato+1.35,\ptAy) node[font=\footnotesize] {\ldots};
            
             \filldraw[fill=white,draw=black] (\ptAx+\lato+.6+2.5,\ptAy+.3) circle (.25cm);
              \filldraw[fill=white,draw=black] (\ptAx+\lato+.6+2.5,\ptAy) circle (.35cm);
                 
             \filldraw[fill=white,draw=black] (\ptAx+\lato-.8,\ptAy-.3) rectangle (\ptAx+\lato,\ptAy+.3);
             
             \draw (\ptAx+\lato+.6+1.5,\ptAy) node[font=\footnotesize] {$2$};
             \draw (\ptAx+\lato+.6+2.5,\ptAy) node[font=\footnotesize] {$1$};
             \draw (\ptAx+\lato+.6,\ptAy) node[font=\footnotesize] {$N$};
             \draw (\ptAx+\lato-.4,\ptAy) node[font=\footnotesize] {$N$+1};

             \draw (\ptAx-.5,\ptAy+.075-\latoA) --  (\ptAx+.6,\ptAy+.075-\latoA) ;
             \draw (\ptAx-.5,\ptAy-.075-\latoA) --  (\ptAx+.6,\ptAy-.075-\latoA) ;
	    \draw (\ptAx-2.5,\ptAy+.075-\latoA) --  (\ptAx-1,\ptAy+.075-\latoA) ;
             \draw (\ptAx-2.5,\ptAy-.075-\latoA) --  (\ptAx-1,\ptAy-.075-\latoA) ;
 	    
	    \filldraw[fill=white,draw=black] (\ptAx,\ptAy+.3-\latoA) circle (.25cm);
             \filldraw[fill=white,draw=black] (\ptAx,\ptAy-\latoA) circle (.35cm);

             \filldraw[fill=white,draw=black] (\ptAx-1.5,\ptAy+.3-\latoA) circle (.25cm);
             \filldraw[fill=white,draw=black] (\ptAx-1.5,\ptAy-\latoA) circle (.35cm);
                   
             \draw (\ptAx-.72,\ptAy-\latoA) node[font=\footnotesize] {\ldots};
            
             \filldraw[fill=white,draw=black] (\ptAx-2.5,\ptAy+.3-\latoA) circle (.25cm);
             \filldraw[fill=white,draw=black] (\ptAx-2.5,\ptAy-\latoA) circle (.35cm);

                  \filldraw[fill=white,draw=black] (\ptAx+1.55,\ptAy-\latoA) circle (.25cm);
                   
             \filldraw[fill=white,draw=black] (\ptAx+.6,\ptAy-.3-\latoA) rectangle (\ptAx+1.4,\ptAy+.3-\latoA);
             
             \draw (\ptAx-2.5,\ptAy-\latoA) node[font=\footnotesize] {$1$};
             \draw (\ptAx-1.5,\ptAy-\latoA) node[font=\footnotesize] {$2$};
             \draw (\ptAx,\ptAy-\latoA) node[font=\footnotesize] {$N$};
             \draw (\ptAx+1,\ptAy-\latoA) node[font=\footnotesize] {$N$+1};

             \draw (\ptAx+\lato-.8,\ptAy+.075-\latoA) --  (\ptAx+\lato+1.1,\ptAy+.075-\latoA) ;
             \draw (\ptAx+\lato-.8,\ptAy-.075-\latoA) --  (\ptAx+\lato+1.1,\ptAy-.075-\latoA) ;            
             \draw (\ptAx+\lato+1.6,\ptAy+.075-\latoA) --  (\ptAx+\lato+.6+2.5,\ptAy+.075-\latoA) ;
             \draw (\ptAx+\lato+1.6,\ptAy-.075-\latoA) --  (\ptAx+\lato+.6+2.5,\ptAy-.075-\latoA) ;
             \filldraw[fill=white,draw=black] (\ptAx+\lato+.6,\ptAy+.3-\latoA) circle (.25cm);
             \filldraw[fill=white,draw=black] (\ptAx+\lato+.6,\ptAy-\latoA) circle (.35cm);
                       
              \filldraw[fill=white,draw=black] (\ptAx+\lato+.6+1.5,\ptAy+.3-\latoA) circle (.25cm);
              \filldraw[fill=white,draw=black] (\ptAx+\lato+.6+1.5,\ptAy-\latoA) circle (.35cm);
                 
              \draw (\ptAx+\lato+1.35,\ptAy-\latoA) node[font=\footnotesize] {\ldots};
            
             \filldraw[fill=white,draw=black] (\ptAx+\lato+.6+2.5,\ptAy+.3-\latoA) circle (.25cm);
              \filldraw[fill=white,draw=black] (\ptAx+\lato+.6+2.5,\ptAy-\latoA) circle (.35cm);

                   \filldraw[fill=white,draw=black] (\ptAx+\lato-.95,\ptAy-\latoA) circle (.25cm);
                 
             \filldraw[fill=white,draw=black] (\ptAx+\lato-.8,\ptAy-.3-\latoA) rectangle (\ptAx+\lato,\ptAy+.3-\latoA);
             
             \draw (\ptAx+\lato+.6+1.5,\ptAy-\latoA) node[font=\footnotesize] {$2$};
             \draw (\ptAx+\lato+.6+2.5,\ptAy-\latoA) node[font=\footnotesize] {$1$};
             \draw (\ptAx+\lato+.6,\ptAy-\latoA) node[font=\footnotesize] {$N$};
             \draw (\ptAx+\lato-.4,\ptAy-\latoA) node[font=\footnotesize] {$N$+1};

             \def\ptBx{-1.5}
             \def\ptBy{+1}
             \def\lato {5}
             \def\hor{7}
             \def\spazio{.2}
             
       
         \draw[line width=.6pt ,<->
         ]  (\ptBx-1.2,\ptBy-1.2) -- (\ptBx-1.2,\ptBy-\lato+1.2);
         

  	 \draw[line width=.6pt,<->]  (\ptBx+\hor-1.2,\ptBy-1.2) -- (\ptBx+\hor-1.2,\ptBy-\lato+1.2);
         

             \draw[line width=.6pt,<->]  (\ptBx+3*\spazio,\ptBy) -- (\ptBx+\lato-4*\spazio,\ptBy);
             \draw[line width=.6pt,<->]  (\ptBx+3*\spazio,\ptBy-\lato) -- (\ptBx+\lato-4*\spazio,\ptBy-\lato);

             \node[scale=.95] at (\ptBx-5.5*\spazio,\ptBy+0.5*\spazio) {$T[SU(\Num)]$};
             \node[scale=.95] at (\ptBx-6.8*\spazio,\ptBy-\lato-0.5*\spazio) {$FFT[SU(\Num)]$};
             \node[scale=.95] at (\ptBx+\lato+5.5*\spazio,\ptBy+0.45*\spazio) {$\chk{T[SU(\Num)]}$};
	     \node[scale=.95] at (\ptBx+\lato+7*\spazio,\ptBy-\lato-0.5*\spazio) {$FF\chk{T[SU(\Num)]}$};
             
             \end{tikzpicture}        
\ee

Horizontal
arrows correspond to mirror dualities, while vertical arrows are
new dualities which, as we will explain in more details, can be regarded as a generalization of 
Aharony duality \cite{Aharony:1997gp} to a quiver tail. In section \ref{tsun_sec}, 
we discuss the map between operators across the four dual frames. 
Particularly interesting is the way nilpotent orbits are mapped under Flip-Flip duality. 

In section \ref{difo_gen} we
show equality of the partition functions on $S^3_b$.
Contrary to the case of $4d$ and $3d$ Seiberg-like
dualities, where the equality of localised partition functions reduces to 
well studied integral identities, developed in the mathematical literature by \cite{Spiridonov:2009za}, \cite{Rains:2006dfy} and \cite{Fokko},
for $T[SU(\Num)]$ and its dual partition functions
there are no analogous results. So we follow the strategy
of~\cite{Bullimore:2014awa} regarding our partition functions as
eigenfunctions of a set of Hamiltonians.

In section \ref{defsec} we consider the effect of deforming $T[SU(N+1)]$ by a
linear monopole superpotential as in~\cite{Benvenuti:2017kud},
\cite{Giacomelli:2017vgk}. Applying the monopole
duality of~\cite{Benini:2017dud}, we show that this deformation has the
effect of confining sequentially all the nodes but the last one. The
result is a $U(N)$ theory with $(N+1)$ flavors and several gauge
singlets, which we call theory $A$. We then follow the
monopole deformation across the duality frames and obtain four new dual theories. 
This is the inner $ABCD$ square shown in the picture below.

\begin{center}
             \begin{tikzpicture}[>=stealth]
            
             \def\ptAx{-3}
             \def\ptAy{-.5}
             \def\lato {3}
             \def\hor{6}
             \def\spazio{.3}
       
             \draw[line width=.6pt,<->]  (\ptAx+0.5*\lato+2*\spazio,\ptAy) -- (\ptAx+1.5*\lato-\spazio,\ptAy); 
             \draw[line width=.6pt,<->]  (\ptAx+1.5*\spazio,\ptAy-2*\spazio) --  (\ptAx+1.5*\spazio,\ptAy-\lato+2*\spazio);
             \draw[line width=.6pt,<->]  (\ptAx+2.55*\lato,\ptAy-2*\spazio) -- (\ptAx+2.55*\lato,\ptAy-\lato+2*\spazio);
             \draw[line width=.6pt,<->]  (\ptAx+0.5*\lato+2*\spazio,\ptAy-\lato) -- (\ptAx+1.5*\lato-\spazio,\ptAy-\lato); 

             \draw (\ptAx,\ptAy+.075) --  (\ptAx+.6,\ptAy+.075) ;
             \draw (\ptAx,\ptAy-.075) --  (\ptAx+.6,\ptAy-.075) ;
             \filldraw[fill=white,draw=black] (\ptAx,\ptAy) circle (.35cm);
             \filldraw[fill=white,draw=black] (\ptAx+.6,\ptAy-.3) rectangle (\ptAx+1.4,\ptAy+.3);
             \draw (\ptAx,\ptAy) node[font=\footnotesize] {$N$};
             \draw (\ptAx+1,\ptAy) node[font=\footnotesize] {$N$+1};
             \draw (\ptAx-.55,\ptAy+.04) node[font=\footnotesize] {$\vdots$};
             \foreach \y in {-.29,+.28} \filldraw (\ptAx-.55,\ptAy +\y) circle (.025cm);

             \draw (\ptAx,\ptAy-\lato+.075) --  (\ptAx+.6,\ptAy-\lato+.075) ;
             \draw (\ptAx,\ptAy-\lato-.075) --  (\ptAx+.6,\ptAy-\lato-.075) ;
             \filldraw[fill=white,draw=black] (\ptAx,\ptAy-\lato) circle (.35cm);
             \filldraw[fill=white,draw=black] (\ptAx+1.45,\ptAy-\lato) circle (.25cm);
             \filldraw[fill=white,draw=black] (\ptAx+.6,\ptAy-\lato-.3) rectangle (\ptAx+1.4,\ptAy-\lato+.3);
             \draw (\ptAx,\ptAy-\lato) node[font=\footnotesize] {1};
             \draw (\ptAx+1,\ptAy-\lato) node[font=\footnotesize] {$N$+1};
              \draw (\ptAx-.55,\ptAy-\lato+.04) node[font=\footnotesize] {$\vdots$};
             \foreach \y in {-.29,+.28}  \filldraw (\ptAx-.55,\ptAy-\lato +\y)  circle (.025cm);

             \draw (\ptAx+\hor+1,\ptAy+.075) --  (\ptAx+\hor-.6,\ptAy+.075) ;
             \draw (\ptAx+\hor+1,\ptAy-.075) --  (\ptAx+\hor-.6,\ptAy-.075) ;
             
              \filldraw[fill=white,draw=black] (\ptAx+\hor-.6,\ptAy-.3) rectangle (\ptAx+\hor-1.2,\ptAy+.3);
              \draw (\ptAx+\hor-.9,\ptAy) node[font=\footnotesize] {1};
             
             \draw (\ptAx+\hor,\ptAy+.075) --  (\ptAx+\hor+1,\ptAy+.075) ;
             \draw (\ptAx+\hor,\ptAy-.075) --  (\ptAx+\hor+1,\ptAy-.075) ;
             \draw (\ptAx+\hor+.075,\ptAy) --  (\ptAx+\hor+.075,\ptAy-1) ;
	     \draw (\ptAx+\hor-.075,\ptAy) --  (\ptAx+\hor-.075,\ptAy-1) ; 
	    \filldraw[fill=white,draw=black] (\ptAx+\hor-.3,\ptAy-.6) rectangle (\ptAx+\hor+.3,\ptAy-1.2);
       	     \draw (\ptAx+\hor,\ptAy-.9) node[font=\footnotesize] {1}; 
             
             \filldraw[fill=white,draw=black] (\ptAx+\hor,\ptAy+.3) circle (.25cm);
             \filldraw[fill=white,draw=black] (\ptAx+\hor,\ptAy) circle (.355cm);
             \draw (\ptAx+\hor,\ptAy) node[font=\footnotesize] {$N$};
             \filldraw[fill=white,draw=black] (\ptAx+\hor+1,\ptAy+.3) circle (.25cm);
             \filldraw[fill=white,draw=black] (\ptAx+\hor+1,\ptAy) circle (.355cm);
             \draw (\ptAx+\hor+1,\ptAy) node[font=\footnotesize] {$N$-1};
             
              \draw (\ptAx+\hor+1.62,\ptAy) node[font=\footnotesize] {$\ldots$};
             
             \filldraw[fill=white,draw=black] (\ptAx+\hor+2.55,\ptAy) circle (.25cm);
             \filldraw[fill=white,draw=black] (\ptAx+\hor+2.25,\ptAy) circle (.355cm);
             \draw (\ptAx+\hor+2.25,\ptAy) node[font=\footnotesize] {1};

             \draw (\ptAx+\hor+1,\ptAy-\lato+.075) --  (\ptAx+\hor-.6,\ptAy-\lato+.075) ;
             \draw (\ptAx+\hor+1,\ptAy-\lato-.075) --  (\ptAx+\hor-.6,\ptAy-\lato-.075) ;
             
              \filldraw[fill=white,draw=black] (\ptAx+\hor-.6,\ptAy-\lato-.3) rectangle (\ptAx+\hor-1.2,\ptAy-\lato+.3);
               \draw (\ptAx+\hor-.9,\ptAy-\lato) node[font=\footnotesize] {1};
             
              \filldraw[fill=white,draw=black] (\ptAx+\hor,\ptAy-\lato+.3) circle (.25cm);
	     \filldraw[fill=white,draw=black] (\ptAx+\hor,\ptAy-\lato) circle (.35cm);
	      \draw (\ptAx+\hor,\ptAy-\lato) node[font=\footnotesize] {1};
             \filldraw[fill=white,draw=black] (\ptAx+\hor+1,\ptAy-\lato+.3) circle (.25cm);
             \filldraw[fill=white,draw=black] (\ptAx+\hor+1,\ptAy-\lato) circle (.35cm);
             \draw (\ptAx+\hor+1,\ptAy-\lato) node[font=\footnotesize] {1};
             
              \draw (\ptAx+\hor+1.62,\ptAy-\lato) node[font=\footnotesize] {$\ldots$};

             \draw (\ptAx+\hor+2.25,\ptAy-\lato+.075) --  (\ptAx+\hor+3,\ptAy-\lato+.075) ;
             \draw (\ptAx+\hor+2.25,\ptAy-\lato-.075) --  (\ptAx+\hor+3,\ptAy-\lato-.075) ; 
              
             \filldraw[fill=white,draw=black] (\ptAx+\hor+2.27,\ptAy-\lato+.3) circle (.25cm);           
             \filldraw[fill=white,draw=black] (\ptAx+\hor+2.27,\ptAy-\lato) circle (.35cm);
             \draw (\ptAx+\hor+2.27,\ptAy-\lato) node[font=\footnotesize] {1};
                          
             \filldraw[fill=white,draw=black] (\ptAx+\hor+2.9,\ptAy-\lato-.3) rectangle (\ptAx+\hor+3.5,\ptAy-\lato+.3);
             
             \draw (\ptAx+\hor+3.2,\ptAy-\lato) node[font=\footnotesize] {1};
             
             \draw (\ptAx+\hor+3.8,\ptAy-\lato+.04) node[font=\footnotesize] {$\vdots$};
             \foreach \y in {-.29,+.28} \filldraw (\ptAx+\hor+3.8,\ptAy-\lato +\y) circle (.025cm);

             \def\ptBx{-4.5}
             \def\ptBy{+1.5}
             \def\latox {12}
             \def\latoy {7}

             \def\spazio{.2}
             
            \draw[line width=.6pt,<->]  (\ptBx+7*\spazio,\ptBy) -- (\ptBx+\latox-7.5*\spazio,\ptBy); 
            \draw[line width=.6pt,<->]  (\ptBx,\ptBy-2*\spazio) -- (\ptBx,\ptBy-\latoy+2*\spazio);
            \draw[line width=.6pt,<->]  (\ptBx+\latox,\ptBy-2*\spazio) -- (\ptBx+\latox,\ptBy-\latoy+2*\spazio);
            \draw[line width=.6pt,<->]  (\ptBx+7*\spazio,\ptBy-\latoy) -- (\ptBx+\latox-7.5*\spazio,\ptBy-\latoy); 

             \node[scale=.95] at (\ptBx,\ptBy+0.5*\spazio) {$T[SU(\Num)]$};
             \node[scale=.95] at (\ptBx-\spazio,\ptBy-\latoy-0.5*\spazio) {$FFT[SU(\Num)]$};
             \node[scale=.95] at (\ptBx+\latox,\ptBy+0.45*\spazio) {$\chk{T[SU(\Num)]}$};
	     \node[scale=.95] at (\ptBx+\latox+\spazio,\ptBy-\latoy-0.5*\spazio) {$FF\chk{T[SU(\Num)]}$};
 
             
             \draw[line width=.6pt,->]  (\ptBx+2*\spazio,0.8*\ptBy)--(\ptAx-0.75*\spazio,\ptAy+3*\spazio); 
             \draw[line width=.6pt,->]  (\ptBx+\latox-3*\spazio,0.8*\ptBy)--(\ptAx+1.2*\hor+6*\spazio,\ptAy+3*\spazio); 
             \draw[line width=.6pt,->]  (\ptBx+2*\spazio,\ptBy-\latoy+2*\spazio)--(\ptAx-0.75*\spazio,\ptAy-0.5*\latoy-1*\spazio); 
             \draw[line width=.6pt,->]  (\ptBx+\latox-3*\spazio,\ptBy-\latoy+2*\spazio)--(\ptAx+1.2*\hor+6*\spazio,\ptAy-0.5*\latoy-1*\spazio);

             \node at (\ptAx+1.2*\spazio,\ptAy+4*\spazio) {$A$};
             \node at (\ptAx+1.2*\hor+4*\spazio,\ptAy+4*\spazio) {$B$};
             \node at (\ptAx+1.2*\hor+4*\spazio,\ptAy-0.5*\latoy-1.75*\spazio) {$C$};
             \node at (\ptAx+1.2*\spazio,\ptAy-0.5*\latoy-1.75*\spazio) {$D$};

             \end{tikzpicture}        
\end{center}
 
Interestingly enough, the horizontal lines in the $ABCD$ diagram also correspond to mirror dualities, 
while the vertical line connecting $A$ and $D$ is precisely Aharony duality.
Mirror symmetry relates $A$ and $B$ in very much the same way of \cite{Giacomelli:2017vgk}, 
but the connection among theories $A$, $D$ and $C$ is more involved. In particular, 
the original monopole deformation on $T[SU(\Num)]$ translates in $FF\chk{T[SU(\Num)]}$ 
into a nilpotent vev for the Higgs branch flipping fields. 
By studying in detail the low energy theory on such nilpotent vev we show that it corresponds to the abelian quiver $C$ in the picture.
Then we obtain the same theory by performing piece-wise mirror symmetry to theory $D$.

In section \ref{spd} we move on to the construction of our $3d$ spectral dualities.
Several ingredients goes into it. Firstly, 
we realise our $3d$ theories $\mathcal{T}_{\mathcal{X}}$ as codimension-two
defect theories coupled to a (trivial) $5d$ $\mathcal{N}=1$ bulk theory. 
In particular, $\mathcal{T}_{\mathcal{X}}$ is generated by
the so-called Higgsing prescription \cite{Gaiotto:2012xa}, i.e.\ by turning on appropriate
vevs in $5d$ $\mathcal{N}=1$ linear quiver theories which can be geometrically
engineered by $(p,q)$-brane webs of NS5 and D5' branes.
This higgsed configuration involves D3-branes, stretching between  NS5 and D5' branes, on which our defect theories live,
and it is a variation of the construction of  \cite{Hanany:2004ea}.
Secondly, we also realise our $5d$ linear quiver theory by the compactification
of M-theory on a toric CY three-fold $\mathfrak{X}$ (with the toric
diagram given by the $(p,q)$-web). This leads to direct interpretation of the $5d$ instanton partition function as the refined
topological string partition function $Z^{\mathfrak{X}}_{\mathrm{inst}, \mathrm{top}}$.
In this language, the Higgsing prescription amounts
to tuning the values of the K\"ahler parameters of the CY
$\mathfrak{X}$ to special quantized values in order to obtain the ``Higgsed
CY'' $\mathcal{X}$, and typically it reduces the instanton partition
function to an instanton-vortex partition function for the coupled
$3d$--$5d$ system \cite{Mironov:2009qt,Kozcaz:2010af,Dimofte:2010tz,Dorey:2011pa,Nieri:2013vba,Gaiotto:2014ina}.

 In the cases we are going to consider we
have a ``complete Higgsing'', by which we mean that starting from
$5d$ we are left with a $3d$ theory coupled to a \emph{trivial} $5d$
theory (free hypers). In this way we can unambiguously identify the Higgsed
instanton-vortex/topological string partition function as the vortex
partition function of the defect theory $ Z^{\alpha_0}_{\mathrm{vort},
  \mathcal{T}_{\mathcal{X}}}=Z^{\mathcal{X}}_{\mathrm{vort},
  \mathrm{top}}$.
More precisely, we can relate the $D^2\times S^1$ partition function
of theory $\mathcal{T}_{\mathcal{X}}$ (known as the holomorphic block
${\mathcal B}^{\alpha_0}_{\mathcal{T}_{\mathcal{X}}}$) evaluated in a
certain vacuum $\alpha_0$ to the partition function of the Higgsed
topological string: \footnote{For illustration, we have dropped a prefactor depending on contact terms, however keep track of it when presenting the
  actual matching of partition functions in
  section \ref{sec:spectr-dual-from}.}
\begin{equation}
  \boxed{{\mathcal B}^{\alpha_0}_{\mathcal{T}_{\mathcal{X}}}= Z^{\mathcal{X}}_{\mathrm{top}}}
\end{equation}

Our $5d$ linear quiver theories admits a  spectral dual description which can be 
 equivalently stated as the invariance of the topological string
partition function $Z^{\mathfrak{X}}_{\mathrm{top}}=Z^{\mathfrak{X}'}_{\mathrm{top}}$, 
where we denote by $\mathfrak{X}'$ the fiber-base dual CY. Of course $\mathfrak{X}$ and $\mathfrak{X}'$ are just
two equivalent description of the same toric CY, hence the equality of
the partition functions.
Therefore, our main idea is to combine Higgsing and  fiber-base duality 
to obtain new $3d$ spectral pairs which descend from $5d$. 
Summarizing, we first follow the Higgsing process on $\mathfrak{X}$ down to $\mathcal{X}$, which yields a $3d$
theory $\mathcal{T}_{\mathcal{X}}$. Then we follow this same Higgsing on the
fiber-base dual, $\mathfrak{X}'$ down to ${ \mathcal{X}'}$,  and we obtain
another $3d$ theory $\mathcal{T}_{{ \mathcal{X}}'}$ which we call the
$3d$ \emph{spectral dual} theory. 
The fiber-base invariance of the Higgsed
topological string partition function 
implies the equality of the holomorphic blocks ${\mathcal
  B}^{\alpha_0}_{\mathcal{T}_{\mathcal{X}}}= {\mathcal
  B}^{\alpha_0}_{\mathcal{T}_{\mathcal{X}'}}$ of the $3d$ spectral dual theories $\mathcal{T}_{\mathcal{X}}$ and $\mathcal{T}_{\mathcal{X}'}$.

In this paper we propose  two examples of spectral dual theories and together with the topological string construction, 
we support our proposals by providing purely field theory arguments. 
A third example of spectral duality has been recently discussed in~\cite{Nieri:2018pev}.

The first spectral dual pair we construct follows from the duality between
$FFT[SU(\Num)]\leftrightarrow\chk{T[SU(\Num)]}$, 
which can be understood as lying on the SE-NW diagonal
of the $T[SU(\Num)]$ diagram.
We denote with $FT[SU(\Num)]$
the $T[SU(\Num)]$ theory with an extra set of singlets which flip only the
moment map operator on the Higgs branch. Then, upon flipping the SE-NW $T[SU(\Num)]$ frames, 
we obtain the  spectral dual pair,
$FT[SU(\Num)]\leftrightarrow\chk{FT[SU(\Num)]}$.  Notice that
$FT[SU(\Num)]$ has been realised previously in~\cite{Zenkevich:2017ylb} as a defect theory 
in the square $(p,q)$-web with $(N+1)$ D5' and NS5 branes, 
and there it was also shown that indeed the equality
of the holomorphic blocks ${\mathcal B}^{\alpha_0}_ {FT[SU(\Num)]}=
{\mathcal B}^{\alpha_0}_{\chk{FT[SU(\Num)]}}$ follows via Higgsing
from the equality of the topological string partition functions for
the fiber-base dual diagrams.

The second spectral dual pair $\mathcal{T}\leftrightarrow
\mathcal{T}'$ is obtained within the $ABCD$ diagram by flipping the SE-NW diagonal $D\leftrightarrow B$. 
We discuss the operator map and check the equality of
the sphere partition functions. We then show how $\mathcal{T}$ and
$\mathcal{T'}$, can be realised as defect theories inside spectral
dual $5d$ theories and obtain their holomorphic blocks ${\mathcal
  B}^{\alpha_0}_{\mathcal{T}}$, ${\mathcal
  B}^{\alpha_0}_{\mathcal{T}'}$ by tuning the K\"ahler parameters in
the fiber-base dual CYs.  Again we prove that our 3d spectral duality 
descends from fiber-base duality in topological strings.

The main novelty of our construction is that it provides two completely independent and quantitative tools to check spectral dualities. Indeed, it is quite remarkable that both a field theory computation, i.e. the localized supersymmetric partition function, and the refined vertex on the topological string side exactly agree.

The spectral dual pairs we have constructed are related by various flips to mirror dual pairs,
and can be realised with brane setup related by various rotations to the $\mathcal{N}=4$ configuration 
relevant for mirror dualities.
It would be interesting to understand better the interplay between  these type of dualities.

Our construction can be generalized in a variety of ways. For example, 
we might consider more general Higgsing patterns in the toric diagrams, 
corresponding to more general vevs for the $T[SU(\Num)]$ tail, or even more interestingly,
corresponding to coupled $3d-5d$ systems in which the $5d$ theory is non trivial. 
As in the examples we have proposed, performing fiber-base duality 
on a generic Higgsing pattern will produce a new duality for the $3d-5d$ system.

In this paper we have focused on spectral duality, or fiber-base duality, 
which is just one element of the S-duality group of the $(p,q)$-web. 
It would be interesting to investigate the interplay between Higgsing and the action of the other elements. Some investigations along these lines have been proposed in \cite{Nieri:2018pev}.

\section{$T[SU(\Num)]$ dualities}\label{tsun_sec}

$T[SU(\Num)]$ is the $3d$ $\mathcal{N}=4$ quiver theory  arising from the study of $S$-duality 
and Dirichlet boundary conditions in four-dimensional $\mathcal{N}=4$ SYM \cite{Gaiotto:2008ak}:
\be\label{quiverT}
             \begin{tikzpicture}
		
		\def\ox{0}
		\def\oy{0}
		\def\radius{.3}
		\def\lunghBif{1.5}
		\def\yshift{.1}
		
		\foreach \intero in {0,1}   \draw[line width=.6pt]    (\ox+\intero*\lunghBif,\oy+\yshift) --  (\ox+\lunghBif+\intero*\lunghBif,\oy+\yshift) ;
		\foreach \intero in {0,1}   \draw[line width=.6pt]    (\ox+\intero*\lunghBif,\oy-\yshift) --  (\ox+\lunghBif+\intero*\lunghBif,\oy-\yshift) ;
		
		\draw[line width=.6pt,dashed]    (\ox+2*\lunghBif,\oy+\yshift) --  (\ox+0.6*\lunghBif+2*\lunghBif,\oy+\yshift) ;
		\draw[line width=.6pt,dashed]    (\ox+2*\lunghBif,\oy-\yshift) --  (\ox+0.6*\lunghBif+2*\lunghBif,\oy-\yshift) ;		
		
		\draw[line width=.6pt,dashed]    (\ox+3*\lunghBif,\oy+\yshift) --  (\ox+0.5*\lunghBif+3*\lunghBif,\oy+\yshift) ;
		\draw[line width=.6pt,dashed]    (\ox+3*\lunghBif,\oy-\yshift) --  (\ox+0.5*\lunghBif+3*\lunghBif,\oy-\yshift) ;	
	
		\draw[line width=.6pt]    (\ox+3.5*\lunghBif,\oy+\yshift) --  (\ox+4.5*\lunghBif,\oy+\yshift) ;	
		\draw[line width=.6pt]    (\ox+3.5*\lunghBif,\oy-\yshift) --  (\ox+4.5*\lunghBif,\oy-\yshift) ;	
		\foreach \intero in {1,2}   	\node        at  (\ox+\intero*\lunghBif,\oy+5*\yshift) 	[circle,inner sep=2mm,draw=black,fill=white,thick]   {}    ;
		\node	at (\ox+.05,\oy+5*\yshift) 	[circle,inner sep=2mm,draw=black,fill=white,thick]   {}    ;
		\foreach \intero in {2,3}           \node	at (\ox-\lunghBif+\intero*\lunghBif,\oy) 	[circle,inner sep=2mm,draw=black,fill=white,thick]   {$\intero$}    ;
		\node	at (\ox+.05,\oy) 	[circle,inner sep=2mm,draw=black,fill=white,thick]   {$1$}    ;

	         \node        at  (\ox+3.5*\lunghBif,\oy+5*\yshift) 	[circle,inner sep=2mm,draw=black,fill=white,thick]   {}    ;
		 \node	at (\ox+3.5*\lunghBif,\oy) 	[circle,inner sep=1.6mm,draw=black,fill=white,thick]   {$N$}    ;
		 \node	at (\ox+4.6*\lunghBif,\oy) 	[minimum height=.8cm, draw=black,fill=white,thick]   {$N+1$}   ;
		 		
	      \end{tikzpicture}
\ee
Each one of the $N$ round gauge nodes, labelled by its rank $k=1,\ldots,N$, is associated to a vector multiplet 
decomposed into an $\mathcal{N}=2$ vector multiplet and an adjoint chiral field $\Phi_{k}$, represented by a loop.
Bifundamental chiral fields $Q_{ab}$, and antichiral fields $\tilde{Q}_{\tilde{a}\tilde{b}}$ are represented by lines connecting 
adjacent nodes and  pair up into hypermultiplets\footnote{In our conventions the bifundamentals $Q^{(k,k+1)}_{ab}$ 
transform in the reps $\Box\otimes\overline{\Box}$ of $U(k)\times U(k+1)$, and the bifundamental 
$\tilde{Q}^{(k,k+1)}_{\tilde{a}\tilde{b}}$ transform in the reps $\Box\otimes\overline{\Box}$ of $U(k+1)\times U(k)$ }.
The $N+1$ rectangular node is ungauged. In the quiver rapresentation, the flavor note is what we call `head' of $T[SU(\Num)]$. 
In $\mathcal{N}=2$ notation the superpotential of the theory is 
\bea\label{superPini}
W^T[\Phi,\mathbb{Q}]\equiv \sum_{k=1}^N  \Tr_k \left[ \Phi_{k} \left( \Tr_{k+1} \mathbb{Q}^{(k,k+1)}-\Tr_{k-1} \mathbb{Q}^{(k-1,k)} \right) \right]
\eea
where we defined the matrix of bifundamentals $\mathbb{Q}^{(L,R)}=Q^{(L,R)}_{ab} \widetilde{Q}^{(L,R)}_{\tilde{a}\tilde{b}}$,
labelled by the pair $(L,R)$ attached to the link between a left ($L$) and a right ($R$) node. 
On the first node $\mathbb{Q}^{(0,1)}=0$. 
Traces $\Tr_k$ are taken in the adjoint of $U(k)$. 

The global symmetry of $T[SU(\Num)]$ is $SU(\Num)_{\mathrm{flavor}}\times SU(\Num)_{\mathrm{top}}$. 
The flavor symmetry $SU(\Num)_{\mathrm{flavor}}$ rotates the fundamental hypers at the end of the tail.
The non abelian $SU(\Num)_{\mathrm{top}}$ is an IR symmetry, and 
the UV Lagrangian only manifests a $U(1)_{\mathrm{top}}^N$ topological 
symmetry, coming from the dual photons on the gauge nodes.
For each Cartan in the flavor symmetry group and each $U(1)_{\mathrm{top}}$ 
we can turn on a real masses, $M_p$ and $T_p$, respectively. 

The R-symmetry of a $3d$ $\mathcal{N}=4$ theory is $SU(2)_C \times SU(2)_H$ with Cartans 
$U(1)_C\subset SU(2)_C$ and $ U(1)_H\subset SU(2)_H$. We will work with a family of $\mathcal{N}=2^*$ 
theories obtained by introducing a real mass parameter for the anti-diagonal combination $U(1)_A=C-H$ \cite{Tong:2000ky}.
We take the UV R-charge equal to the combination $R_0 = C + H$. In the IR, the R-symmetry 
can mix with other abelian symmetries, but since the topological symmetry is non-abelian, 
$R_0$ will only mix with $U(1)_A$. Thus we introduce a trial R-charge, defined by $R=C+H+\alpha (C-H)$ for some $\alpha\in \mathbb{R}$. 
For the bifundamental fields we find $R=\tfrac{1-\alpha}{2}\equiv r$, in agreement with the assignment $C=0$, $H=\tfrac{1}{2}$.
For the adjoint fields $R[\Phi_{k}]=2(1-r)$ iff the superpotential has R-charge $2$.
Notice also that $R[\Phi_{k}]=1+\alpha=2(1-r)$ is consistent with $C=1$ and $H=0$.
The exact value of $r$ can be fixed by F-extremization \cite{Jafferis:2010un}.

We define the gauge invariant $(N+1)\times (N+1)$ meson matrix:
\be\label{meson_def_ori}
\cQ_{ij}\equiv \Tr_N \mathbb{Q}^{(N,N+1)}, \qquad \qquad R[\cQ_{ij}]=2r\,.
\ee
The dynamics might impose additional relations on $\cQ_{ij}$, thus restricting the set of generators of the Higgs branch (HB) chiral ring. 
Classical relations follows from the F-terms, and for $T[SU(\Num)]$ the F-terms of the fields $\Phi_k$
imply that $\cQ$ is nilpotent \cite{Gaiotto:2008ak}. The argument goes as follows: 
$\cQ=\widetilde{Q}^{(N,N+1)}Q^{(N,N+1)}$ has rank at most $N$ by definition. Then, the F-term of $\Phi_N$ can be used to rewrite
\bea
\cQ^2&=&\widetilde{Q}^{(N,N+1)}Q^{(N,N+1)}\tilde{Q}^{(N,N+1)}Q^{(N,N+1)}\\
	&=&\big(\widetilde{Q}^{(N,N+1)}\widetilde{Q}^{(N-1,N)}\big)\big(Q^{(N-1,N)}Q^{(N,N+1)}\big)
\eea
which implies $\cQ^2$ has at most rank $N-1$. Iterating this computation we find that certainly $\cQ^{N+1}=0$. 
The Higgs branch is related to the nilpotent cone $\mathcal{N}$ for matrices in $SL(N+1,\mathbb{C})$. 
This space can be organized as the union of all the orbits $S\cdot\mathbb{J}_{\underline{\lambda}}\cdot S^{-1}$ 
where $S\in SL(\Num,\mathbb{C})$ and $\mathbb{J}$ is the Jordan form associated to a partition $\underline{\lambda}$ 
of $n$, see \cite{Cabrera:2016vvv} for a review on related topics.

The meson $\cQ_{}$ comes along with the moment map operator $\Pi^{\mathcal{Q}}$, which is better suited to 
describe global symmetries of the theory. Indeed, $\Pi^{\mathcal{Q}}$ is the half-BPS primary in a supermultiplet 
which contains conserved global currents. In our case, $\Pi^{\mathcal{Q}}$ is defined as 
\be
\Pi^{\mathcal{Q}}\equiv \mathcal{Q} -\frac{1}{N+1}\Tr\mathcal{Q}\ .
\ee

Coulomb branch (CB) operators can be obtained from  $\Tr\,\Phi_k$ and monopole operators $\Mon^{f_1\ldots f_N}$
carrying $f_i$ units of flux for the topological $U(1)$ on the $i$-th node.
The R-charge of a (BPS) monopole operator is determined by the R-charges of all the fermions $\psi$ of the theory by the formula:
\be
R[\Mon^{f_1,\ldots f_N}] = - \frac{1}{2} \sum_{\text{fermions }\psi} R[\psi] \, \big| \rho_\psi(f_1,\ldots f_N) \big| \;,
\ee 
where $\rho_\psi(f_1,\ldots f_N) $ is the monopole charge of $\psi$ \cite{Borokhov:2002cg,Gaiotto:2008ak,Benini:2011cma}.
\footnote{
Example: Consider a $U(N)$ theory with $2N$ flavors $Q, \widetilde Q$, an adjoint $\Phi\in U(N)$, and superpotential 
$\cW=\Phi Q \tilde Q$. The monopoles $\Mon^{\pm 1}$ have 
$R[\Mon^{\pm 1}]=2N(1-R_Q)+(N-1)(1-R_\Phi)-(N-1)$, which in our case becomes $R[\Mon^{\pm 1}]=2-2r$.} 
We find that monopole operators defined by a string of fluxes of the form $[0^n (\pm1)^m 0^p]$, 
where $0$ and $1$ are repeated with integer multiplicities $n$, $m$, and $p$ 
constrained by $n+m+p=N$, have the same R-charge of the adjoint fields, i.e. $R[\Phi_k]=2(1-r)$. 
These monopole operators are $N(N+1)$ and together with the $\Phi_{k=1,..N}$ 
can be arranged into a $(N+1)\times (N+1)$ matrix, analogous to the meson matrix.
For $N=3$ this matrix reads
\be\label{1monopolematrix}
\mathcal{M}_{ij}\equiv 
\left(\begin{array}{ccccccc}
0									&\ & \Mon^{[1,0,0]} 							&\ & \Mon^{[1,1,0]} 		&\ & \Mon^{[1,1,1]}	\\[.15cm]	
\Mon^{[\textrm{-}1,0,0]}      				&\ & 0									&\ & \Mon^{[0,1,0]} 		&\ & \Mon^{[0,1,1]}  \\[.15cm]	
\Mon^{[\textrm{-}1,\textrm{-}1,0]}			&\ &  	\Mon^{[0,\textrm{-}1,0]}				&\ & 	0				&\ & \Mon^{[0,0,1]} 	 \\[.15cm]		
\Mon^{[\textrm{-}1,\textrm{-}1,\textrm{-}1]}		&\ &  	\Mon^{[0,\textrm{-}1,\textrm{-}1]}		&\ &\Mon^{[0,0,\textrm{-}1]}				&\ & 0 
\end{array}\right)+ \sum_{i=1}^{3} \Tr\Phi_i \mathcal{D}_i
\ee
where $\mathcal{D}_i$ are traceless diagonal generators of $SU(\Num)_{\mathrm{top}}$. 
The generators of the CB chiral ring can be obtained from such an $\mathcal{M}_{ij}$ upon imposing further relations. 

In the rest of the paper we will refer to a matrix assembled as in \eqref{1monopolematrix}, 
as the monopole matrix of the theory under consideration.

The moment map $\Pi^\mathcal{Q}$ and the monopole matrix belong to the adjoint of $SU(\Num)$.

\subsection{Mirror Simmetry}\label{Mirror_dual_sec}

It is well known that $T[SU(\Num)]$ is self-dual under mirror symmetry \cite{Gaiotto:2008ak}. 
The dual theory, hereafter $\chk{T[SU(\Num)]}$, has quiver diagram
\be\label{quiverTcheck}
             \begin{tikzpicture}
		
		\def\ox{3}
		\def\oy{0}
		\def\radius{.3}
		\def\lunghBif{1.5}
		\def\yshift{.1}
		
		\foreach \intero in {1,2}   \draw[line width=.6pt]    (\ox+\intero*\lunghBif,\oy+\yshift) --  (\ox+\lunghBif+\intero*\lunghBif,\oy+\yshift) ;
		\foreach \intero in {1,2}   \draw[line width=.6pt]    (\ox+\intero*\lunghBif,\oy-\yshift) --  (\ox+\lunghBif+\intero*\lunghBif,\oy-\yshift) ;
		
		\draw[line width=.6pt,dashed]    (\ox-0.5*\lunghBif,\oy+\yshift) --  (\ox+0.2*\lunghBif+0*\lunghBif,\oy+\yshift) ;
		\draw[line width=.6pt,dashed]    (\ox-0.5*\lunghBif,\oy-\yshift) --  (\ox+0.2*\lunghBif+0*\lunghBif,\oy-\yshift) ;		
		
		\draw[line width=.6pt,dashed]    (\ox+0.4*\lunghBif,\oy+\yshift) --  (\ox+1*\lunghBif+0*\lunghBif,\oy+\yshift) ;
		\draw[line width=.6pt,dashed]    (\ox+0.4*\lunghBif,\oy-\yshift) --  (\ox+1*\lunghBif+0*\lunghBif,\oy-\yshift) ;	
	
		\draw[line width=.6pt]    (\ox-1.2*\lunghBif,\oy+\yshift) --  (\ox-0.5*\lunghBif,\oy+\yshift) ;	
		\draw[line width=.6pt]    (\ox-1.2*\lunghBif,\oy-\yshift) --  (\ox-0.5*\lunghBif,\oy-\yshift) ;	
		\foreach \intero in {1,2}   	\node        at  (\ox+\intero*\lunghBif,\oy+5*\yshift) 	[circle,inner sep=2mm,draw=black,fill=white,thick]   {}    ;
		\node        at  (\ox+2.95*\lunghBif,\oy+5*\yshift) 	[circle,inner sep=2mm,draw=black,fill=white,thick]   {}    ;
		\foreach \intero in {2,3}           \node	at (\ox+4*\lunghBif-\intero*\lunghBif,\oy) 	[circle,inner sep=2mm,draw=black,fill=white,thick]   {$\intero$}    ;
		\node	at (\ox+2.95*\lunghBif,\oy) 	[circle,inner sep=2mm,draw=black,fill=white,thick]   {$1$};

	         \node        at  (\ox-0.4*\lunghBif,\oy+5*\yshift) 	[circle,inner sep=2mm,draw=black,fill=white,thick]   {}    ;
		 \node	at (\ox-0.4*\lunghBif,\oy) 	[circle,inner sep=1.6mm,draw=black,fill=white,thick]   {$N$}    ;
		 \node	at (\ox-1.5*\lunghBif,\oy) 	[minimum height=0.8cm,draw=black,fill=white,thick]   {$N+1$}   ;
		 		
	      \end{tikzpicture}
\ee
and the same field content as $T[SU(\Num)]$. In $\chk{T[SU(\Num)]}$ we denote
the adjoint chirals by $\checkPhi_k$, the monopoles operators by ${\Non}_{\ f_1\ldots f_N}$, 
and the bifundamental fields by $P_{ab}$ and $\widetilde{P}_{\tilde{a}\tilde{b}}$.
The indexes $k$ and $f_1\ldots f_N$ have the same meaning as in $T[SU(\Num)]$.
We introduce the matrix $\mathbb{P}^{(L,R)}=P^{(L,R)}_{ab} \tilde{P}^{(L,R)}_{\tilde{a}\tilde{b}}$ 
for each pair of nodes $(L,R)$. Then the dual the superpotential reads
\be
W^{\chk{T}}=W^T[ \checkPhi, \mathbb{P} ] \,.
\ee

Mirror symmetry exchanges the Higgs and Coulomb branch. 
Therefore the bifundamental fields have now R-charge $R[{P}_{ab}]=1-r$. Consequently the monopole operators have $R$-charge $R[\Non_{\ ij}]=2r$. It follows from the superpotential that $R[\Omega_k]=2r$ for any $k$.

On the Higgs branch we define the meson $\mathcal{P}$ and its moment map $\Pi^{\mathcal{P}}$. The meson is
\be
\mathcal{P}_{ij}\equiv\Tr_N \mathbb{P}^{(N+1,N)}, \qquad \qquad R[\mathcal{P}_{ij}]=2-2r\,.
\ee

On the Coulomb branch we consider the monopole matrix $\mathcal{N}_{ij}$, which similarly to the previous section, is obtained from
$\Tr\checkPhi_k$ and from monopole operators with fluxes valued in $[0^n1^m0^p]$. For $N=3$ we have 
\be\label{1monopolematrix}
\mathcal{N}_{ij}\equiv 
\left(\begin{array}{ccccccc}
0									&\ & \Non^{[1,0,0]} 							&\ & \Non^{[1,1,0]} 		&\ & \Non^{[1,1,1]}	\\[.15cm]	
\Non^{[\textrm{-}1,0,0]}      				&\ & 0									&\ & \Non^{[0,1,0]} 		&\ & \Non^{[0,1,1]}  \\[.15cm]	
\Non^{[\textrm{-}1,\textrm{-}1,0]}			&\ &  	\Non^{[0,\textrm{-}1,0]}				&\ & 	0				&\ & \Non^{[0,0,1]} 	 \\[.15cm]		
\Non^{[\textrm{-}1,\textrm{-}1,\textrm{-}1]}		&\ &  	\Non^{[0,\textrm{-}1,\textrm{-}1]}		&\ &\Non^{[0,0,\textrm{-}1]}				&\ & 0 
\end{array}\right)+ \sum_{i=1}^{3} \Tr\checkPhi_i \mathcal{D}_i
\ee
where again $\mathcal{D}_i$ are traceless diagonal generators of $SU(\Num)$. 
%
%
%
%

Mirror symmetry exchanges
\be
\cM_{ij}\leftrightarrow \Pi^{\mathcal{P}}_{ij}\qquad;\qquad \Pi^{\mathcal{Q}}_{ij}\leftrightarrow \cN_{\ ij}\,.
\ee
and therefore HB and CB. 

The fact that $T[SU(N+1)$ is self-dual under mirror symmetry can be neatly derived from the IIB brane engineering of the $T[SU(N+1)]$.
More precisely, $T[SU(N+1)]$ can be understood as the low energy theory of a system of D3 branes suspended between 
$(N+1)$ D5 and NS5 branes \cite{Hanany:1996ie}. The brane configuration is summarized in Table~\ref{tabe:1} and goes as follows. 
The NS5 extend along directions 012789, and the D5 branes
along directions 012456. The NS5 and D5 branes
are separated in the third direction, where  $(N+1)$ D3 branes are
stretched in between, so that each NS5 brane is connected to a distinct D5 brane. 
These D3 branes extend along directions 0123, but since they are bounded in the third direction by D5 and NS5 branes, 
the low energy dynamics on their wordlvolume is three-dimensional. In fact it is precisely the
$T[SU(N+1)]$ theory. 
The $R$-symmetry group factors $SU(2)_C$ and $ SU(2)_H$ correspond to the rotation
symmetry of the NS5 and D5 branes in the directions transverse to the
D3 branes, i.e.\ to $SO(3)_{456}$ and $SO(3)_{789}$
respectively. 
The action of mirror symmetry, which exchanges CB and
HB, is precisely that of IIB $S$-duality, which exchanges the NS5 and D5
branes (leaving the system invariant). 
Equivalently, one can think of this transformation as the
exchange of the 456 and 789 directions.

\begin{table}[h]
  \label{tabe:1}
  \centering
  \begin{tabular}{c|c|c|c|c|c|c|c|c|c|c}
      & $0$ & $1$ & $2$ & $3$ & $4$ & $5$ & $6$ & $7$ & $8$ & $9$\\
      \hline
      NS5 & $-$ & $-$ & $-$ &  &  &  &  & $-$ & $-$ & $-$\\
      D5 & $-$ & $-$ & $-$ &  & $-$ & $-$ & $-$ &  &  & \\
      D3 & $-$ & $-$ & $-$ & $-$ &  &  &  &  &  &  
\end{tabular}
\caption{The brane setup giving rise to the 3d $\mathcal{N}=4$ $T[SU(N+1)]$ gauge theory.}
\end{table}

\subsection{Flip-Flip duals} \label{flipflip_sec}

In this section we propose new duals for $T[SU(\Num)]$  and its mirror. We name them flip-flip dualities for reasons that will become soon clear.


Let us begin by describing the flip-flip dual of  $T[SU(\Num)]$, which we denote by $FFT[SU(\Num)]$. 
This theory has the content of $T[SU(\Num)]$ plus two extra sets of fields, the {\it flipping fields}. 
We represent $FFT[SU(\Num)]$ by the quiver
\be\label{quiverFFT}
             \begin{tikzpicture}
		
		\def\ox{0}
		\def\oy{0}
		\def\radius{.3}
		\def\lunghBif{1.5}
		\def\yshift{.1}
		
		\foreach \intero in {0,1}   \draw[line width=.6pt]    (\ox+\intero*\lunghBif,\oy+\yshift) --  (\ox+\lunghBif+\intero*\lunghBif,\oy+\yshift) ;
		\foreach \intero in {0,1}   \draw[line width=.6pt]    (\ox+\intero*\lunghBif,\oy-\yshift) --  (\ox+\lunghBif+\intero*\lunghBif,\oy-\yshift) ;
		
		\draw[line width=.6pt,dashed]    (\ox+2*\lunghBif,\oy+\yshift) --  (\ox+0.6*\lunghBif+2*\lunghBif,\oy+\yshift) ;
		\draw[line width=.6pt,dashed]    (\ox+2*\lunghBif,\oy-\yshift) --  (\ox+0.6*\lunghBif+2*\lunghBif,\oy-\yshift) ;		
		
		\draw[line width=.6pt,dashed]    (\ox+3*\lunghBif,\oy+\yshift) --  (\ox+0.5*\lunghBif+3*\lunghBif,\oy+\yshift) ;
		\draw[line width=.6pt,dashed]    (\ox+3*\lunghBif,\oy-\yshift) --  (\ox+0.5*\lunghBif+3*\lunghBif,\oy-\yshift) ;	
	
		\draw[line width=.6pt]    (\ox+3.5*\lunghBif,\oy+\yshift) --  (\ox+4.5*\lunghBif,\oy+\yshift) ;	
		\draw[line width=.6pt]    (\ox+3.5*\lunghBif,\oy-\yshift) --  (\ox+4.5*\lunghBif,\oy-\yshift) ;	
		\foreach \intero in {1,2}   	\node        at  (\ox+\intero*\lunghBif,\oy+5*\yshift) 	[circle,inner sep=2mm,draw=black,fill=white,thick]   {}    ;
		\node	at (\ox+.05,\oy+5*\yshift) 	[circle,inner sep=2mm,draw=black,fill=white,thick]   {}    ;
		\node        at  (\ox+5.2*\lunghBif,\oy)										[circle,inner sep=2.75mm,draw=black,fill=white,thick]   {}    ;	
		\foreach \intero in {2,3}           \node	at (\ox-\lunghBif+\intero*\lunghBif,\oy) 	[circle,inner sep=2mm,draw=black,fill=white,thick]   {$\intero$}    ;
		\node	at (\ox+.05,\oy) 	[circle,inner sep=2mm,draw=black,fill=white,thick]   {$1$}    ;

	         \node        at  (\ox+3.5*\lunghBif,\oy+5*\yshift) 	[circle,inner sep=2mm,draw=black,fill=white,thick]   {}    ;
		 \node	at (\ox+3.5*\lunghBif,\oy) 	[circle,inner sep=1.6mm,draw=black,fill=white,thick]   {$N$}    ;
		 \node	at (\ox+4.6*\lunghBif,\oy) 	[minimum height=.8cm, draw=black,fill=white,thick]   {$N+1$}   ;
		 
\end{tikzpicture}
\ee
where the horizontal loops attached on the flavor node indicate the addition of flipping fields.
We have adjoint chiral fields $\FFPhi_k$,  bifundamental fields $R_{ab}$ and $\widetilde{R}_{\tilde{a}\tilde{b}}$, with $R[R_{ab}]=R[\widetilde{R}_{\tilde{a}\tilde{b}}]=1-r$, and
monopoles operators. Out of them, we define the meson $\mathcal{R}_{ij}$, its moment map $\Pi^{\mathcal{R}}_{ij}$, and the monopole matrix $\m_{\,ij}$. 
The flipping fields are elementary fields, singlet of gauge groups, and transform respectively in the adjoint of $SU(\Num)_{\mathrm{flavor}}$ and $SU(\Num)_{top.}$, in particular they are traceless. 
These are denoted by $\Flip^{\mathcal{R}}_{ij} $ and  $\Flip^{\m}_{ij}$, since they will couple to $\mathcal{R}_{ij}$ and $\m_{\,ij}$ in the superpotential:
\be\label{superP_FFT}
W^{FFT}=W^{T}[ \FFPhi, \mathbb{R} ] - \Flip^{\mathcal{R}}_{ij} \,\Pi^{\mathcal{R}}_{ij}- {\m}_{\,ij} \Flip^{\m}_{ij} \,.
\ee  
From $W^{FFT}$ we deduce the R-charge assigment,  $R[\Flip^{\m}_{ij}]=2-2r$ and  $R[\Flip^{\mathcal{R}}_{ij}]=2r$.
We then discover that the flip-flip duality between $T[SU(\Num)]$ and $FFT[SU(\Num)]$ maps:
\be
\Pi^{\cQ}_{} \leftrightarrow \Flip^{\mathcal{R}}_{} \qquad;\qquad \mathcal{M}_{}\leftrightarrow \Flip^{\m}_{} \,.
\ee

The F-terms of $\Flip^{\mathcal{R}}_{ij} $ and  $\Flip^{\m}_{ij}$ imply $\Pi^{\mathcal{R}}_{ij}=0$
and ${\m}_{\,ij}=0$. As a result, the HB and CB will now be described by $\Flip^{\mathcal{R}}_{ij}$ and $\Flip^{\m}_{ij}$, respectively. 
In this sense, $\Flip^{\mathcal{R}}_{ij}$ is the flip of the moment map, i.e. the meson $\mathcal{R}_{ij}$, and $\Flip^{\m}_{ij}$ that of the monopole matrix. 

It is interesting to look at the description of the HB.
The F-terms of the bifundamentals $R^{(k,k+1)}$ and $\widetilde{R}^{(k,k+1)}$ imply the equations
\bea
\widetilde{R}^{(k,k+1)}\Theta_{k}   =   \Theta_{k+1}   \widetilde{R}^{(k,k+1)} \label{HB_flip1}\\
\Theta_k R^{(k,k+1)}=R^{(k,k+1)}\Theta_{k+1}     \label{HB_flip2}
\eea
where we defined $\Theta_{N+1}\equiv\Flip^{\mathcal{R}}$ so to have a uniform notation in \eqref{HB_flip1} and \eqref{HB_flip2}.
The F-terms of the diagonal component of the monopole matrix give $\Tr\Theta_k=0$ for all $k\le N$, in particular $\Theta_1=0$.
Furthermore, it is always possible to use $SU(N)_{gauge}\times SU(\Num)_{\mathrm{flavor}}$ to put one of the bifundamentals on the last link in a diagonal form. 
For concreteness we take, 
\be\label{gauge_bif_flipflip}
\langle \widetilde{R}^{(N,N+1)}\rangle=\left[\begin{array}{cccc}   v_1 & 0    & \ldots & 0 \\
												0     & v_2 & \ldots & 0 \\
												0     & 0     & \ldots & 0\\
												0     & 0 & \ldots & v_N \\
												0     & 0 & \ldots & 0 \end{array}\right]\,.
\ee 
with arbitrary $v_i$. The constraint $\Pi^{\mathcal{R}}_{ij}=0$ trivializes \eqref{HB_flip2}.  
Let us discuss the case $N=1$ to start with. 
From  \eqref{HB_flip1} we find
\be
0=\widetilde{R}^{(1,2)}\Theta_{1} = \Theta_{2} \left[\begin{array}{c}v_1 \\ 0 \end{array}\right]\,,
\ee
therefore $\widetilde{R}^{(1,2)}$ is in the kernel of $\Theta_{2}$. 
The flipping fields $\Theta_2$ are in the adjoint of $SU(2)$ and we shall take $\Tr\Theta_2=0$. 
It follows that a traceless matrix in the adjoint with a one dimensional kernel can be put into the form, 
\be\label{exemp_2}
\Theta_2=\left[\begin{array}{cc} 0 & \ \theta_{2} \\ 0 & 0 \end{array}\right].
\ee 
i.e. $\Theta_2$ is nilpotent. More in general, we can use a recursive 
argument to show that $\Theta_{N+1}$ can be taken to be nilpotent.  
So let us assume that $\Theta_N$ is nilpotent, and  
consider the matrix $R^{(N,N+1)}_{aux}$ such that 
$R^{(N,N+1)}_{aux} \langle \widetilde{R}^{(N,N+1)}\rangle= \mathbb{I}_{N\times N}$. 
This matrix can be explicitly constructed in the gauge \eqref{gauge_bif_flipflip}.
Then \eqref{HB_flip1} becomes
\be\label{aux_nilpontent}
\Theta_N=R^{(N,N+1)}_{aux}\Theta_{N+1} \langle  \widetilde{R}^{(N,N+1)} \rangle\ .
\ee
Considering $\Theta_N$ is in its Jordan Form we introduce a basis $\{\vec{w}_i\}$ such that: 
\bea
\Theta_N \vec{w}_1&=&0\nn\\
\Theta_N \vec{w}_i&=&\sum_{j<i} c^j_i \vec{w}_j\qquad i\ge 2
\eea
for given coefficients $c^j_i$ which depend on the partition associated to $\Theta_{N}$ as nilpotent matrix. 
Equations \eqref{aux_nilpontent} now imply the relations
\bea
R^{(N,N+1)}_{aux}\Theta_{N+1} \langle  \widetilde{R}^{(N,N+1)} w_1 \rangle  &=& 0 \nn\\
R^{(N,N+1)}_{aux}  \Theta_{N+1} \langle  \widetilde{R}^{(N,N+1)}  w_i  \rangle &=&\sum_{j<i} c^j_i w_{j}   .
\label{all_equa_nilp}
\eea
Since $\vec{w}_{i=1,..N}$ is a basis, the span of $\vec{u}_i=\langle  \widetilde{R}^{(N,N+1)} \vec{w}_i\rangle $ 
is by construction an $N$-dimensional subspace in $N+1$ dimensions. 
The solution of \eqref{all_equa_nilp} is then 
\bea
\Theta_{N+1} \langle  \widetilde{R}^{(N,N+1)} w_1 \rangle &=& \theta_2 K \nn\\
\Theta_{N+1} \langle  \widetilde{R}^{(N,N+1)}  w_i \rangle&=& \sum_{j<i} c^j_i  \langle  \widetilde{R}^{(N,N+1)}  w_{j}  \rangle + \theta_{i+1} K \qquad i\ge 2
\label{solu_Flip_nilp}
\eea
where $K$ parametrizes the one dimensional kernel of $R^{(N,N+1)}_{aux}$, and the coefficients $\theta_i$ are arbitrary. 
It is straightforward to plug \eqref{solu_Flip_nilp} back into \eqref{all_equa_nilp} and check that the equations are satisfied by using $R^{(N,N+1)}_{aux} \langle \widetilde{R}^{(N,N+1)}\rangle= \mathbb{I}_{N\times N}$. 
Moreover, this relation implies that $K$ and the set of vectors $\{\vec{u}_i\}$ are independent, and therefore we can use them to span a basis in $N+1$ dimensions. 
Thus we fix completely $\Theta_{N+1}$ by specifying its action on such an basis, i.e. by adding  $\Theta_{N+1} K$ to the list in \eqref{solu_Flip_nilp}. The nilpotent solution is given by 
\bea
\Theta_{N+1} K &=& 0 \nn\\
\Theta_{N+1} \langle  \widetilde{R}^{(N,N+1)} w_1 \rangle &=& \theta_{2} K \nn\\
\Theta_{N+1} \langle  \widetilde{R}^{(N,N+1)}  w_i \rangle&=& \sum_{j<i} c^j_i  \langle  \widetilde{R}^{(N,N+1)}  w_{j}  \rangle + \theta_{i+1} K \qquad i\ge 2
\eea
where the $\theta_{j}$ play a role analogous to $\theta_{2}$ appearing in \eqref{exemp_2}. 
%

The outcome of our computation is interesting for two reasons: 
On one hand we obtained nilpotent solutions for the vev of the flipping fields, 
which in turn supports our duality, i.e. our identification $\Pi^{\cQ}_{} \leftrightarrow \Flip^{\mathcal{R}}_{}$. 
On the other hand, this nilpotency condition on the flipping fields shows up in a totally opposite way 
compared to the case of the meson $\cQ$ in $T[SU(\Num)]$.
As reviewed in the Introduction, in order to show that $\cQ$ is nilpotent in $T[SU(\Num)]$ we used the $F$-term 
constraints starting from the head of the tail back to the first gauge node. For the flipping fields, instead, we used 
$F$-term constraints recursively from the first gauge node up to the head of $FFT[SU(\Num)]$.
We will have more to say about this in Section \ref{nilpot_higg}.

In the case of $N=1$, our duality relates $T[SU(2)]$ to a $U(1)$ theory with two flavors and various singlets. 
This case can be understood as a version of the Aharony duality, as we now argue.
Let us recall that Aharony duality maps $\mathcal{N}=2$ SQED  
theory with two electric flavors $(Q_i,\tilde Q_j)$ and no superpotential, to an abelian theory 
with two magnetic flavors  $(q_i,\tilde q_j)$, and extra singlets $M_{ij}$ and $S^\pm$. 
The magnetic superpotential is non trivial: $W_{magn.}= \sum_{ij} M_{ij}\, q_i\tilde q_j+  V^-S^++V^+S^-$. 
In our language, $M_{ij}$ and $S^\pm$ are "flipping" fields for the magnetic mesons $q_i\tilde q_j$, 
and for the dual monopoles $V^\pm$, respectively. 
Notice that $M_{ij}$ belongs to the adjoint of $U(2)$, so it is not yet our flipping field.
In order to get $T[SU(2)]$ out of the electric side of Aharony duality, 
we introduce an extra singlet $\phi$, and we add a cubic superpotential  
of the form $W_{el.}=\phi \sum_i Q_i \tilde Q_i$. 
This is indeed the tail superpotential for $T[SU(2)]$. 
Adding a corresponding singlet field $\phi'$ also on the magneric side, Aharony duality 
maps our deformation to the mass term $\phi' \Tr(M)$.
Integrating out these two fields in the full magnetic superpotential, we obtain:
$$W_{magn.}'=\left(M -\tfrac{\Tr(M)}{2}\mathbb{I}\right)_{ij}\,q_i\tilde q_j+  V^-S^++V^+S^-.$$ 
At this point the meson $q_i \tilde{q}_j$ can be replaced by its moment map without changing $W_{magn.}'$. Then, 
$W_{magn.}'$ will be precisely what turns out to be the superpotential of $FFT[SU(2)]$. The expression of $W^{FFT}$ in this case is, 
\bea
W^{FFT}&=&W^{T}[ \FFPhi, \mathbb{R} ] -\mathcal{R}_{ij}  \Flip^{\mathcal{R}}_{ij} - {\m}_{\,ij} \Flip^{\m}_{ij}\nn\\
&=&\theta \sum_i r_i\tilde r_i-\Flip^{\mathcal{R}}_{ij} \Pi^{\mathcal{R}}_{ij}   -\m^+\Flip^{\m}_{+}-\m^-\Flip^{\m}_{-}-\theta \Flip^{\m}_3,
\eea
where $\theta \Flip^{\m}_3$ is the coupling due to the $\sigma_3$ 
generator in \eqref{1monopolematrix}. Similarly for $\m^{\pm}\Flip^{\m}_{\pm}$. 
Integrating out $\Flip^{\m}_3$, we recover $W_{magn.}'$ upon a trivial field redefinition.

The Flip-Flip duality on the mirror side works in a similar fashion: 
The starting point is $\chk{T[SU(\Num)]}$ and its quiver diagram \eqref{quiverTcheck}. 
The quiver diagram of $\chk{FFT[SU(\Num)]}$ is essentially \eqref{quiverTcheck}, 
except for the flavor node on which the new flipping fields are attached. 
On $\chk{FFT[SU(\Num)]}$ we will use the following notation:
${\Psi}_k$ for the adjoint chirals, $S_{ab}$ and $\widetilde{S}_{\tilde{a}\tilde{b}}$ for the bifundamental fields, with $R[S_{ab}]=R[\widetilde{S}_{\tilde{a}\tilde{b}}]=r$, $\mathcal{S}_{ij}$  for the mesons and $\n_{\,ij}$ for the monopole matrix. The flipping fields are denoted by $\Flip^{\mathcal{S}}_{ij}$ and $\Flip^{\n}_{ij}$. 
The superpotential is
\be
W^{\chk{FFT}}=W^T[ {\Psi}, \mathbb{S} ] - \Flip^{\mathcal{S}}_{ij}\, \Pi^{\mathcal{S}}_{ij}  - {\n}_{\,ij} \Flip^{\n}_{ij} \,,
\ee
from which we read the R-charges $R[\Flip^{\n}_{ij}]=2r$ and $R[\Flip^{\mathcal{S}}_{ij}]=2-2r$. 
According to the Flip-Flip duality between $\chk{T[SU(\Num)]}$ and $\chk{FFT[SU(\Num)]}$ the operators are mapped as
\be
\Pi^{\mathcal{P}}_{} \leftrightarrow \Flip^{\mathcal{S}}_{} \qquad;\qquad \mathcal{N}_{}\leftrightarrow \Flip^{\n}_{} \,.
\ee

\subsection{A commutative diagram}
We can represent our four dualities through the following commutative diagram:
\be
\label{tffweb}
             \begin{tikzpicture}
             \def\ptAx{-2 }
             \def\ptAy{.2}
             \def\lato {5}
             \def\latoA{3.5}
          
             \def\spazio{.3}

             \draw (\ptAx-.5,\ptAy+.075) --  (\ptAx+.6,\ptAy+.075) ;
             \draw (\ptAx-.5,\ptAy-.075) --  (\ptAx+.6,\ptAy-.075) ;
	    \draw (\ptAx-2.5,\ptAy+.075) --  (\ptAx-1,\ptAy+.075) ;
             \draw (\ptAx-2.5,\ptAy-.075) --  (\ptAx-1,\ptAy-.075) ;
 	    
	    \filldraw[fill=white,draw=black] (\ptAx,\ptAy+.3) circle (.25cm);
             \filldraw[fill=white,draw=black] (\ptAx,\ptAy) circle (.35cm);

             \filldraw[fill=white,draw=black] (\ptAx-1.5,\ptAy+.3) circle (.25cm);
             \filldraw[fill=white,draw=black] (\ptAx-1.5,\ptAy) circle (.35cm);
                   
             \draw (\ptAx-.72,\ptAy) node[font=\footnotesize] {\ldots};
            
             \filldraw[fill=white,draw=black] (\ptAx-2.5,\ptAy+.3) circle (.25cm);
             \filldraw[fill=white,draw=black] (\ptAx-2.5,\ptAy) circle (.35cm);

             \filldraw[fill=white,draw=black] (\ptAx+.6,\ptAy-.3) rectangle (\ptAx+1.4,\ptAy+.3);
             
             \draw (\ptAx-2.5,\ptAy) node[font=\footnotesize] {$1$};
             \draw (\ptAx-1.5,\ptAy) node[font=\footnotesize] {$2$};
             \draw (\ptAx,\ptAy) node[font=\footnotesize] {$N$};
             \draw (\ptAx+1,\ptAy) node[font=\footnotesize] {$N$+1};

             \draw (\ptAx+\lato-.8,\ptAy+.075) --  (\ptAx+\lato+1.1,\ptAy+.075) ;
             \draw (\ptAx+\lato-.8,\ptAy-.075) --  (\ptAx+\lato+1.1,\ptAy-.075) ;            
             \draw (\ptAx+\lato+1.6,\ptAy+.075) --  (\ptAx+\lato+.6+2.5,\ptAy+.075) ;
             \draw (\ptAx+\lato+1.6,\ptAy-.075) --  (\ptAx+\lato+.6+2.5,\ptAy-.075) ;
             \filldraw[fill=white,draw=black] (\ptAx+\lato+.6,\ptAy+.3) circle (.25cm);
             \filldraw[fill=white,draw=black] (\ptAx+\lato+.6,\ptAy) circle (.35cm);
                       
              \filldraw[fill=white,draw=black] (\ptAx+\lato+.6+1.5,\ptAy+.3) circle (.25cm);
              \filldraw[fill=white,draw=black] (\ptAx+\lato+.6+1.5,\ptAy) circle (.35cm);
                 
              \draw (\ptAx+\lato+1.35,\ptAy) node[font=\footnotesize] {\ldots};
            
             \filldraw[fill=white,draw=black] (\ptAx+\lato+.6+2.5,\ptAy+.3) circle (.25cm);
              \filldraw[fill=white,draw=black] (\ptAx+\lato+.6+2.5,\ptAy) circle (.35cm);
                 
             \filldraw[fill=white,draw=black] (\ptAx+\lato-.8,\ptAy-.3) rectangle (\ptAx+\lato,\ptAy+.3);
             
             \draw (\ptAx+\lato+.6+1.5,\ptAy) node[font=\footnotesize] {$2$};
             \draw (\ptAx+\lato+.6+2.5,\ptAy) node[font=\footnotesize] {$1$};
             \draw (\ptAx+\lato+.6,\ptAy) node[font=\footnotesize] {$N$};
             \draw (\ptAx+\lato-.4,\ptAy) node[font=\footnotesize] {$N$+1};

             \draw (\ptAx-.5,\ptAy+.075-\latoA) --  (\ptAx+.6,\ptAy+.075-\latoA) ;
             \draw (\ptAx-.5,\ptAy-.075-\latoA) --  (\ptAx+.6,\ptAy-.075-\latoA) ;
	    \draw (\ptAx-2.5,\ptAy+.075-\latoA) --  (\ptAx-1,\ptAy+.075-\latoA) ;
             \draw (\ptAx-2.5,\ptAy-.075-\latoA) --  (\ptAx-1,\ptAy-.075-\latoA) ;
 	    
	    \filldraw[fill=white,draw=black] (\ptAx,\ptAy+.3-\latoA) circle (.25cm);
             \filldraw[fill=white,draw=black] (\ptAx,\ptAy-\latoA) circle (.35cm);

             \filldraw[fill=white,draw=black] (\ptAx-1.5,\ptAy+.3-\latoA) circle (.25cm);
             \filldraw[fill=white,draw=black] (\ptAx-1.5,\ptAy-\latoA) circle (.35cm);
                   
             \draw (\ptAx-.72,\ptAy-\latoA) node[font=\footnotesize] {\ldots};
            
             \filldraw[fill=white,draw=black] (\ptAx-2.5,\ptAy+.3-\latoA) circle (.25cm);
             \filldraw[fill=white,draw=black] (\ptAx-2.5,\ptAy-\latoA) circle (.35cm);

                  \filldraw[fill=white,draw=blue] (\ptAx+1.55,\ptAy-\latoA) circle (.25cm);
                   
             \filldraw[fill=white,draw=black] (\ptAx+.6,\ptAy-.3-\latoA) rectangle (\ptAx+1.4,\ptAy+.3-\latoA);
             
             \draw (\ptAx-2.5,\ptAy-\latoA) node[font=\footnotesize] {$1$};
             \draw (\ptAx-1.5,\ptAy-\latoA) node[font=\footnotesize] {$2$};
             \draw (\ptAx,\ptAy-\latoA) node[font=\footnotesize] {$N$};
             \draw (\ptAx+1,\ptAy-\latoA) node[font=\footnotesize] {$N$+1};

             \draw (\ptAx+\lato-.8,\ptAy+.075-\latoA) --  (\ptAx+\lato+1.1,\ptAy+.075-\latoA) ;
             \draw (\ptAx+\lato-.8,\ptAy-.075-\latoA) --  (\ptAx+\lato+1.1,\ptAy-.075-\latoA) ;            
             \draw (\ptAx+\lato+1.6,\ptAy+.075-\latoA) --  (\ptAx+\lato+.6+2.5,\ptAy+.075-\latoA) ;
             \draw (\ptAx+\lato+1.6,\ptAy-.075-\latoA) --  (\ptAx+\lato+.6+2.5,\ptAy-.075-\latoA) ;
             \filldraw[fill=white,draw=black] (\ptAx+\lato+.6,\ptAy+.3-\latoA) circle (.25cm);
             \filldraw[fill=white,draw=black] (\ptAx+\lato+.6,\ptAy-\latoA) circle (.35cm);
                       
              \filldraw[fill=white,draw=black] (\ptAx+\lato+.6+1.5,\ptAy+.3-\latoA) circle (.25cm);
              \filldraw[fill=white,draw=black] (\ptAx+\lato+.6+1.5,\ptAy-\latoA) circle (.35cm);
                 
              \draw (\ptAx+\lato+1.35,\ptAy-\latoA) node[font=\footnotesize] {\ldots};
            
             \filldraw[fill=white,draw=black] (\ptAx+\lato+.6+2.5,\ptAy+.3-\latoA) circle (.25cm);
              \filldraw[fill=white,draw=black] (\ptAx+\lato+.6+2.5,\ptAy-\latoA) circle (.35cm);

                   \filldraw[fill=white,draw=red] (\ptAx+\lato-.95,\ptAy-\latoA) circle (.25cm);
                 
             \filldraw[fill=white,draw=black] (\ptAx+\lato-.8,\ptAy-.3-\latoA) rectangle (\ptAx+\lato,\ptAy+.3-\latoA);
             
             \draw (\ptAx+\lato+.6+1.5,\ptAy-\latoA) node[font=\footnotesize] {$2$};
             \draw (\ptAx+\lato+.6+2.5,\ptAy-\latoA) node[font=\footnotesize] {$1$};
             \draw (\ptAx+\lato+.6,\ptAy-\latoA) node[font=\footnotesize] {$N$};
             \draw (\ptAx+\lato-.4,\ptAy-\latoA) node[font=\footnotesize] {$N$+1};

             \def\ptBx{-1.5}
             \def\ptBy{+1}
             \def\lato {5}
             \def\hor{7}
             \def\spazio{.2}
             
       
         \draw[line width=.6pt,<->]  (\ptBx-1.2,\ptBy-1.2) -- (\ptBx-1.2,\ptBy-\lato+1.2);
         \draw (\ptBx-1.2-.5,\ptBy-1.7) node[font=\footnotesize,blue] {$\Pi^{\mathcal{Q}}$};
         \draw (\ptBx-1.2+.4,\ptBy-1.7) node[font=\footnotesize,red] {$\mathcal{M}$};
         
         \draw (\ptBx-1.2-.5,\ptBy-1.7-1.5) node[font=\footnotesize,blue] {$F^{\mathcal{R}}$};
         \draw (\ptBx-1.2+.4,\ptBy-1.7-1.53) node[font=\footnotesize,red] {$F^{\m}$};

  	 \draw[line width=.6pt,<->]  (\ptBx+\hor-1.2,\ptBy-1.2) -- (\ptBx+\hor-1.2,\ptBy-\lato+1.2);
         \draw (\ptBx+\hor-1.2-.5,\ptBy-1.7) node[font=\footnotesize,red] {$\Pi^{\mathcal{P}}$};
         \draw (\ptBx+\hor-1.2+.4,\ptBy-1.7) node[font=\footnotesize,blue] {$\mathcal{N}$};
         
         \draw (\ptBx+\hor-1.2-.5,\ptBy-1.7-1.5) node[font=\footnotesize,red] {$F^{\mathcal{S}}$};
         \draw (\ptBx+\hor-1.2+.4,\ptBy-1.7-1.53) node[font=\footnotesize,blue] {$F^{\n}$};

             \draw[line width=.6pt,<->]  (\ptBx+3*\spazio,\ptBy) -- (\ptBx+\lato-4*\spazio,\ptBy);
             \draw[line width=.6pt,<->]  (\ptBx+3*\spazio,\ptBy-\lato) -- (\ptBx+\lato-4*\spazio,\ptBy-\lato);

             \node[scale=.95] at (\ptBx-5.5*\spazio,\ptBy+0.5*\spazio) {$T[SU(\Num)]$};
             \node[scale=.95] at (\ptBx-6.8*\spazio,\ptBy-\lato-0.5*\spazio) {$FFT[SU(\Num)]$};
             \node[scale=.95] at (\ptBx+\lato+5.5*\spazio,\ptBy+0.45*\spazio) {$\chk{T[SU(\Num)]}$};
	     \node[scale=.95] at (\ptBx+\lato+7*\spazio,\ptBy-\lato-0.5*\spazio) {$FF\chk{T[SU(\Num)]}$};
             
             \end{tikzpicture}        
\ee

Horizontal arrows connect mirror dual theories while vertical arrows connect flip-flip dual theories.

We stress an important property of the commutative diagram: If we turn off real axial mass deformations, 
both ${T[SU(\Num)]}$ and $\chk{T[SU(\Num)]}$ are strictly $\mathcal{N}=4$ theories, and our duality web implies that both
${FFT[SU(\Num)]}$ and $\chk{FFT[SU(\Num)]}$ have to acquire an emergent  $\mathcal{N}=4$ symmetry in the
IR (even though their UV superpotentials preserve only $\mathcal{N}=2$).
In section \ref{difo} we provide further evidence about the duality web hence of the emergent $\mathcal{N}=4$ by showing that 
the partition functions of  the four dual theories are all equal, as function of the fugacities 
for the global symmetries. 
A first indication of this fact comes from F-extremization \cite{Jafferis:2010un}. 
Indeed, when we extremize the partition functions we set to zero the fugacities for the non-abelian symmetries, since these can't mix with the $R$-charge. 
But, as we will see later, if we turn-off the non-abelian fugacities, the contribution of the two sets of flipping fields cancel-out, hence the extremal $R$-charges for ${FFT[SU(\Num)]}$ 
and $T[SU(\Num)]$ are the both equal to $1/2$, which is the $\mathcal{N}=4$ value.

\section{Deformations of the commutative diagram}\label{defsec}

In this section we consider a certain monopole deformation of $T[SU(\Num)]$ and follow its RG-flow across the commutative diagram. 
This computation offers an interesting and novel consistency check about the (mother) $T[SU(\Num)]$ commutative diagram, and produces
another set of dual theories, named $ABCD$, themselves organized as a (daughter)
commutative diagram. The final picture is presented in section \ref{ABCD_sec} and summarized as follows:\\[.1cm] 
\be
             \begin{tikzpicture}
             \def\ptAx{0}
             \def\ptAy{0}
             \def\lato {2}
             \def\spazio{.3}
       
             \draw[line width=.6pt,<->]  (\ptAx+\spazio,\ptAy) -- (\ptAx+\lato-\spazio,\ptAy);
             \draw[line width=.6pt,<->]  (\ptAx,\ptAy-\spazio) -- (\ptAx,\ptAy-\lato+\spazio);
             \draw[line width=.6pt,<->]  (\ptAx+\lato,\ptAy-\spazio) -- (\ptAx+\lato,\ptAy-\lato+\spazio);
             \draw[line width=.6pt,<->]  (\ptAx+\spazio,\ptAy-\lato) -- (\ptAx+\lato-\spazio,\ptAy-\lato);

             \node at (\ptAx,\ptAy) {$A$};
             \node at (\ptAx+\lato-0.1*\spazio,\ptAy-0.05*\spazio) {$B$};
             \node at (\ptAx+\lato,\ptAy-\lato+0.1*\spazio) {$C$};
              \node at (\ptAx,\ptAy-\lato+0.1*\spazio) {$D$};

             \def\ptBx{-1}
             \def\ptBy{+1}
             \def\lato {4}
             \def\spazio{.2}
             
             \draw[line width=.6pt,<->]  (\ptBx+\spazio,\ptBy) -- (\ptBx+\lato-\spazio,\ptBy);
             \draw[line width=.6pt,<->]  (\ptBx,\ptBy-\spazio) -- (\ptBx,\ptBy-\lato+\spazio);
             \draw[line width=.6pt,<->]  (\ptBx+\lato,\ptBy-\spazio) -- (\ptBx+\lato,\ptBy-\lato+\spazio);
             \draw[line width=.6pt,<->]  (\ptBx+\spazio,\ptBy-\lato) -- (\ptBx+\lato-\spazio,\ptBy-\lato);

             \node[scale=.95] at (\ptBx-5.5*\spazio,\ptBy+0.5*\spazio) {$T[SU(\Num)]$};
             \node[scale=.95] at (\ptBx-6.8*\spazio,\ptBy-\lato-0.5*\spazio) {$FFT[SU(\Num)]$};
             \node[scale=.95] at (\ptBx+\lato+5.5*\spazio,\ptBy+0.45*\spazio) {$\chk{T[SU(\Num)]}$};
	     \node[scale=.95] at (\ptBx+\lato+7*\spazio,\ptBy-\lato-0.5*\spazio) {$FF\chk{T[SU(\Num)]}$};
 
             
             \draw[line width=.6pt,->]  (\ptBx+\spazio,0.8*\ptBy)--(\ptAx-0.75*\spazio,\ptAy+0.75*\spazio); 
             \draw[line width=.6pt,->]  (\ptBx+\lato-\spazio,0.8*\ptBy)--(\ptAx+0.5*\lato+0.75*\spazio,\ptAy+0.75*\spazio); 
             \draw[line width=.6pt,->]  (\ptBx+1*\spazio,\ptBy-\lato+0.75*\spazio)--(\ptAx-1.1*\spazio,\ptAy-0.5*\lato-1*\spazio); 
             \draw[line width=.6pt,->]  (\ptBx+\lato-1*\spazio,\ptBy-\lato+0.75*\spazio)--(\ptAx+0.5*\lato+0.7*\spazio,\ptAy-0.5*\lato-1*\spazio);

             \end{tikzpicture}        
\ee      
~\\[.1cm]       
The monopole deformation we are interested in turns on the following components of the monopole matrix $\mathcal{M}_{ij}$,
\be\label{lin_mon_sup}
\mathcal{L}^T_{\{1,\ldots,N-1\}}=\Mon^{[10\cdots00]}+\Mon^{[010\cdots00]}+\cdots+\Mon^{[00\cdots10]} \,.
\ee
The last gauge node is underformed.\footnote{This condition is relevant for the stability of the IR dualities \cite{Giacomelli:2017vgk}.} 
We denote the deformed superpotential in $T[SU(\Num)]$ by $W_{def}^T$, namely
\be
W_{def}^T=W^T+\mathcal{L}^T_{\{1,\ldots,N-1\}}
\ee
More generally we will define $W_{def}^{\chk{T}}$ and $W_{def}^{FF\chk{T}}$ for theories $B$ and $C$, respectively.


%

\subsection{Theory A: Monopole deformed $T[SU(\Num)]$}\label{theory_A_sec}
The quiver diagram for $T[SU(\Num)]$ was introduced in section \ref{tsun_sec},
\be\label{tail_quiver}
             \begin{tikzpicture}
		
		\def\ox{0}
		\def\oy{0}
		\def\radius{.3}
		\def\lunghBif{1.5}
		\def\yshift{.1}
		
		\foreach \intero in {0,1}   \draw[line width=.6pt]    (\ox+\intero*\lunghBif,\oy+\yshift) --  (\ox+\lunghBif+\intero*\lunghBif,\oy+\yshift) ;
		\foreach \intero in {0,1}   \draw[line width=.6pt]    (\ox+\intero*\lunghBif,\oy-\yshift) --  (\ox+\lunghBif+\intero*\lunghBif,\oy-\yshift) ;
		
		\draw[line width=.6pt,dashed]    (\ox+2*\lunghBif,\oy+\yshift) --  (\ox+0.6*\lunghBif+2*\lunghBif,\oy+\yshift) ;
		\draw[line width=.6pt,dashed]    (\ox+2*\lunghBif,\oy-\yshift) --  (\ox+0.6*\lunghBif+2*\lunghBif,\oy-\yshift) ;		
		
		\draw[line width=.6pt,dashed]    (\ox+3*\lunghBif,\oy+\yshift) --  (\ox+0.5*\lunghBif+3*\lunghBif,\oy+\yshift) ;
		\draw[line width=.6pt,dashed]    (\ox+3*\lunghBif,\oy-\yshift) --  (\ox+0.5*\lunghBif+3*\lunghBif,\oy-\yshift) ;	
	
		\draw[line width=.6pt]    (\ox+3.5*\lunghBif,\oy+\yshift) --  (\ox+4.5*\lunghBif,\oy+\yshift) ;	
		\draw[line width=.6pt]    (\ox+3.5*\lunghBif,\oy-\yshift) --  (\ox+4.5*\lunghBif,\oy-\yshift) ;	
		\foreach \intero in {1,2}   	\node        at  (\ox+\intero*\lunghBif,\oy+5*\yshift) 	[circle,inner sep=2mm,draw=black,fill=white,thick]   {}    ;
		\node	at (\ox+.05,\oy+5*\yshift) 	[circle,inner sep=2mm,draw=black,fill=white,thick]   {}    ;
		\foreach \intero in {2,3}           \node	at (\ox-\lunghBif+\intero*\lunghBif,\oy) 	[circle,inner sep=2mm,draw=black,fill=white,thick]   {$\intero$}    ;
		\node	at (\ox+.05,\oy) 	[circle,inner sep=2mm,draw=black,fill=white,thick]   {$1$}    ;

	         \node        at  (\ox+3.5*\lunghBif,\oy+5*\yshift) 	[circle,inner sep=2mm,draw=black,fill=white,thick]   {}    ;
		 \node	at (\ox+3.5*\lunghBif,\oy) 	[circle,inner sep=1.6mm,draw=black,fill=white,thick]   {$N$}    ;
		 \node	at (\ox+4.6*\lunghBif,\oy) 	[minimum height=.8cm, draw=black,fill=white,thick]   {$N+1$}   ;
		 		
	      \end{tikzpicture}
\ee
It will be convenient to decompose the adjoint fields on a basis of hermitian generators of $U(k)$, namely $\Phi_{k}=\sum \phi_{k}^a T^a$, 
and extract from the superpotential \eqref{superPini}, the abelian components, defined hereafter as,
\bea
W^{T}\supset \cW_{k}&\equiv&\tfrac{1}{k} \Tr\Phi_{k} \left[ \Tr_k\Tr_{k+1} \mathbb{Q}^{(k,k+1)} - \Tr_k\Tr_{k-1}  \mathbb{Q}^{(k-1,k)} \right]
\eea
The reason is that abelian and non-abelian components decouple.\footnote{Consider $T_{}^0=\mathbb{I}$, the identity. Thus, $\Tr(T_{}^0A)=\Tr A$ for any matrix $A$, 
and it follows that $\Tr\Phi_{k}=k\phi_{k}^0$, i.e. $\phi_{k}^{a=0}=\frac{1}{k} \Tr\Phi_{k}$.}

In the presence of the monopole deformation $\mathcal{L}_{\{1,\ldots,N-1\}}$, we can burn $\Mon^{[10\cdots00]}$ on the first gauge node and dualize the fields as follows:
\be\label{sara_benve_beni_1}
U(1)\oplus\,2\ {\rm flavors\ and}\ W=\Mon^{+} \ \leftrightarrow\ \,4\oplus1\ {\rm singlets}\ M_{ij}\oplus\gamma\ {\rm and}\ W=\gamma\det M
\ee
where the magnetic fields $M_{ij}$ replace the electric meson. 
This is the first instance of a family of electric-magnetic dualities introduced in \cite{Benini:2017dud}:
\be\label{sara_benve_beni}
\, U(N_c)\oplus\,N_{flav.}\ {\rm and}\ W=\Mon^{+} \ \leftrightarrow\ N_f^2\oplus1\ {\rm singlets}\ M_{ij}\oplus\gamma\ {\rm and}\ W=\gamma\det M
\ee
The map \eqref{sara_benve_beni} does not include adjoint fields, which instead are present on the quiver tail. However, on the first node, $\Phi_1$ is just a singlet, thus it can be taken into account afterwards. 
Similarly, we add the coupling $\gamma\det M$ on top of $W^T$. 

Since the magnetic dual of a $U(1)$ gauge theory with two flavors is a Wess-Zumino model, the $U(1)$ dynamics has confined in the IR.  
The presence of $W^T$ allows for a sequence of iterations of this procedure, where at each step the duality \eqref{sara_benve_beni} is used for an increasing value of $N_c$. 
This is the content of the sequential confinement  introduced by \cite{Benvenuti:2017kud}. Here we generalize it to the case of $T[SU(\Num)]$, building on previous work done in \cite{Giacomelli:2017vgk}.
Before presenting results for the final low energy theory, we discuss in detail the confinement of the first two nodes.

\subsection*{Move \# 1}

Consider the restriction of  $T[SU(\Num)]$ to the first and the second gauge node. Locally, the theory is described by the quiver
\be\label{Quiver1}
             \begin{tikzpicture}
		
		\def\ox{0}
		\def\oy{0}
		\def\radius{.3}
		\def\lunghBif{1.5}
		\def\yshift{.1}
	
		\foreach \intero in {0,1}   \draw[line width=.6pt]    (\ox+\intero*\lunghBif,\oy+\yshift) --  (\ox+\lunghBif+\intero*\lunghBif,\oy+\yshift) ;
		\foreach \intero in {0,1}   \draw[line width=.6pt]    (\ox+\intero*\lunghBif,\oy-\yshift) --  (\ox+\lunghBif+\intero*\lunghBif,\oy-\yshift) ;
		\foreach \intero in {0,1}  	       \node         at  (\ox+\intero*\lunghBif,\oy+5*\yshift) 	[circle,inner sep=2mm,draw=black,fill=white,thick]   {}    ;
		\foreach \intero in {1,2}              \node	at (\ox-\lunghBif+\intero*\lunghBif,\oy) 	[circle,inner sep=2mm,draw=black,fill=white,thick]   {$\intero$}    ;
								\node	at (\ox+2*\lunghBif,\oy) 	        [rectangle,inner sep=2.5mm,draw=black,fill=white,thick]   {$3$}   ;
	      \end{tikzpicture}
\ee
with superpotential 
\bea
W_{def}^{T[SU(3)]}&= & \cW_{1}+\cW_{2} + \sum_{a=1}^3 \phi_{2}^a\, \Tr_2\left[ T^a \left( \Tr_3\mathbb{Q}^{(2,3)} -\Tr_1 \mathbb{Q}^{(1,2)} \right)\right] + \Mon^{[10\cdots00]}\notag\\[.2cm]
\cW_{1}+\cW_{2}&=&  \Tr\Phi_{1} \left[\Tr_1\Tr_2 \mathbb{Q}^{(1,2)} \right]+\tfrac{1}{2} \Tr\Phi_{2} \left[ \Tr_2\Tr_3\mathbb{Q}^{(2,3)} -\Tr_2\Tr_1 \mathbb{Q}^{(1,2)} \right] 
\label{superP1}
\eea
where in \eqref{superP1}  we have specified the abelian component. Notice the property  $\Tr_1\Tr_2 \mathbb{Q}^{(1,2)}=\Tr_2\Tr_1 \mathbb{Q}^{(1,2)}$, i.e. we can commute the two traces. 

We use the monopole duality \eqref{sara_benve_beni_1} on the first gauge node. 
Accordingly, we replace the electric meson, $\Tr_1\mathbb{Q}^{(1,2)}\rightarrow M_{2}$, where $M_{2}$ is in the adjoint of $U(2)$, the second gauge node. 
In the dual theory the superpotential has become: 
\bea
W^{T[SU(3)]}_{def}= \cW_{1}+\cW_{2} + \sum_{a=1}^3 \phi_{2}^a \, \Tr_2\,\left[ T^a \left( \Tr_3\mathbb{Q}^{(2,3)} -M^{(2)}  \right)\right] +\gamma_2 \det M_{2} + \Mon^{[01\cdots00]}\qquad \label{superP2}
\eea
where the abelian superpotential reads,
\bea
\cW_{1}+\cW_{2}=  \Tr\Phi_{1} \Tr M_{2}+\ \tfrac{1}{2} \Tr\Phi_{2}\, \left[ \Tr_2\Tr_3\mathbb{Q}^{(2,3)} -\Tr M_2 \right]  \label{superP3}
\eea
The interaction term, $\gamma_2 \det M_{2}$, is part of the duality map.
It is convenient to rotate the abelian adjoints to
\bea
\varphi_2^{-}\equiv  \left(\Tr\Phi_{1} - \tfrac{1}{2} \Tr\Phi_{2} \right) \qquad 
\varphi_2^{+}\equiv \left(\Tr\Phi_{1} + \tfrac{1}{2}\Tr\Phi_{2} \right)
\eea
in such a way that  
\bea\label{superP4}
\cW_{1}+\cW_{2} &=& \varphi_2^{-}  \left( \Tr M_{2}  - \tfrac{1}{2} \,\Tr_2\Tr_3\mathbb{Q}^{(2,3)} \right)+  \tfrac{1}{ 2} \varphi_2^+\, \Tr_2\Tr_3\mathbb{Q}^{(2,3)}
\eea
An important remark is that $M_{2}$ is an elementary field in the dual theory.
Then the F-term of $\varphi_2^{-}$ and $\phi_{a=1,2,3}$,  determine a vev for $M_{2}$. In particular, $\langle M_{2} \rangle$ depends on $\Tr_3\mathbb{Q}^{(2,3)}$ as follows
\be\label{F-term2}
 \Tr\langle M_{2} \rangle - \tfrac{1}{2} \,\Tr_2\Tr_3\mathbb{Q}^{(2,3)} =0, \qquad \Tr_2\left[ T^{a=1,2,3} \left( \langle M_{2} \rangle -\Tr_3\mathbb{Q}^{(2,3)}  \right)\right] =0
\ee
Equations \eqref{F-term2} imply that $\langle M_{2} \rangle$ has the same non abelian components of 
$\Tr_3\mathbb{Q}^{(2,3)}$ but differ by a factor of $\tfrac{1}{2}$ in the abelian component. 
In matrix form, the solution is 
\be
\langle M_{2} \rangle = \Tr_3\mathbb{Q}^{(2,3)}  - \tfrac{\Tr_2 \Tr_3\mathbb{Q}^{(2,3)} }{2*2}\, \mathbb{I}_{2\times 2}\,
\ee
Expanding the superpotential around $M_{2}=\langle M_{2}\rangle+\delta M_{2}$, 
we find mass terms for $\varphi_2^-$, $\delta M_{2}$, and for the non abelian adjoint fields $\phi^a$.  This is obvious from \eqref{superP2} and \eqref{superP4}. 
Below a common mass scale, all these fields can be integrated out. 
As a result, the second node has now only a light $U(1)$ adjoint scalar $\varphi_{2}^+$, and the bifundamentals on the $(2,3)$ link. 
On the vacuum $\langle M_{2}\rangle$ there is a novel effective superpotential, which we determine in the next paragraph.

To proceed further, we would like to express $\det \langle M_{2}\rangle $ in terms of traces over matrices in the adjoint of $U(3)$. The reason is that
a matrix in the adjoint of $U(3)$ plays the role of the meson matrix for $T[SU(3)]$. Thinking about iterating the duality \eqref{sara_benve_beni} on node $(2)$, this rewriting is clearly necessary.  
To achieve the desired result, we first expand
\be
\det M_{2}= \tfrac{1}{2}\left[ \Tr M_{2} \right]^2 - \tfrac{1}{2} \Tr\left[M_{2} M_{2}\right]
\ee
Then, we rewrite
\be
\Tr\left[M_{2} M_{2}\right] = \Tr_2 \left[ \Tr_3 \mathbb{Q}^{(2,3)}\cdot\Tr_3 \mathbb{Q}^{(2,3)} \right] - \tfrac{\Tr_2 \Tr_3\mathbb{Q}^{(2,3)}}{2} \ \Tr_2\left[  \Tr_3 \mathbb{Q}^{(2,3)} \right] +  \tfrac{(\Tr_2 \Tr_3\mathbb{Q}^{(2,3)})^2}{8} \quad
\ee
Finally, two additional manipulations: In the abelian case we interchange the traces in the obvious way, $\Tr_2 \Tr_3 \mathbb{Q}^{(2,3)}=\Tr_3 \Tr_2 \mathbb{Q}^{(2,3)}$. In the non-abelian case, we notice the property
\bea
\Tr_2 \left[ \Tr_3 \mathbb{Q}^{(2,3)}\cdot\Tr_3 \mathbb{Q}^{(2,3)} \right]  
&=& \sum_{x,y=1}^2 \sum_{n=1}^3 Q_{xn} \tilde{Q}_{ny} \sum_{m=1}^3 Q_{y m }\tilde{Q}_{mx} \nn \\
&=& \sum_{m,n=1}^3 \sum_{x=1}^2 Q_{xn}\tilde{Q}_{mx} \sum_{y=1}^2 Q_{y m } \tilde{Q}_{ny} \nn\\
&=& \Tr_3\left[ \Tr_2 \mathbb{Q}^{(2,3)}\cdot \Tr_2\mathbb{Q}^{(2,3)}\right] \nn
\eea
The resulting theory has the following quiver diagram, 
\be\label{eff_quiver12}
             \begin{tikzpicture}
		
		\def\ox{0}
		\def\oy{0}
		\def\radius{.3}
		\def\lunghBif{1.5}
		\def\yshift{.1}
	
	         \node at (\ox-3*\radius,\oy-.05) {$\gamma_2,$};
		\foreach \intero in {0}   \draw[line width=.6pt]    (\ox+\intero*\lunghBif,\oy+\yshift) --  (\ox+\lunghBif+\intero*\lunghBif,\oy+\yshift) ;
		\foreach \intero in {0}   \draw[line width=.6pt]    (\ox+\intero*\lunghBif,\oy-\yshift) --  (\ox+\lunghBif+\intero*\lunghBif,\oy-\yshift) ;

		\foreach \intero in {0}  	       \node         at  (\ox+\intero*\lunghBif,\oy+5*\yshift) 	[circle,inner sep=2mm,draw=blue,fill=white,thick]   {}    ;
		\foreach \intero in {1}                \node	        at (\ox-\lunghBif+\intero*\lunghBif,\oy) 	[circle,inner sep=2mm,draw=black,fill=white,thick]   {$2$}    ;
						     	 	\node	at (\ox+1*\lunghBif,\oy) 	        [rectangle,inner sep=2.5mm,draw=black,fill=white,thick]   {$3$}   ;
	      \end{tikzpicture}
\ee
where the blue loop stands now for $\varphi_{2}^+$, instead of the full adjoint, and we remind ourselves of $\gamma_2$ by displaying it on the l.h.s of the diagram. The
effective superpotential associated to \eqref{eff_quiver12} is, 
\bea
W^{T[SU(3)]}_{eff}=  
\tfrac{1 }{2} \varphi_{2}^+\, \Tr_3\Tr_2\mathbb{Q}^{(2,3)} +
\tfrac{1}{2}\gamma_2\left[ -\Tr_3\left[ \big(\Tr_2 \mathbb{Q}^{(2,3)}\big)^2 \right] +\tfrac{5}{8} \big(\Tr_3\Tr_2 \mathbb{Q}^{(2,3)}\big)^2 \right]+\Mon^{[01\cdots00]}\notag\\
\label{final23}
\eea
Had we chosen $N=2$, there would be no monopole deformation in \eqref{final23}. 
Renaming $\Tr_2 \mathbb{Q}^{(2,3)}$ as the meson matrix $\mathcal{Q}$ introduced in Sec.~\ref{tsun_sec}, this would be the final result. 

%

\subsection*{Move \# 2}
We glue \eqref{eff_quiver12} back to $T[SU(\Num)]$, and move forward.  
On nodes $(2)$ and $(3)$, the theory is now described by the modified quiver
\be
\label{Quiver2}
             \begin{tikzpicture}
		
		\def\ox{0}
		\def\oy{0}
		\def\radius{.3}
		\def\lunghBif{1.5}
		\def\yshift{.1}
	
	         \node at (\ox-3*\radius,\oy-.05) {$\gamma_2,$};
	
		\foreach \intero in {0,1}   \draw[line width=.6pt]    (\ox+\intero*\lunghBif,\oy+\yshift) --  (\ox+\lunghBif+\intero*\lunghBif,\oy+\yshift) ;
		\foreach \intero in {0,1}   \draw[line width=.6pt]    (\ox+\intero*\lunghBif,\oy-\yshift) --  (\ox+\lunghBif+\intero*\lunghBif,\oy-\yshift) ;
							       \node         at  (\ox+0*\lunghBif,\oy+5*\yshift) 	[circle,inner sep=2mm,draw=blue,fill=white,thick]   {}    ;
		\foreach \intero in {1}  	       \node         at  (\ox+\intero*\lunghBif,\oy+5*\yshift) 	[circle,inner sep=2mm,draw=black,fill=white,thick]   {}    ;
		\foreach \intero in {2,3}              \node	at (\ox-2*\lunghBif+\intero*\lunghBif,\oy) 	[circle,inner sep=2mm,draw=black,fill=white,thick]   {$\intero$}    ;
								\node	at (\ox+2*\lunghBif,\oy) 	        [rectangle,inner sep=2.5mm,draw=black,fill=white,thick]   {$4$}   ;
	      \end{tikzpicture}
\ee
The superpotential includes the terms
\bea
W^{T[SU(4)]}_{def}&\supset &W^{T[SU(3)]}_{eff}+\cW_{3} + \sum_{a=1}^8 \phi^a\, \Tr_3\,\left[ T^a \left( \Tr_4\mathbb{Q}^{(3,4)} -\Tr_2\mathbb{Q}^{(2,3)} \right)\right] 
\eea
The gauged matter content attached at node $(2)$ is again of the form \eqref{sara_benve_beni}, plus singlets.  
We dualize by replacing $\Tr_2\mathbb{Q}^{(2,3)}\rightarrow M_{3}$ and add the superpotential term $\gamma_3\det M_{3}$. 
As before we study abelian and non abelian contributions separately. In the abelian sector we find, 
\bea
\tfrac{1}{2} \varphi_2^+\, \Tr_3M_{3} +\tfrac{1}{3} \Tr\Phi_{3} \left[ \Tr_3\Tr_4\mathbb{Q}^{(3,4)} -\Tr_3 M_{3} \right] ,
\eea
which upon performing the rotation 
\bea
\varphi_{3}^- = \tfrac{1}{2}\varphi_{2}^+  -\tfrac{1}{3} \Tr\Phi_{3},\qquad
\varphi_{3}^+ = \tfrac{3}{2}\varphi_{2}^+ +  \Tr\Phi_3,
\eea
becomes
\bea
\varphi_{3}^-\left( \, \Tr M_3-\tfrac{1}{2} \Tr_3\Tr_4\mathbb{Q}^{(3,4)} \right) + \tfrac{1}{2*3}\varphi_{3}^+\Tr_3\Tr_4\mathbb{Q}^{(3,4)} .
\eea
Very much as in move\,\#1, the F-terms of $\phi_3^{a=1,..8}$ and $\varphi_{3}^-$ 
imply that $\langle M_{3}\rangle $ has the same non abelian components of $\Tr_4\mathbb{Q}^{(3,4)}$ but differs in the trace. 
The solution for $\langle M_{3} \rangle$ is 
\be
\langle M_{3} \rangle = \Tr_4\mathbb{Q}^{(3,4)}  - \tfrac{\Tr_3 \Tr_4\mathbb{Q}^{(3,4)} }{3*2}\, \mathbb{I}_{3\times 3}\, 
\ee
By integrating out the massive fields, $\varphi_{3}^-$, $\phi^{a=1,..8}_{3}$, and fluctuations of $\delta M_{3}$, we obtain a low energy theory with light $\varphi^+_{3}$ and bifundamentals $\mathbb{Q}^{(3,4)}$,
\be
             \begin{tikzpicture}
		
		\def\ox{0}
		\def\oy{0}
		\def\radius{.3}
		\def\lunghBif{1.5}
		\def\yshift{.1}

	         \node at (\ox-4.5*\radius,\oy-.05) {$\gamma_2,\gamma_3,$};

		\foreach \intero in {0}   \draw[line width=.6pt]    (\ox+\intero*\lunghBif,\oy+\yshift) --  (\ox+\lunghBif+\intero*\lunghBif,\oy+\yshift) ;
		\foreach \intero in {0}   \draw[line width=.6pt]    (\ox+\intero*\lunghBif,\oy-\yshift) --  (\ox+\lunghBif+\intero*\lunghBif,\oy-\yshift) ;

		\foreach \intero in {0}  	       \node         at  (\ox+\intero*\lunghBif,\oy+5*\yshift) 	[circle,inner sep=2mm,draw=blue,fill=white,thick]   {}    ;
		\foreach \intero in {1}                \node	        at (\ox-\lunghBif+\intero*\lunghBif,\oy) 	[circle,inner sep=2mm,draw=black,fill=white,thick]   {$3$}    ;
						     	 	\node	at (\ox+1*\lunghBif,\oy) 	        [rectangle,inner sep=2.5mm,draw=black,fill=white,thick]   {$4$}   ;
	      \end{tikzpicture}
\ee
As in the previous case, we would like to express the superpotential couplings which are linear in $\gamma_2$ and $\gamma_3$, in terms of traces over matrices in the adjoint of $U(4)$.
When $\det M_{3}$ is expanded out in $\Tr_3$, both $\gamma_2$ and $\gamma_3$ terms can be rearranged by using the following formulas,
\bea
%
\Tr_3\left[ \big(M_{3}\big)^2 \right] &=& \Tr_4 \left[ \big(\Tr_3\mathbb{Q}^{(3,4)}\big)^2 \right] - \tfrac{3(\Tr_3 \Tr_4\mathbb{Q}^{(3,4)} )^2}{4*k=3} \qquad \\
%
\Tr_3\left[ \big(M_{3}\big)^3 \right]&=&\Tr_4 \left[ \big( \Tr_3\mathbb{Q}^{(3,4)}\big)^3  \right] - 
\tfrac{3\Tr_3 \Tr_4\mathbb{Q}^{(3,4)} }{2*k=3} \Tr_4\left[ \big(\Tr_3\mathbb{Q}^{(3,4)}\big)^2\right] + \tfrac{5(\Tr_3 \Tr_4\mathbb{Q}^{(3,4)} )^3}{72}\qquad 
\eea
For the couplings to $\gamma_2$ we obtain
\bea
 -\tfrac{1}{2}\Tr_3\left[ \big( M_{3}\big)^2 \right] +\tfrac{5}{16} \big(\Tr_3 M_{3}\big)^2 &=&  -\tfrac{1}{2} \Tr_4 \left[ \big(\Tr_3\mathbb{Q}^{(3,4)}\big)^2 \right] +\tfrac{13(\Tr_3 \Tr_4\mathbb{Q}^{(3,4)} )^2}{64} \nn\\
 &\equiv& p_{2,3}[\Tr_4, \Tr_3\mathbb{Q}^{(3,4)}]
\label{p23}
\eea
and for the couplings to $\gamma_3$
\bea
 \det M_{3} &=& \tfrac{1}{3} \Tr_3\left[\big(M^{(3)} \big)^3\right] -\tfrac{\Tr_3 \Tr_4\mathbb{Q}^{(3,4)} /2}{2} \Tr_3 \left[\big(M^{(3)} \big)^2\right] +\tfrac{(\Tr_3 \Tr_4\mathbb{Q}^{(3,4)} )^3/2^3}{6} \nn\\
 &=& \tfrac{1}{3} \Tr_4\left[ \big( \Tr_3\mathbb{Q}^{(3,4)}\big)^3 \right] - \tfrac{5\Tr_3 \Tr_4\mathbb{Q}^{(3,4)} }{12}  \Tr_4 \left[ \big( \Tr_3\mathbb{Q}^{(3,4)}\big)^2 \right] + \tfrac{23(\Tr_3 \Tr_4\mathbb{Q}^{(3,4)} )^3}{216} \nn\\
 &\equiv&  p_{3,3}[\Tr_4, \Tr_3\mathbb{Q}^{(3,4)}] 
\label{p33}
\eea
In both cases, the final results can be expressed in terms of polynomials $p_{3,3}$ and $p_{2,3}$ in the variable $\Tr_3\mathbb{Q}^{(3,4)}$, which is indeed in the adjoint of $U(4)$.  
Collecting these contributions, the effective superpotential is determined by 
\be
W^{T[SU(4)]}_{eff}= \tfrac{1}{6 } \varphi_3^+\ \Tr_4 \Tr_3\mathbb{Q}^{(3,4)} +\gamma_2 \, p_{2,3}[\Tr_4, \Tr_3\mathbb{Q}^{(3,4)}] +\gamma_3\, p_{3,3}[\Tr_4, \Tr_3\mathbb{Q}^{(3,4)}] + \ldots\qquad
\ee
where $\ldots$ stands for the remaining monopole superpotential $\mathcal{L}_{\{3,\ldots,N-1\}}$. 

\subsubsection*{Duality moves: from  \# $1$ up to \# $N-1$}\label{generic-seq-conf}
Repeating the reasoning in move \#$1$, and \#$2$, we proceed up to  \#$N-1$. The final gauge theory, which we refer to as theory $A$, has quiver diagram
 \be\label{quiver_thA}
             \begin{tikzpicture}
		
		\def\ox{0}
		\def\oy{0}
		\def\radius{.3}
		\def\lunghBif{1.5}
		\def\yshift{.1}
	
	         \node at (\ox-5*\radius,\oy-.05) {$\gamma_2,\ldots \gamma_N,$};
	
		\foreach \intero in {0}   \draw[line width=.6pt]    (\ox+\intero*\lunghBif,\oy+\yshift) --  (\ox+\lunghBif+\intero*\lunghBif,\oy+\yshift) ;
		\foreach \intero in {0}   \draw[line width=.6pt]    (\ox+\intero*\lunghBif,\oy-\yshift) --  (\ox+\lunghBif+\intero*\lunghBif,\oy-\yshift) ;

		\foreach \intero in {0}  	       \node         at  (\ox+\intero*\lunghBif,\oy+5*\yshift) 	[circle,inner sep=2mm,draw=blue,fill=white,thick]   {}    ;
		\foreach \intero in {1}                \node	        at (\ox-\lunghBif+\intero*\lunghBif,\oy) 	[circle,inner sep=2mm,draw=black,fill=white,thick]   {$N$}    ;
						     	 	\node	at (\ox+1.1*\lunghBif,\oy) 	        [minimum height=.8cm, draw=black,fill=white,thick]   {$N+1$}   ;
	      \end{tikzpicture}
\ee
and superpotential 
\bea\label{finalSuperP}
W_A= \frac{1}{N! } \varphi_N^+ \, \Tr\mathcal{Q}  + \sum_{m=2}^{N } \gamma_m\, p_{m, N}\left[ \Tr,\mathcal{Q} \right]
\eea
where $\Tr_{N}\mathbb{Q}^{(N,N+1)}=\mathcal{Q}$ is the meson matrix, and $p_{m, n}$ are polynomials generalizing \eqref{p23} and \eqref{p33} at each step.
Recall that we did not turn on the monopole superpotential on the last gauge node, therefore \eqref{finalSuperP} is the final result.
Let us summarize the sequential confinement up to  \#$N-1$. In the order:\\[.2cm]
1) After each dualization, labelled hereafter by $k-1$, we derived an equation for $\langle M_{k}\rangle $ which we solved explicitly. 
In each case, the non abelian components of $\langle M_k\rangle $ are fixed by the F-terms of $\phi^{a=1,\ldots k^2-1}$ to be equal to $\Tr_{k+1} \mathbb{Q}^{(k,k+1)}$.
As in move \#$1$, and \#$2$, the abelian equation turns out to be always:
\be
\Tr\langle M_{k}\rangle = \tfrac{1}{2} \Tr_{k}\Tr_{k+1} \mathbb{Q}^{(k,k+1)}
\ee
The solution for $\langle M_k\rangle $ is
\be\label{meson_sol_k}
\langle M_{k}\rangle =\Tr_{k+1} \mathbb{Q}^{(k,k+1)} - \tfrac{ \Tr_{k}\Tr_{k+1} \mathbb{Q}^{(k,k+1)} }{2k} \mathbb{I}_{k\times k}
\ee
~\\
2)
Having found the solution \eqref{meson_sol_k}, we integrate the massive fields at node $k$ and we write the superpotential for the light fields. These are 
$\Tr_{k} \mathbb{Q}^{(k,k+1)}$, $\varphi_+^{(k)}$ and the collection of $\{\gamma_{m}\}_{m=2}^{k}$. 
This step is the most involved, since it requires rearranging the expression of $\{\det_m\}_{m=2}^k$ in terms of traces. The final result is packaged into the polynomials $p_{m,k}$. 
The structure of traces of such polynomials is fixed, i.e. by construction it coincides with that of $\det_m$ in its Laplace expansion. In particular,
\bea
p_{m,k}[\Tr_{k+1}, O]&=& \sum_{ n_1,\ldots n_m} c_{ \{n_1,\ldots n_m\} }^{k} \left[ \Tr_{k+1}\,O^1\right]^{n_1}\left[ \Tr_{k+1}\,O^2\right]^{n_2} \ldots \left[ \Tr_{k+1}\, O^{m} \right]^{n_m}  
\eea
where $O=\Tr_{k}\mathbb{Q}^{(k,k+1)}$, and the sum runs over all 
m-tuple ${n_1,\ldots, n_m}\ge0$ which solve the constraint $\sum_{l=1}^m l n_l =m$.\footnote{
The expression of $c_{ \{n_1,\ldots n_m\} }^{k}$ in the case of $\det_m$ is: 
$
c^k_{n_1,\ldots n_m}\rightarrow (-)^m \prod_{l=1}^m \tfrac{(-l)^{-n_{l}} }{n_{l}!} 
$, and it is independent of $k$.}
From the original $\det_m$ formula, the polynomials $p_{m,k}$ inherit the property of having degree $m$ in $O$. 
However, powers of $\langle M_k\rangle$ will produce an admixture of powers of $O$, therefore a generic coefficients $c^k$ will depend on $k$ in a non trivial way. Only 
the top element $ \Tr\, O^{m}$ has coefficient fixed to be $c_{\{0^{m-1},1\}}^{k}= (-1)^{m-1}/m$ from the original $\det_m$ formula. 
Perhaps, the best description of the coefficients $c^k$ is given in terms of recursion relations. For illustration we quote some simple examples:
\be
\begin{array}{ccccl}
c^1_{\{1,1\}}&=&+\tfrac{1}{2}&,  \qquad & c^k_{\{1,1\}}=\tfrac{1}{4}c^{k-1}_{\{1,1\}}+\tfrac{3}{8k} \qquad \forall k\ge 2 \\[.4cm]
c^2_{\{1,1,0\}}&=&-\tfrac{1}{2}&,\qquad & c^k_{\{1,1,0\}}=\tfrac{1}{2}c^{k-1}_{\{1,1,0\}}-\tfrac{1}{2k} \quad \forall k\ge 3 \\[.1cm]
c^2_{\{1,0,0\}}&=&+\tfrac{1}{6}&,\qquad & c^k_{\{1,0,0\}}=\tfrac{1}{8} c^{k-1}_{\{1,0,0\}} +\tfrac{5}{24k^2} - \tfrac{3}{8k} c^{k-1}_{\{1,1,0\}}\quad \forall k\ge 3\\[.4cm]
c^{m-1}_{\{1,0^{m-3},1,0\}}&=&\tfrac{(-)^m}{2}&,\qquad & c^{k}_{\{1,0^{m-3},1,0\}}= \tfrac{1}{m-1}c^{k-1}_{\{1,0^{m-3},1,0\}}+\tfrac{(-)^m}{2k} \quad \forall k\ge m\ge 3
\end{array}
\ee
In particular, the first three recursions determine $p_{2,k}$ and $p_{3,k}$ for any $k$.

\subsubsection*{Final remarks}

It is important to emphasize some features of the superpotential $W_A$.
A gauge theory $U(N)$ with $N+1$ flavors and no superpotential would have flavor symmetry $SU(N+1)_{\mathrm{flavor}}\times SU(N+1)_{\mathrm{flavor}}$. 
This is reduced to a single $SU(N+1)_{\mathrm{flavor}}$ because of the superpotential. Even in the absence of $\gamma_m$ contributions, 
the presence of $\varphi^+_N\Tr\cQ$ guarantees the correct amount of flavor symmetry. In this respect, $\varphi^+_N$ plays a distinguished role. 

Since the superpotential has R-charge $2$,
the R-charges of the singlets $\gamma_m$ acquire a dependence on $m$, 
\be
R[\gamma_m]=2(1-mr)\,.
\ee

The F-terms of $\varphi_N^+$ and $\gamma_{m=2,\ldots N}$ imply sequentially that $\Tr\mathcal{Q}^{1\leq k\leq N }=0$. 
Then, from the Cayley-Hamilton theorem it also follows $\Tr \mathcal{Q}^{N+1}=0$.\footnote{
Recall that $\det\langle\mathcal{Q}\rangle=0$ because $\langle\mathcal{Q}\rangle$ has at most rank $N$.}
This set of conditions is in fact equivalent to the statement that $\langle \cQ\,\rangle $ is nilpotent.  
At this point it is useful to redefine $\varphi_N^+=N!\gamma_1$ and simplify $W_A$ by invoking chiral 
stability arguments  \cite{Benvenuti:2017lle}. This amounts to drop terms containing $\Tr\mathcal{Q}$. The final form of the superpotential is then
\be
W_A=  -\sum_{m=1}^{N }\frac{(-)^{m} }{m} \ \gamma_m\ \Tr[\,\mathcal{Q}^m] \,,
\ee
In our discussion there will be no difference between these two versions of $W_A$. However, 
we should note that this prescription amounts to drop multi-trace contributions 
to the effective superpotentials, which might affect other details of the theory.

\subsection{Theory B: Monopole deformations on the mirror  }\label{theoryB}

In this section we follow the monopole deformation in the mirror frame $\chk{T[SU(\Num)]}$, which is represented by the quiver diagram below,
\be
             \begin{tikzpicture}
		
		\def\ox{3}
		\def\oy{0}
		\def\radius{.3}
		\def\lunghBif{1.5}
		\def\yshift{.1}
		
		\foreach \intero in {1,2}   \draw[line width=.6pt]    (\ox+\intero*\lunghBif,\oy+\yshift) --  (\ox+\lunghBif+\intero*\lunghBif,\oy+\yshift) ;
		\foreach \intero in {1,2}   \draw[line width=.6pt]    (\ox+\intero*\lunghBif,\oy-\yshift) --  (\ox+\lunghBif+\intero*\lunghBif,\oy-\yshift) ;
		
		\draw[line width=.6pt,dashed]    (\ox-0.5*\lunghBif,\oy+\yshift) --  (\ox+0.2*\lunghBif+0*\lunghBif,\oy+\yshift) ;
		\draw[line width=.6pt,dashed]    (\ox-0.5*\lunghBif,\oy-\yshift) --  (\ox+0.2*\lunghBif+0*\lunghBif,\oy-\yshift) ;		
		
		\draw[line width=.6pt,dashed]    (\ox+0.4*\lunghBif,\oy+\yshift) --  (\ox+1*\lunghBif+0*\lunghBif,\oy+\yshift) ;
		\draw[line width=.6pt,dashed]    (\ox+0.4*\lunghBif,\oy-\yshift) --  (\ox+1*\lunghBif+0*\lunghBif,\oy-\yshift) ;	
	
		\draw[line width=.6pt]    (\ox-2*\lunghBif,\oy+\yshift) --  (\ox-0.5*\lunghBif,\oy+\yshift) ;	
		\draw[line width=.6pt]    (\ox-2*\lunghBif,\oy-\yshift) --  (\ox-0.5*\lunghBif,\oy-\yshift) ;	
		\foreach \intero in {1,2,3}   	\node        at  (\ox+\intero*\lunghBif,\oy+5*\yshift) 	[circle,inner sep=2mm,draw=black,fill=white,thick]   {}    ;
		\foreach \intero in {1,2,3}           \node	at (\ox+4*\lunghBif-\intero*\lunghBif,\oy) 	[circle,inner sep=2mm,draw=black,fill=white,thick]   {$\intero$}    ;

	         \node        at  (\ox-0.4*\lunghBif,\oy+5*\yshift) 	[circle,inner sep=2mm,draw=black,fill=white,thick]   {}    ;
		 \node	at (\ox-0.4*\lunghBif,\oy) 	[circle,inner sep=1.6mm,draw=black,fill=white,thick]   {$N$}    ;
		 \node	at (\ox-1.6*\lunghBif,\oy) 	[minimum height=.8cm, draw=black,fill=white,thick]   {$N+1$}   ;
		 		
	      \end{tikzpicture}
\ee
%
Our notation in section \ref{Mirror_dual_sec} used bifundamentals $P$ and $\tilde{P}$ on each link, and adjoints $\checkPhi$ on each gauge node.
The monopole deformation $\mathcal{L}^T_{\{1,\ldots,N-1\}}$ we considered in \eqref{lin_mon_sup} can actually be written, more suggestively, in terms of the Jordan matrix
\be
\mathbb{J}_N\oplus \mathbb{J}_1=\left[ \begin{array}{c|c}\  \mathbb{J}_N \ & \ 0_{1\times N} \\ \hline 0_{N\times 1} & 0_{1\times 1} \end{array}\right],\qquad
\mathbb{J}_N= \underbrace{ \left(\begin{array}{ccccc} 0 &1& \ldots & \ldots & 0 \\ 0 & 0 & 1 & \ldots  & 0  \\ \vdots & \vdots & &  & \vdots  \\ 0 & 0 & \ldots & 0 & 0 \end{array}\right) }_N\ 
\ee
where $\mathbb{J}_k$ is a single Jordan block of size $k$ and zero eigenvalue. 
It follows that $\mathcal{L}^T_{\{1,\ldots,N-1\}}$ is mirror to a nilpotent mass deformation for the meson $\mathcal{P}_{ij}$. By introducing the vectors 
\be
P^{(N+1,N)}=\left(\begin{array}{cccc} p_1,& p_2 ,& \ldots &, p_{N+1}\end{array}\right)\qquad
\widetilde{P}^{(N+1,N)}=\left(\begin{array}{c} \tilde{p}_1\\ \tilde{p}_2\\ \vdots\\  \tilde{p}_{N+1}\end{array}\right)\qquad
\ee 
we find indeed
\bea\label{deformation_Tcheck}
 \mathcal{L}^{\chk{T}}_{\{1,\ldots,N-1\}} &=&\Tr_{N+1}\Big[\, \mathbb{J}_N\oplus \mathbb{J}_1\cdot\mathcal{P}\Big]=\tilde{p}_2\cdot {p}_1+ \ldots \tilde{p}_{N}\cdot {p}_{N-1} \,.
\eea
and the total superpotential is thus
\bea
W_{def}^{\chk{T}}=
\mathcal{L}^{\chk{T}}_{\{1,\ldots,N-1\}} + \sum_{k=1}^N \Tr_k \left[ \checkPhi_{k} \left( \Tr_{k+1} \mathbb{P}^{(k+1,k)}-\Tr_{k-1} \mathbb{P}^{(k,k-1)} \right) \right]\ .
\eea
The discussion next will closely follow \cite{Giacomelli:2017vgk}.

The F-term equations of $\tilde{p}_{a=2,\ldots, N}$ and ${p}_{b=1,\ldots,N-1}$ are non trivial due to the mass deformation. Let us begin from
the F-terms of the fields $\tilde{p}_{a=2,\ldots, N}$, which read
\be
{p}_{a-1}+\checkPhi_N \,{p}_a=0
\ee 
The solution is expressed in terms of ${p}_N$ as follows:
\be\label{vevpb}
\langle {p}_b\rangle =(-)^{N-b}\underbrace{\,\checkPhi_N \cdots \,\checkPhi_N}_{ N-b {\rm\ times} } {p}_N\qquad b=1,\ldots,N-1\ .
\ee
Equivalently, the F-terms of the fields ${p}_{b=1,\ldots,N-1}$ are solved by
\be\label{vevpa}
\langle \tilde{p}_a\rangle = (-)^{a-1}\ \tilde{p}_1\underbrace{\, \checkPhi_N \cdots \,\checkPhi_N}_{a-1{\rm\ times} }  \qquad a=2,\ldots,N
\ee
On the vacuum ${p}_b\rightarrow \langle {p}_b\rangle +\delta{p}_b$, 
and $\tilde{p}_a\rightarrow \langle \tilde{p}_a\rangle +\delta \tilde{p}_a$, 
the pair of field $(\tilde{p}_1,{p}_{N})$, and $(p_{N+1},\tilde{p}_{N+1})$ do not get a mass terms from the deformation \eqref{deformation_Tcheck},
thus they remain in the low energy spectrum.\footnote{Notice that we started with $2N(N+1)$ d.o.f in the bifundamentals $P$ and $\widetilde{P}$. Then we have $N$ mass (terms) for each of the $N-1$ terms in $\mathcal{L}_B$. So $2N(N+1)-2N(N-1)=4N$ fields are light. } 
The effective superpotential for these light fields is
\bea
W^{\chk{T}}_{eff}&= \,(-)^{N-1} \left[ \tilde{p}_1\underbrace{\checkPhi_N \cdots\checkPhi_N }_{N{\rm\ times} } {p}_N\right] &+\ \tilde{p}_{N+1}\checkPhi_N {p}_{N+1} 
\eea
The low energy theory is then described by the quiver, 
\be
             \begin{tikzpicture}
		
		\def\ox{3}
		\def\oy{0}
		\def\radius{.3}
		\def\lunghBif{1.5}
		\def\yshift{.1}
		
		\foreach \intero in {1,2}   \draw[line width=.6pt]    (\ox+\intero*\lunghBif,\oy+\yshift) --  (\ox+\lunghBif+\intero*\lunghBif,\oy+\yshift) ;
		\foreach \intero in {1,2}   \draw[line width=.6pt]    (\ox+\intero*\lunghBif,\oy-\yshift) --  (\ox+\lunghBif+\intero*\lunghBif,\oy-\yshift) ;
		
		\draw[line width=.6pt,dashed]    (\ox-0.5*\lunghBif,\oy+\yshift) --  (\ox+0.2*\lunghBif+0*\lunghBif,\oy+\yshift) ;
		\draw[line width=.6pt,dashed]    (\ox-0.5*\lunghBif,\oy-\yshift) --  (\ox+0.2*\lunghBif+0*\lunghBif,\oy-\yshift) ;		
		
		\draw[line width=.6pt,dashed]    (\ox+0.4*\lunghBif,\oy+\yshift) --  (\ox+1*\lunghBif+0*\lunghBif,\oy+\yshift) ;
		\draw[line width=.6pt,dashed]    (\ox+0.4*\lunghBif,\oy-\yshift) --  (\ox+1*\lunghBif+0*\lunghBif,\oy-\yshift) ;	
	
		\draw[line width=.6pt]    (\ox-1.4*\lunghBif,\oy+\yshift) --  (\ox-0.5*\lunghBif,\oy+\yshift) ;	
		\draw[line width=.6pt]    (\ox-1.4*\lunghBif,\oy-\yshift) --  (\ox-0.5*\lunghBif,\oy-\yshift) ;	
	

		\draw[line width=.6pt,red,thick]    (\ox-0.33*\lunghBif,\oy)--  (\ox-0.33*\lunghBif,\oy-0.9*\lunghBif) ;	
		\draw[line width=.6pt,red,thick]    (\ox-0.47*\lunghBif,\oy) --  (\ox-0.47*\lunghBif,\oy-0.9*\lunghBif) ;		
		
		 \node	at (\ox-0.4*\lunghBif,\oy-0.85*\lunghBif) 	[rectangle,inner sep=2.5mm,draw=black,fill=white,thick]   {$1$}   ;

		\foreach \intero in {1,2,3}   	\node        at  (\ox+\intero*\lunghBif,\oy+5*\yshift) 	[circle,inner sep=2mm,draw=black,fill=white,thick]   {}    ;
		\foreach \intero in {1,2,3}           \node	at (\ox+4*\lunghBif-\intero*\lunghBif,\oy) 	[circle,inner sep=2mm,draw=black,fill=white,thick]   {$\intero$}    ;

	         \node        at  (\ox-0.4*\lunghBif,\oy+5*\yshift) 	[circle,inner sep=2mm,draw=black,fill=white,thick]   {}    ;
		 \node	at (\ox-0.4*\lunghBif,\oy) 	[circle,inner sep=1.6mm,draw=black,fill=white,thick]   {$N$}    ;
		 \node	at (\ox-1.3*\lunghBif,\oy) 	[rectangle,inner sep=2.5mm,draw=black,fill=white,thick]   {$1$}   ;

	      \end{tikzpicture}
\ee
where we have isolated the fields $(d,\tilde{d})\equiv (p_1,\tilde{p}_{N})$ on the bottom of the diagram. Compared to the tail, these fields are the ones with a special superpotential interaction.
The matrix $\mathbb{P}^{(1,N)}$ is now truncated to a one flavor component $p_{N+1}\tilde{p}_{N+1}$, and the total superpotential of theory B is
\bea
W_{B}= \,(-)^{N-1} \left[ \tilde{d}\underbrace{\,\checkPhi_N \cdots\checkPhi_N }_{N{\rm\ times} } {d} \right]  +\ \tilde{p}_{N+1}\checkPhi_N {p}_{N+1} -  \Tr_N\left[ \checkPhi_N  \Tr_{N-1}\mathbb{P}^{(N,N-1)}\right] \notag\\
%
+\sum_{k=1}^{N-1} \Tr_k \left[ \checkPhi_k \left( \Tr_{k+1} \mathbb{P}^{(k+1,k)}-\Tr_{k-1} \mathbb{P}^{(k,k-1)} \right) \right]
\eea

From the expression of $W_B$ we deduce that 
the R-charge assignment is modified compared to $\chk{T[SU(\Num)]}$.
The newly generated term, $\tilde{d}\Omega^Nd$, implies
\bea
R[\Omega_k]=2r;\qquad
R[P]=R[\tilde{P}]=(1-r);\qquad R[d]=R[\tilde{d}]=1-Nr
\eea
where $R[\Omega_k]+R[\mathcal{P}]=2$.
Consequently 
\be
R[\tilde{d}\underbrace{\,\checkPhi_N \cdots\checkPhi_N }_{i{\rm\ times} } {d}]=2(1-r(N-i))\,.
\ee
%
We then have the   following map between the singlets of theory $A$ and dressed mesons of theory $B$:
\bea
\gamma_1\leftrightarrow \tilde{d}\underbrace{\,\checkPhi_N \cdots\checkPhi_N }_{N-1{\rm\ times} } {d}
\qquad;\qquad
\left\{
\begin{array}{clc}%
\gamma_2&\leftrightarrow&\tilde{d}\underbrace{\,\checkPhi_N \cdots\checkPhi_N }_{N-2{\rm\ times} } {d}\\
\vdots\\[.2cm]
\gamma_N&\leftrightarrow&\tilde{d} {d}
\end{array}\right.
\eea
As expected, the duality between $A$ and $B$ is a particular case of the SQCD mirror dual discussed in \cite{Giacomelli:2017vgk} with a minor difference, i.e. we have kept the fields $\gamma_i$ on the side of theory $A$.

\subsection{Theory C: Nilpotent Higgsing from Monopoles}\label{nilpot_higg}

The theory $FF\chk{T[SU(\Num)]}$ introduced section \ref{flipflip_sec} is described by the quiver
\be\label{tail_quiver_FCFH}
             \begin{tikzpicture}
		
		\def\ox{0}
		\def\oy{0}
		\def\radius{.3}
		\def\lunghBif{1.5}
		\def\yshift{.1}
		
		\foreach \intero in {0,1}   \draw[line width=.6pt]    (\ox+\intero*\lunghBif,\oy+\yshift) --  (\ox+\lunghBif+\intero*\lunghBif,\oy+\yshift) ;
		\foreach \intero in {0,1}   \draw[line width=.6pt]    (\ox+\intero*\lunghBif,\oy-\yshift) --  (\ox+\lunghBif+\intero*\lunghBif,\oy-\yshift) ;
		
		\draw[line width=.6pt,dashed]    (\ox+2*\lunghBif,\oy+\yshift) --  (\ox+0.6*\lunghBif+2*\lunghBif,\oy+\yshift) ;
		\draw[line width=.6pt,dashed]    (\ox+2*\lunghBif,\oy-\yshift) --  (\ox+0.6*\lunghBif+2*\lunghBif,\oy-\yshift) ;		
		
		\draw[line width=.6pt,dashed]    (\ox+3*\lunghBif,\oy+\yshift) --  (\ox+0.5*\lunghBif+3*\lunghBif,\oy+\yshift) ;
		\draw[line width=.6pt,dashed]    (\ox+3*\lunghBif,\oy-\yshift) --  (\ox+0.5*\lunghBif+3*\lunghBif,\oy-\yshift) ;	
	
		\draw[line width=.6pt]    (\ox+3.5*\lunghBif,\oy+\yshift) --  (\ox+4.5*\lunghBif,\oy+\yshift) ;	
		\draw[line width=.6pt]    (\ox+3.5*\lunghBif,\oy-\yshift) --  (\ox+4.5*\lunghBif,\oy-\yshift) ;	
		\foreach \intero in {1,2}   	\node        at  (\ox+\intero*\lunghBif,\oy+5*\yshift) 	[circle,inner sep=2mm,draw=black,fill=white,thick]   {}    ;
		\node	at (\ox+.05,\oy+5*\yshift) 	[circle,inner sep=2mm,draw=black,fill=white,thick]   {}    ;
		\node        at  (\ox+5.2*\lunghBif,\oy)										[circle,inner sep=2.5mm,draw=black,fill=white,thick]   {}    ;	
		\foreach \intero in {2,3}           \node	at (\ox-\lunghBif+\intero*\lunghBif,\oy) 	[circle,inner sep=2mm,draw=black,fill=white,thick]   {$\intero$}    ;
		\node	at (\ox+.05,\oy) 	[circle,inner sep=2mm,draw=black,fill=white,thick]   {$1$}    ;

	         \node        at  (\ox+3.5*\lunghBif,\oy+5*\yshift) 	[circle,inner sep=2mm,draw=black,fill=white,thick]   {}    ;
		 \node	at (\ox+3.5*\lunghBif,\oy) 	[circle,inner sep=1.6mm,draw=black,fill=white,thick]   {$N$}    ;
		 \node	at (\ox+4.6*\lunghBif,\oy) 	[minimum height=.8cm, draw=black,fill=white,thick]   {$N+1$}   ;
		 
\end{tikzpicture}
\ee
where we denoted the bifundamentals on each link by $S$ and $\widetilde{S}$, the adjoint chiral on each gauge node by $\Psi$, and finally the 
flipping fields by $\Flip^{\n}_{ij}$ and $\Flip^{\mathcal{S}}_{ij}$. 
The deformation $\chk{ \mathcal{L}}_{\{1,\ldots,N-1\}}$ in $\chk{T[SU(\Num)]}$ maps in $\chk{FFT[SU(\Num)]}$ to
\be\label{deform_in_ff}
\mathcal{L}^{FF\chk{T}}_{\{1,\ldots,N-1\}} = \Tr_{N+1}\Big[\, \mathbb{J}_N\oplus \mathbb{J}_1 \cdot \Flip^{\mathcal{S}}_{ij} \Big]\,.
\ee
The total superpotential is then
\bea
W_{def}^{FF\chk{T}}&=&W^{FF\chk{T}}+\mathcal{L}^{FF\chk{T}}\\
W^{FF\chk{T}}&=& \sum_{k=1}^N \Tr_k \left[ \Psi_k \left( \Tr_{k+1} \mathbb{S}^{(k,k+1)}-\Tr_{k-1} \mathbb{S}^{(k-1,k)} \right) \right] -\mathcal{S}_{ij}  \Flip^{\mathcal{S}}_{ij} - {\n}_{\,ij} \Flip^{\n}_{ij} \notag
\eea
It will convenient to momentarily modify our notation, and
denote $\Flip^{\mathcal{S}}_{ij}$ by $\Psi_{N+1}$. Then $\mathcal{S}_{ij}  \Flip^{\mathcal{S}}_{ij}$ fits with the pattern of the $\mathcal{N}=4$ superpotential, and will allow us to display some recursions in a neat way.

We first consider the F-term of $\Psi_{N+1}$, which set
\be\label{start_rec}
 \Tr_{N} \mathbb{S}^{(N,N+1)}  =+ \mathbb{J}_N\oplus \mathbb{J}_1 
\ee
This equation shows that  the bifundamentals on the last link of the tail acquire a non trivial vev.\footnote{Let us recall that in matrix notation, with standard multiplication, our definitions reduce to $\Tr_k \mathbb{S}^{(k,k+1)}= \widetilde{S}^{(k,k+1)}\cdot S^{(k,k+1)}$, and $\Tr_{k}\mathbb{S}^{(k-1,k)}= S^{(k-1,k)} \cdot \widetilde{S}^{(k-1,k)}$ }
By definition, our bifundamentals are rectangular matrices. 
However, it is convenient to describe the vev in terms of square matrices where we specify which column/row has to be dropped. 
In this way, the solution of \eqref{start_rec} is
\bea\label{vevstep1}
\begin{array}{cll}
\langle \widetilde{S}^{(N,N+1)} \rangle= & \mathbb{J}_N\oplus \mathbb{J}_1&{\rm\qquad drop\ the\ last\ column} \\
\langle S^{(N,N+1)}\rangle = & (\mathbb{J}_1\oplus \mathbb{I}_{N-1})\oplus \mathbb{J}_1&{\rm \qquad drop\ the\ last\ row}
\end{array}
\eea
Up to gauge and flavor rotations, the natural strategy to solve an equation of the form \eqref{start_rec}
is to take $\langle\widetilde{S}\rangle$ equal to the nilpotent vev on the r.h.s, and $\langle S\rangle $ such that the equation is satisfied.
In particular, $\langle S\rangle $ has an identity block of rank $N-1$ \cite{Agarwal:2014rua}.
The solution when $N=1$ reduces to zero, since there is no monopole potential in this case.

%
The $F$-terms of the fields $\Psi_{k\leq N}$ have two types of contributions. One is coming from the superpontential of the tail, i.e. $W^T[ \Psi, \mathbb{S} ]$, 
and a second one originates from the coupling ${\n}_{\,ij} \Flip^{\n}_{ij}$:
Indeed, recall from the definition \eqref{1monopolematrix} that the monopole matrix has traceless diagonal components of the form $\Tr\Psi_k\mathcal{D}_k$. 
We will study a vacuum for which $\langle \Flip^{\n}_{ij}\rangle=0$. Therefore the $F$-term of $\Psi^k$ will be
\be\label{rec_higgs}
\Tr_{k-1} \mathbb{S}^{(k-1,k)} =\Tr_{k+1} \mathbb{S}^{(k,k+1)}
\ee
Reading \eqref{rec_higgs} from right to left, 
we conclude that the nilpotent vev \eqref{vevstep1} propagates along the quiver,  from the last node towards the left. The solution of this recursion is
\bea\label{nilp_vev_tail}
\begin{array}{cll}
\langle \widetilde{S}^{(k-1,k)}\rangle&= \mathbb{J}_1\oplus \mathbb{J}_{k-1}  &{\rm\qquad drop\ the\ first\ column} \\
\langle {S}^{(k-1,k)}\rangle&= \mathbb{J}_1\oplus (\mathbb{J}_1\oplus \mathbb{I}_{k-2})   & {\rm\qquad drop\ the\ first\ row} 
\end{array}
\eea
Note that for $k>2$ the vev $\langle \widetilde{S}^{(k-1,k)}\rangle$ is always next to a maximal Jordan matrix $\mathbb{J}_k$. 
At the terminating value $k=2$ both vevs vanish. These correspond to the bifundamentals on the first link $(1,2)$. 

\subsection*{Nilpotent vevs}
A supersymmetric nilpotent vev should satisfy the F-terms of the bifundamentals, and finally D-terms. In matrix notation, the $F$-terms of the bifundamentals are 
\be
\widetilde{S}^{(k,k+1)}\Psi_k=\Psi_{k+1}\widetilde{S}^{(k,k+1)} \label{FQ}\\
\Psi_k S^{(k,k+1)}=S^{(k,k+1)}\Psi_{k+1} \label{FQt}
\ee
These equations put constraints on the fields $\Psi_{k}$, and 
before proceeding, let us recall that an additional constraint comes from the F-terms equations for the diagonal generators of $\Flip^{
n}_{ij}$, which imply the condition $\Tr\Psi_{k}=0$ for any $k\leq N$.

A trivial solution to \eqref{FQt} is $\langle\Psi_k\rangle=0$ for any $k$. However, this solution will not be consistent with vanishing of D-terms, as we now explain. 
%

Notice that equations \eqref{FQ} and \eqref{FQt} don't fix a solution, rather they impose a constraint on $\Psi_{k+1}$ which depends on $S^{(k,k+1)}$, $\widetilde{S}^{(k,k+1)}$ and $\Psi_k$. 
In this new recursion, the starting point is the beginning of the tail, i.e. the $U(1)$ gauge node, and the first link $(1,2)$. 
Consistency of this recursion requires that the solution in the case of $T[SU(k)]$ uplifits to $T[SU(\Num)]$ for any $k\leq N$. The study of the first few cases will be enough to understand the nilpotent vev in the adjoint sector. 

Consider $T[SU(2)]$. The D-term on the $U(1)$ gauge node is 
\be\label{equaN2higg}
S^{R}S^{R\dagger} - \widetilde{S}^{R\dagger}\widetilde{S}^R = \xi_1,\qquad R=(1,2),
\ee
with $\xi_1$ an FI parameter. This equation reduces always to $\xi_1=0$, because $\langle S^{(1,2)}\rangle=\langle \widetilde{S}^{(1,2)}\rangle =0$. 
The solution is then compatible with $\langle\Psi_1\rangle=0$. From equations \eqref{FQ} and \eqref{FQt} it follows $\langle\Psi_2\rangle=0$, and we are back to the nilpotent vev for the flipping fields we started with, for this case.

Consider now $T[SU(3)]$. The D-term on the $U(2)$ gauge node is 
\be\label{esempioDsu3}
S^{R}S^{R\dagger} - \widetilde{S}^{R\dagger}\widetilde{S}^R + S^{L\dagger}S^{L} - \widetilde{S}^{L}\widetilde{S}^{L\dagger} = [\Psi_2,\Psi^\dagger_2]+\xi_2 \mathbb{I}_2, \qquad \left\{\begin{array}{lcc} R&=&(2,3)\\ L&=&(1,2) \end{array}\right. .
\ee
A short computation shows that the terms labelled by `$R$' cancel each other. 
Since $\langle S^{(1,2)}\rangle=\langle \widetilde{S}^{(1,2)}\rangle =0$, the terms labelled by `$L$' do not contribute. 
The case of $T[SU(3)]$ is again special and the solution $\langle\Psi_2\rangle=0$, $\xi_2=0$ is consistent. In particular $T[SU(2)]$ uplifits to $T[SU(3)]$.
Finally equations \eqref{FQ} and \eqref{FQt} imply the relations 
\be
\Psi_3\widetilde{S}^{(2,3)}=\left[\begin{array}{cc} 0 & 0 \\ 0 & 0 \\ 0 & 0 \end{array}\right],\qquad 
S^{(2,3)}\Psi_3=\left[\begin{array}{ccc} 0 & 0 & 0 \\ 0 & 0 & 0 \end{array}\right]. 
\ee
These equation do not fix all the components of $\langle\Psi_3\rangle$. The trivial solution is possible, but we claim that the actual solution, compatible with $T[SU(4)]$, is a nilpotent vev for $\langle\Psi_3\rangle$, i.e. 
\be\label{Xvev3}
\langle\Psi_3\rangle=\mathbb{J}_2\oplus \mathbb{J}_1\ .
\ee
Again, we are back to the nilpotent vev for the flipping fields we started with. 

The lesson from the previous case is the following: by moving forward to the right of a longer quiver tail we will have to deal with D-term equations of the form
\be
S^{R}S^{R\dagger} - \widetilde{S}^{R\dagger}\widetilde{S}^R + S^{L\dagger}S^{L} - \widetilde{S}^{L}\widetilde{S}^{L\dagger} = [\Psi_k,\Psi^\dagger_k]+\xi_k \mathbb{I}_k, 
\ee
for any $R=(k,k+1)$ and $L=(k-1,k)$. 
We show in Appendix \ref{Higgs} that terms labelled by `$R$' always cancel each other. 
On the other hand, terms labelled by `$L$' do not, and give a non zero commutator $[\Psi_k,\Psi^\dagger_k]$.  

The solution of F- and D-terms induced by the next-to-maximal nilpotent vev \eqref{start_rec} is:
\bea
 \langle\Psi_{k}\rangle = \mathbb{J}_1\oplus\mathbb{J}_{k-1},\qquad \langle \Psi_{N+1} \rangle = \mathbb{J}_{N}\oplus \mathbb{J}_1.
\eea
It is important to point out that $D$-terms equations are automatically solved by the
$SU(2)$ relation, $\rho_k(\sigma_3)=[\Psi_k,\Psi^\dagger_k]$ which follows from the construction of the nilpotent vev. In our case the embedding $\rho_k$ of the $\sigma_3$ element is 
\bea
&&
 \rho_k(\sigma_3)\equiv S^{L\dagger}S^{L} - \widetilde{S}^{L}\widetilde{S}^{L\dagger} ={\rm diag}( 0,-1,0_{k-3},+1), \qquad L=(k-1,k).
\eea
In this solution, the FI parameters $\xi_k$ are zero for any $k$. 
%

The list of scalar vevs includes the real scalars in the vector multiplets, which do not play any role, i.e. $\langle \sigma^k\rangle=0$.

\subsection*{The low energy theory}
Given the nilpotent vevs found in the previous section, we can explicitly study the Higgs mechanism and obtain the massless field content. 

Let us begin from the gauge sector. It is useful to recall that a generic gauge transformation on the quiver acts on the matter fields in the following fashion:
\bea
&&
G[ \{\Psi_{k},S^{(k,k+1)},\widetilde{S}^{(k,k+1)}\}_{k=1}^N ]= \bigoplus_{k=1}^N G_k
\eea
where $G_k$ is the action restricted to a single gauge node $U(k)$. Taking the connection 
\bea
\mathcal{A}_k= \sum_{a=1}^{k^2} g_a^k T^a 
\eea
we will find 
\bea
G_k= \sum_{a=1}^{k^2} g^a_k\, \left(  [T^a,  \Psi_k], - S^{(k-1,k)} T^a, T^a  S^{(k,k+1)} , T^a \widetilde{S}^{(k-1,k)},-  \widetilde{S}^{(k,k+1)}  T^a \right)    
\eea 
where we listed all the different matter representations.

 A broken generator does not leave the vev invariant, therefore
$T^a_{\mathfrak{R}}\langle z\rangle\neq 0$. Here $z$ stands (at least) 
for one among all the fields of the tail and 
the various representation have been indicated by $\mathfrak{R}$. 
Unbroken generators annhilate the vevs. 
The determination of unbroken generators is equivalent to the study of the kernel 
of the mass matrix obtained from the expansion of the covariant derivatives. 
More details on such a matrix are collected in the Appendix \ref{Higgs}.
In conclusion, fixing a basis of $T^a$ we find a solution for the coefficients $g^a_k$, 
which corresponds to a single unbroken generator $\mathcal{A}_k$ for gauge group $U(k)$. Its explicit form is very simple,
\be\label{U1quivergaugeF}
\mathcal{A}_k\equiv 
{\rm diag}(1,\underbrace{0,\ldots,0}_{k-1{\rm\, zeros}}\,)\ .
\ee
As far as the gauge groups are concerned, the quiver \eqref{tail_quiver_FCFH} is Higgsed to\\[.1cm]
\be\label{higgs_quiver_1}
             \begin{tikzpicture}
		
		\def\ox{0}
		\def\oy{0}
		\def\radius{.3}
		\def\lunghBif{1.5}
		\def\yshift{.1}
		
		\draw[line width=.6pt,dashed]    (\ox+2.5*\lunghBif,\oy+\yshift) --  (\ox+3*\lunghBif,\oy+\yshift) ;
		\draw[line width=.6pt,dashed]    (\ox+2.5*\lunghBif,\oy-\yshift) --  (\ox+3*\lunghBif,\oy-\yshift) ;		
		

		\foreach \intero in {1,2,3}           \node	at (\ox-\lunghBif+\intero*\lunghBif,\oy) 	[circle,inner sep=2mm,draw=black,fill=white,thick]   {$1$}    ;

		 \node	at (\ox+3.5*\lunghBif,\oy) 	[circle,inner sep=1.6mm,draw=black,fill=white,thick]   {$1$}    ;
		 
		\node     at   (\ox-\lunghBif+2.7*\lunghBif,\oy-.75)     {$\underbrace{\rule{6.25cm}{0cm}}_{N\ nodes}$};
		 		
	      \end{tikzpicture}
\ee

The next task will be to deduce the massless matter content by the studying the kernel of the mass matrix for all the scalar fields.  
We focus on the chiral multiplets.\footnote{ Real scalar fields $\{ \sigma^1\ldots \sigma^N\}$ in the vector multiples have the same mass matrix as the gauge fields. This follows from unbroken susy and it is obvious from a $4d$ perspective. In the $3d$ Lagrangian is manifests in matter couplings of the type $z^\dagger\sigma^2_{\mathfrak{R}} z$. In particular, on the vacuum $\langle\sigma^k\rangle=0$ there are no non trivial off-diagonal mass terms with chiral multiplets. }  
The mass matrix is hermitian and admits an eigenvector decomposition, which we split into ${ker}\,\oplus\,{ker}^\perp$, where the latter describes massive fields.
The massless sector will contain bifundamentals charged under $(L,R)$ gauge groups, fundamentals and anti-fundamentals charged under a single gauge group, other massless neutral fields, and finally Goldstone bosons. 
%
The `physical' massless fields of the IR theory correspond to those vectors in ${ker}$ which cannot be written only as linear combination of Goldstone bosons. 
On the other hand, a physical configuration might still have components along the directions parametrized by the Goldstone bosons.

The deformation $\mathcal{L}^{FF\chk{T}}$ breaks explicitly the non abelian flavor symmetry, therefore all the Goldstone bosons we will find correspond only to the action of broken gauge generators on the nilpotent vev. In the field variables $S$, $\widetilde{S}$, and $\Psi$, these Goldstone bosons are described by independent field configurations, of the form, 
\bea
&&
G_k= \sum_{a=1}^{k^2-1} g^a_k\, \left(  [T^a, \langle \Psi_k\rangle], -\langle S^{(k-1,k)}\rangle T^a, T^a \langle S^{(k,k+1)} \rangle, T^a\langle \widetilde{S}^{(k-1,k)}\rangle,- \langle \widetilde{S}^{(k,k+1)} \rangle T^a \right)    \nn
\eea
for parameters $\{g^{a=1,\ldots k^2-1}\}_{k=1}^N$ corresponding to broken gauge generators. 

We computed the mass matrix generated from the superpotential. Holomophy implies the existence of $\sum_{k=2}^{N} (k^2-1)=\tfrac{1}{6} (N-1) N (2 N+5)$ complex Goldstone bosons.
The resulting $ker$ can be quite cumbersome at first, but the physical massless fields can be brought to a simple form by taking linear combinations with Goldstone bosons, i.e. setting to zero unwanted components.  
We checked all our computations with computer algebra up to $N=6$.  After all this work is done, we find that most of the final answer can presented in a more intuitive way. 
This is the case for charged fields, as we now argue. IR flipping fields will have instead a more complicated description.

Let us begin from bifundamental fields in the abelian quiver \eqref{higgs_quiver_1}, i.e. fields simultaneously charged under a left and right gauge node.  
%
Considering the UV $S^{(k,k+1)}$, we want to select those components which transform non trivially under $\mathcal{A}_k$ and $\mathcal{A}_{k+1}$, where the gauge field is given explicitly in \eqref{U1quivergaugeF}.
It is simple to see that the first row of $S^{(k,k+1)}$ transforms non trivially under $\mathcal{A}_k$, while the first column transforms non trivially under $\mathcal{A}_{k+1}$. 
%
For $k=1,\ldots N-1$ the low energy bifundamentals, $s^{(k,k+1)}$ and $\tilde{s}^{(k,k+1)}$, embedded into $S^{(k,k+1)}$ and $\widetilde{S}^{(k,k+1)}$, are indeed in the $(1,1)$ entry,
\be
\label{bif_nilp_IR}
S^{(k,k+1)}\rightarrow \left[ \begin{array}{c|l} s^{(k,k+1)} & \ \ 0_{1\times k} \\ \hline 0_{k-1\times 1} & \ \ 0_{k-1\times k} \end{array}\right]\qquad
\widetilde{S}^{(k,k+1)}\rightarrow \left[ \begin{array}{c|l} \tilde{s}^{(k,k+1)}& \ \ 0_{1\times k-1} \\ \hline 0_{k\times 1} & \ \ 0_{k\times k-1} \end{array}\right]
\ee
The nilpotent vev \eqref{nilp_vev_tail} restricted on the the first two gauge nodes is vanishing. Indeed both $S^{(1,2)}=(s^{(1,2)},f_1)$ and $\widetilde{S}^{(1,2)}=(\widetilde{S}^{(1,2)},\tilde{f}_1)^T$ are massless in the IR. However the fields $f_1$ and $\tilde{f}_1$ are not charged under the $U(1)$ on the second gauge node, so they become a pair of fundamental/antifundamental attached to the first gauge node. 
On the flavor node at the end of the tail we also find a pair of fundamental/antifundamental fields, $f_N$ and $\tilde{f}_N$. 
In the case of $S^{(N,N+1)}$ we know from previous discussion that $f_N$ will lie on the first row, since this transforms under $\mathcal{A}_N$. However, differently from the bifundamentals $S^{(k,k+1)}$ in \eqref{bif_nilp_IR}, the location of $f_N$ inside $S^{(N,N+1)}$ depends on the form of the nilpontent vev \eqref{vevstep1}.
A similar reasoning holds for $\tilde{f}_N\subset\widetilde{S}^{(N,N+1)}$. The solution is,
\be\label{bif_nilp_IR_fl}
S^{(N,N+1)}\rightarrow \left[ \begin{array}{c|l}  \ 0_{1\times N} &\ \ f_N \\ \hline 0_{N-1\times N} & \ \ 0_{N-1\times 1} \end{array}\right]\qquad
\widetilde{S}^{(N,N+1)}\rightarrow \left[ \begin{array}{c|l} 0_{N\times 1}& \ \ 0_{1\times N-1} \\ \hline \rule{0pt}{.4cm}\tilde{f}_N & \ \ 0_{N\times N-1} \end{array}\right]
\ee
The IR quiver theory until now is described by the diagram
\be
             \begin{tikzpicture}
		
		\def\ox{0}
		\def\oy{0}
		\def\radius{.3}
		\def\lunghBif{1.5}
		\def\yshift{.1}
		
		\foreach \intero in {0,1}   \draw[line width=.6pt]    (\ox+\intero*\lunghBif,\oy+\yshift) --  (\ox+\lunghBif+\intero*\lunghBif,\oy+\yshift) ;
		\foreach \intero in {0,1}   \draw[line width=.6pt]    (\ox+\intero*\lunghBif,\oy-\yshift) --  (\ox+\lunghBif+\intero*\lunghBif,\oy-\yshift) ;
		
		\draw[line width=.6pt,dashed]    (\ox+2*\lunghBif,\oy+\yshift) --  (\ox+0.6*\lunghBif+2*\lunghBif,\oy+\yshift) ;
		\draw[line width=.6pt,dashed]    (\ox+2*\lunghBif,\oy-\yshift) --  (\ox+0.6*\lunghBif+2*\lunghBif,\oy-\yshift) ;		
		
		\draw[line width=.6pt,dashed]    (\ox+3*\lunghBif,\oy+\yshift) --  (\ox+0.5*\lunghBif+3*\lunghBif,\oy+\yshift) ;
		\draw[line width=.6pt,dashed]    (\ox+3*\lunghBif,\oy-\yshift) --  (\ox+0.5*\lunghBif+3*\lunghBif,\oy-\yshift) ;	
	
		\draw[line width=.6pt]    (\ox+3.5*\lunghBif,\oy+\yshift) --  (\ox+4.5*\lunghBif,\oy+\yshift) ;	
		\draw[line width=.6pt]    (\ox+3.5*\lunghBif,\oy-\yshift) --  (\ox+4.5*\lunghBif,\oy-\yshift) ;	
		 \node	at (\ox+.1*\lunghBif,\oy) 	[rectangle,inner sep=2.5mm,draw=black,fill=white,thick]   {$1$}   ;
		\foreach \intero in {2,3}           \node	at (\ox-\lunghBif+\intero*\lunghBif,\oy) 	[circle,inner sep=2mm,draw=black,fill=white,thick]   {$1$}    ;

		 \node	at (\ox+3.5*\lunghBif,\oy) 	[circle,inner sep=2mm,draw=black,fill=white,thick]   {$1$}    ;
		 \node	at (\ox+4.4*\lunghBif,\oy) 	[rectangle,inner sep=2.5mm,draw=black,fill=white,thick]   {$1$}   ;
		\node     at   (\ox-\lunghBif+3.25*\lunghBif,\oy-.75)     {$\underbrace{\rule{4.6cm}{0cm}}_{N\ nodes}$};
		 		
	      \end{tikzpicture}
\ee 

We move on to the study of neutral fields.
The simplest case is represented by singlet fields on each gauge node, which we denote by $\psi_k$. These are given by the following embedding: $\Psi_k=\mathrm{diag}(\psi_k,0^{k-1})$, $k=1,\ldots N$, as it could have been anticipated. 
The analysis of IR massless fields originating from the flipping fields $\Psi_{N+1}$ is less straighforward, 
and there is a novelty: It is not possible to localize such fields on components of $\Psi_{N+1}$, but
the corresponding vectors in $ker$ will have components on both $\Psi_{N+1}$ and $\Psi_{i\leq N}$. 
Furthermore, the latter cannot be eaten up by taking linear combinations with Goldstone bosons.

We begin by assuming $\Psi_{N+1}\in U(N+1)$ for simplicity, and we will obtain the case 
$\Psi_{N+1}\in SU(\Num)$, which is of interest for our duality web, by a minor modification. 
This procedure is instructive since it will have a counterpart in the next section.

Let us introduce first the IR fields $\Gamma_{i=2,\ldots N}$. In terms of components of $\Psi_{N+1}$, we find
\bea\label{conf_Psi_nilpot_0}
\Psi_{N+1}\supset \left[\begin{array}{ccccl|c} 0 & \Gamma_2 & \Gamma_3 & \ldots & \Gamma_N & 0 \\ 
				       0 &       0 &            \Gamma_2 & \ldots & \Gamma_{N-1} & 0 \\
				       \vdots & & & & \vdots & \vdots\\
				       \vdots & & & & \Gamma_2 & 0\\
				       0 &0 & 0&0 &0 &  0 \\
				       \hline 0 & 0 & 0 & 0 & 0 & 0\end{array}\right] &=&\left[  \Gamma_{2} \mathbb{J}_{N}\oplus \Gamma_{3} \mathbb{J}^2_{N} \ldots \oplus \Gamma_{N} \mathbb{J}^{N-1}_{N}\right]\oplus \mathbb{J}_1 
\eea
where the rewriting on the r.h.s makes manifest that these fields parametrize nilpotent directions inside $\Psi_{N+1}$. The configuration \eqref{conf_Psi_nilpot_0} extends on 
the UV adjoint fields $\Psi_{i\leq N}$, as follows 				       
\bea\label{conf_Psi_nilpot_1}			       
\Psi_{N+1-k}\supset \mathbb{J}_1\oplus \left[  \Gamma_{2} \mathbb{J}^{1+k}_{N-k}\oplus \Gamma_{3} \mathbb{J}^{2+k} _{N-k} \ldots \oplus \Gamma_{N-k} \mathbb{J}^{N-k-1}_{N-k}\right] \qquad k=1,\ldots N-2.	       
\eea
Matrices are multiplied a number of times defined by the upper index, i.e.  $\mathbb{J}^{\#}_{N-k}= \mathbb{J}^{}_{N-k}\cdots \mathbb{J}^{}_{N-k}$ $\#$ times. 
For example, notice that $\Gamma_2$ extends backwards to $\Psi_3$, $\Gamma_3$ extends backwards to $\Psi_4$, and so on. ($\Psi_1$ and $\Psi_2$ vanish in this case.) On a similar footing we find the field $\Delta$, which has the following UV embedding,
\be\label{conf_Psi_nilpot_2}
\Psi_{N+1}\supset \Delta\left[ \mathbb{I}_{N}\oplus\mathbb{J}_1\right],\qquad \Psi_{N+1-k}\supset \Delta\left[\mathbb{J}_1\oplus \mathbb{I}_{N-k}\right]\ .
\ee
Finally, there are three other fields, $\Sigma_{\pm}$ and $\delta$. These ones are localized on specific components of $\Psi_{N+1}$, 
\be
\left[\begin{array}{cc|c}  0_{1\times N-1} & 0 & \Sigma_+ \\ 
						      0_{2\times N-1} & 0 & 0\\
						      \vdots & \vdots & \vdots\\
						      0_{N\times N-1} & 0 & 0\\
						    \hline 0_{1\times N-1} & \Sigma_- & \delta
				                     \end{array}\right] 
\ee

The IR flipping fields presented so far have been obtained by refining the output of $ker$. We now explain how to see explicitly that these fields do not get a mass term from the superpotential. 
For each link $(L,R)$ of the quiver, consider the mass terms coming from fluctuations $\delta\Psi_L$, $\delta\Psi_R$ and $\delta S^{(L,R)}$, on top of the nilpotent vev. 
We find
\bea\label{mass_term_Gamma}
\delta W^{FF\chk{T}} &\supset&  \Tr\Big[ \ \left( \langle \widetilde{S}^{(L,R)}\rangle \delta \Psi_L - \delta\Psi_R \langle \widetilde{S}^{(L,R)}\rangle \right) \delta S^{(L,R)}+  \nn\\
		& & \qquad \left(  \delta \Psi_L \langle {S}^{(L,R)}\rangle - \langle {S}^{(L,R)}\rangle \delta\Psi_R \right) \delta\widetilde{S}^{(L,R)} \Big]
\eea
We have to show that when looking at the components of $\delta\Psi_{k\leq N+1}$ parametrized by $\Gamma_{i=2,..}$ or $\Delta$,
each term in \eqref{mass_term_Gamma} vanishes for any $(L,R)$. Indeed, because of the form of the nilpotent vev \eqref{nilp_vev_tail}, the equation
\bea
\langle \widetilde{S}^{(k,k+1)}\rangle \delta \Psi_k = \delta\Psi_{k+1} \langle \widetilde{S}^{(k,k+1)}\rangle\qquad k=1,\ldots N,\\ 
\delta \Psi_k \langle {S}^{(k,k+1)}\rangle =  \langle {S}^{(k,k+1)}\rangle \delta\Psi_{k+1}\qquad k=1,\ldots N.  
\eea
are solved precisely by \eqref{conf_Psi_nilpot_0}-\eqref{conf_Psi_nilpot_1} and \eqref{conf_Psi_nilpot_2}.
Moreover, on the link $(N,N+1)$, this same computation shows that the directions parametrized by $\Sigma_{\pm}$ and $\delta$ are also massless,
since when we multiply by the vev, these matrix elements are shifted either to the right, or to the bottom, by two units, i.e. they go out of the equations. 

The case of interest, $\Psi_{N+1}\in SU(N+1)$, is quite simple to deduce. Indeed only the IR fields $\Delta$ and $\delta$ parametrize directions which overlap with the identity. 
Therefore, out of these two, we should consider the traceless combination and drop the other. We associate to such a combination the IR field $\Gamma_1$, whose UV embedding is
\be
\Psi_{N+1}\supset\frac{\Gamma_1}{\sqrt{N^2+N}} \left[\begin{array}{cc} \mathbb{I}_{N\times N} & 0 \\  0 & - N \end{array}\right] \qquad  \Psi_{N+1-k}\supset \frac{ \Gamma_1}{ \sqrt{N^2+N}} \left[\,\mathbb{J}_1\oplus \mathbb{I}_{N-k}\right]\ .
\ee
\\

Collecting all the fields, the final low energy theory, which we denote by Theory C, is
described by the abelian quiver
\be
             \begin{tikzpicture}
		
		\def\ox{0}
		\def\oy{0}
		\def\radius{.3}
		\def\lunghBif{1.5}
		\def\yshift{.1}
		
		\foreach \intero in {0,1}   \draw[line width=.6pt]    (\ox+\intero*\lunghBif,\oy+\yshift) --  (\ox+\lunghBif+\intero*\lunghBif,\oy+\yshift) ;
		\foreach \intero in {0,1}   \draw[line width=.6pt]    (\ox+\intero*\lunghBif,\oy-\yshift) --  (\ox+\lunghBif+\intero*\lunghBif,\oy-\yshift) ;
		
		\draw[line width=.6pt,dashed]    (\ox+2*\lunghBif,\oy+\yshift) --  (\ox+0.6*\lunghBif+2*\lunghBif,\oy+\yshift) ;
		\draw[line width=.6pt,dashed]    (\ox+2*\lunghBif,\oy-\yshift) --  (\ox+0.6*\lunghBif+2*\lunghBif,\oy-\yshift) ;		
		
		\draw[line width=.6pt,dashed]    (\ox+3*\lunghBif,\oy+\yshift) --  (\ox+0.5*\lunghBif+3*\lunghBif,\oy+\yshift) ;
		\draw[line width=.6pt,dashed]    (\ox+3*\lunghBif,\oy-\yshift) --  (\ox+0.5*\lunghBif+3*\lunghBif,\oy-\yshift) ;	
	
		\draw[line width=.6pt]    (\ox+3.5*\lunghBif,\oy+\yshift) --  (\ox+4.5*\lunghBif,\oy+\yshift) ;	
		\draw[line width=.6pt]    (\ox+3.5*\lunghBif,\oy-\yshift) --  (\ox+4.5*\lunghBif,\oy-\yshift) ;	
		 \node	at (\ox+.1*\lunghBif,\oy) 	[rectangle,inner sep=2.5mm,draw=black,fill=white,thick]   {$1$}   ;
		\foreach \intero in {1,2}   	\node        at  (\ox+\intero*\lunghBif,\oy+5*\yshift) 	[circle,inner sep=2mm,draw=black,fill=white,thick]   {}    ;
		\foreach \intero in {2,3}           \node	at (\ox-\lunghBif+\intero*\lunghBif,\oy) 	[circle,inner sep=2mm,draw=black,fill=white,thick]   {$1$}    ;

	         \node        at  (\ox+3.5*\lunghBif,\oy+5*\yshift) 	[circle,inner sep=2mm,draw=black,fill=white,thick]   {}    ;
		 \node	at (\ox+3.5*\lunghBif,\oy) 	[circle,inner sep=2mm,draw=black,fill=white,thick]   {$1$}    ;
		 \node	at (\ox+4.4*\lunghBif,\oy) 	[rectangle,inner sep=2.5mm,draw=black,fill=white,thick]   {$1$}   ;
		 \node       at (\ox+6.4*\lunghBif,\oy) {$\Gamma_1,\ldots \Gamma_N, \,\Sigma_{\pm}$};
		\node     at   (\ox-\lunghBif+3.25*\lunghBif,\oy-.75)     {$\underbrace{\rule{4.6cm}{0cm}}_{N\ nodes}$};
		 		
	      \end{tikzpicture}
\ee 
A simple counting shows that we have determined $4 (N + 1)$ chiral fields.

\subsection{The $ABCD$ of monopole deformed $T[SU(\Num)]$}\label{ABCD_sec}

In this section we define theory $D$ as the Aharony dual of theory $A$, and we show that the mirror of theory $D$ {is precisely} theory $C$. 
Quite remarkably, mirror symmetry between $T[SU(\Num)]\leftrightarrow \chk{T[SU(\Num)]}$ and $FFT[SU(\Num)]\leftrightarrow FF\chk{T[SU(\Num)]}$ descends to mirror symmetry between $AB$ and $CD$.
The IR commutative diagram, initiated from $T[SU(\Num)]$ through the monopole deformation, is thus complete:

\begin{center}
             \begin{tikzpicture}[>=stealth]
            
             \def\ptAx{-3}
             \def\ptAy{-.5}
             \def\lato {3}
             \def\hor{6}
             \def\spazio{.3}
       
             \draw[line width=.6pt,<->]  (\ptAx+0.5*\lato+2*\spazio,\ptAy) -- (\ptAx+1.5*\lato-\spazio,\ptAy); 
             \draw[line width=.6pt,<->]  (\ptAx+1.5*\spazio,\ptAy-2*\spazio) --  (\ptAx+1.5*\spazio,\ptAy-\lato+2*\spazio);
             \draw[line width=.6pt,<->]  (\ptAx+2.55*\lato,\ptAy-2*\spazio) -- (\ptAx+2.55*\lato,\ptAy-\lato+2*\spazio);
             \draw[line width=.6pt,<->]  (\ptAx+0.5*\lato+2*\spazio,\ptAy-\lato) -- (\ptAx+1.5*\lato-\spazio,\ptAy-\lato); 

             \draw (\ptAx,\ptAy+.075) --  (\ptAx+.6,\ptAy+.075) ;
             \draw (\ptAx,\ptAy-.075) --  (\ptAx+.6,\ptAy-.075) ;
             \filldraw[fill=white,draw=black] (\ptAx,\ptAy) circle (.35cm);
             \filldraw[fill=white,draw=black] (\ptAx+.6,\ptAy-.3) rectangle (\ptAx+1.4,\ptAy+.3);
             \draw (\ptAx,\ptAy) node[font=\footnotesize] {$N$};
             \draw (\ptAx+1,\ptAy) node[font=\footnotesize] {$N$+1};
             \draw (\ptAx-.55,\ptAy+.09) node[font=\footnotesize] {$\vdots$};
             \foreach \y in {-.29,+.28} \filldraw (\ptAx-.55,\ptAy +\y) circle (.025cm);

             \draw (\ptAx,\ptAy-\lato+.075) --  (\ptAx+.6,\ptAy-\lato+.075) ;
             \draw (\ptAx,\ptAy-\lato-.075) --  (\ptAx+.6,\ptAy-\lato-.075) ;
             \filldraw[fill=white,draw=black] (\ptAx,\ptAy-\lato) circle (.35cm);
             \filldraw[fill=white,draw=black] (\ptAx+1.45,\ptAy-\lato) circle (.25cm);
             \filldraw[fill=white,draw=black] (\ptAx+.6,\ptAy-\lato-.3) rectangle (\ptAx+1.4,\ptAy-\lato+.3);
             \draw (\ptAx,\ptAy-\lato) node[font=\footnotesize] {1};
             \draw (\ptAx+1,\ptAy-\lato) node[font=\footnotesize] {$N$+1};
              \draw (\ptAx-.55,\ptAy-\lato+.09) node[font=\footnotesize] {$\vdots$};
             \foreach \y in {-.29,+.28}  \filldraw (\ptAx-.55,\ptAy-\lato +\y)  circle (.025cm);

             \draw (\ptAx+\hor+1,\ptAy+.075) --  (\ptAx+\hor-.6,\ptAy+.075) ;
             \draw (\ptAx+\hor+1,\ptAy-.075) --  (\ptAx+\hor-.6,\ptAy-.075) ;
             
              \filldraw[fill=white,draw=black] (\ptAx+\hor-.6,\ptAy-.3) rectangle (\ptAx+\hor-1.2,\ptAy+.3);
              \draw (\ptAx+\hor-.9,\ptAy) node[font=\footnotesize] {1};
             
             \draw (\ptAx+\hor,\ptAy+.075) --  (\ptAx+\hor+1,\ptAy+.075) ;
             \draw (\ptAx+\hor,\ptAy-.075) --  (\ptAx+\hor+1,\ptAy-.075) ;
             \draw (\ptAx+\hor+.075,\ptAy) --  (\ptAx+\hor+.075,\ptAy-1) ;
	     \draw (\ptAx+\hor-.075,\ptAy) --  (\ptAx+\hor-.075,\ptAy-1) ; 
	    \filldraw[fill=white,draw=black] (\ptAx+\hor-.3,\ptAy-.6) rectangle (\ptAx+\hor+.3,\ptAy-1.2);
       	     \draw (\ptAx+\hor,\ptAy-.9) node[font=\footnotesize] {1}; 
             
             \filldraw[fill=white,draw=black] (\ptAx+\hor,\ptAy+.3) circle (.25cm);
             \filldraw[fill=white,draw=black] (\ptAx+\hor,\ptAy) circle (.355cm);
             \draw (\ptAx+\hor,\ptAy) node[font=\footnotesize] {$N$};
             \filldraw[fill=white,draw=black] (\ptAx+\hor+1,\ptAy+.3) circle (.25cm);
             \filldraw[fill=white,draw=black] (\ptAx+\hor+1,\ptAy) circle (.355cm);
             \draw (\ptAx+\hor+1,\ptAy) node[font=\footnotesize] {$N$-1};
             
              \draw (\ptAx+\hor+1.62,\ptAy) node[font=\footnotesize] {$\ldots$};
             
             \filldraw[fill=white,draw=black] (\ptAx+\hor+2.55,\ptAy) circle (.25cm);
             \filldraw[fill=white,draw=black] (\ptAx+\hor+2.25,\ptAy) circle (.355cm);
             \draw (\ptAx+\hor+2.25,\ptAy) node[font=\footnotesize] {1};

             \draw (\ptAx+\hor+1,\ptAy-\lato+.075) --  (\ptAx+\hor-.6,\ptAy-\lato+.075) ;
             \draw (\ptAx+\hor+1,\ptAy-\lato-.075) --  (\ptAx+\hor-.6,\ptAy-\lato-.075) ;
             
              \filldraw[fill=white,draw=black] (\ptAx+\hor-.6,\ptAy-\lato-.3) rectangle (\ptAx+\hor-1.2,\ptAy-\lato+.3);
               \draw (\ptAx+\hor-.9,\ptAy-\lato) node[font=\footnotesize] {1};
             
              \filldraw[fill=white,draw=black] (\ptAx+\hor,\ptAy-\lato+.3) circle (.25cm);
	     \filldraw[fill=white,draw=black] (\ptAx+\hor,\ptAy-\lato) circle (.35cm);
	      \draw (\ptAx+\hor,\ptAy-\lato) node[font=\footnotesize] {1};
             \filldraw[fill=white,draw=black] (\ptAx+\hor+1,\ptAy-\lato+.3) circle (.25cm);
             \filldraw[fill=white,draw=black] (\ptAx+\hor+1,\ptAy-\lato) circle (.35cm);
             \draw (\ptAx+\hor+1,\ptAy-\lato) node[font=\footnotesize] {1};
             
              \draw (\ptAx+\hor+1.62,\ptAy-\lato) node[font=\footnotesize] {$\ldots$};

             \draw (\ptAx+\hor+2.25,\ptAy-\lato+.075) --  (\ptAx+\hor+3,\ptAy-\lato+.075) ;
             \draw (\ptAx+\hor+2.25,\ptAy-\lato-.075) --  (\ptAx+\hor+3,\ptAy-\lato-.075) ; 
              
             \filldraw[fill=white,draw=black] (\ptAx+\hor+2.27,\ptAy-\lato+.3) circle (.25cm);           
             \filldraw[fill=white,draw=black] (\ptAx+\hor+2.27,\ptAy-\lato) circle (.35cm);
             \draw (\ptAx+\hor+2.27,\ptAy-\lato) node[font=\footnotesize] {1};
                          
             \filldraw[fill=white,draw=black] (\ptAx+\hor+2.9,\ptAy-\lato-.3) rectangle (\ptAx+\hor+3.5,\ptAy-\lato+.3);
             
             \draw (\ptAx+\hor+3.2,\ptAy-\lato) node[font=\footnotesize] {1};
             
             \draw (\ptAx+\hor+3.8,\ptAy-\lato+.09) node[font=\footnotesize] {$\vdots$};
             \foreach \y in {-.29,+.28} \filldraw (\ptAx+\hor+3.8,\ptAy-\lato +\y) circle (.025cm);

             \def\ptBx{-4.5}
             \def\ptBy{+1.5}
             \def\latox {12}
             \def\latoy {7}

             \def\spazio{.2}
             
            \draw[line width=.6pt,<->]  (\ptBx+7*\spazio,\ptBy) -- (\ptBx+\latox-7.5*\spazio,\ptBy); 
            \draw[line width=.6pt,<->]  (\ptBx,\ptBy-2*\spazio) -- (\ptBx,\ptBy-\latoy+2*\spazio);
            \draw[line width=.6pt,<->]  (\ptBx+\latox,\ptBy-2*\spazio) -- (\ptBx+\latox,\ptBy-\latoy+2*\spazio);
            \draw[line width=.6pt,<->]  (\ptBx+7*\spazio,\ptBy-\latoy) -- (\ptBx+\latox-7.5*\spazio,\ptBy-\latoy); 

             \node[scale=.95] at (\ptBx,\ptBy+0.5*\spazio) {$T[SU(\Num)]$};
             \node[scale=.95] at (\ptBx-\spazio,\ptBy-\latoy-0.5*\spazio) {$FFT[SU(\Num)]$};
             \node[scale=.95] at (\ptBx+\latox,\ptBy+0.45*\spazio) {$\chk{T[SU(\Num)]}$};
	     \node[scale=.95] at (\ptBx+\latox+\spazio,\ptBy-\latoy-0.5*\spazio) {$FF\chk{T[SU(\Num)]}$};
 
             
             \draw[line width=.6pt,->]  (\ptBx+2*\spazio,0.8*\ptBy)--(\ptAx-0.75*\spazio,\ptAy+3*\spazio); 
             \draw[line width=.6pt,->]  (\ptBx+\latox-3*\spazio,0.8*\ptBy)--(\ptAx+1.2*\hor+6*\spazio,\ptAy+3*\spazio); 
             \draw[line width=.6pt,->]  (\ptBx+2*\spazio,\ptBy-\latoy+2*\spazio)--(\ptAx-0.75*\spazio,\ptAy-0.5*\latoy-1*\spazio); 
             \draw[line width=.6pt,->]  (\ptBx+\latox-3*\spazio,\ptBy-\latoy+2*\spazio)--(\ptAx+1.2*\hor+6*\spazio,\ptAy-0.5*\latoy-1*\spazio);

             \node at (\ptAx+1.2*\spazio,\ptAy+4*\spazio) {$A$};
             \node at (\ptAx+1.2*\hor+4*\spazio,\ptAy+4*\spazio) {$B$};
             \node at (\ptAx+1.2*\hor+4*\spazio,\ptAy-0.5*\latoy-1.75*\spazio) {$C$};
             \node at (\ptAx+1.2*\spazio,\ptAy-0.5*\latoy-1.75*\spazio) {$D$};

             \end{tikzpicture}        
\end{center}

~\\[.1cm]
Let us remind that theory $A$ is $U(N)$ SQCD with $N+1$ flavors coupled to additional singlets $\gamma_{m}$ through the superpotential 
\be
W_A=  -\sum_{m=1}^{N }\frac{(-)^{m} }{m} \ \gamma_m\ \Tr_{N+1}[\,\mathcal{Q}^m] \,.
\ee
We apply the Aharony duality \cite{Aharony:1997gp} to Theory ${A}$ and obtain Theory ${D}$, a $U(1)$ gauge theory with $N+1$ flavors $U_{i}$ and $\tilde{U}_j$, 
flipping fields $\Flip^{\,\mathcal{U}}_{\,ij}$ for the meson $\mathcal{U}_{ij}=U_i\tilde{U}_j$, 
and flipping fields $\sigma_{\pm}$ for the $U(1)$ monopoles $\m^{\pm}$. (For simplicity we borrow from the $FFT[SU(\Num)]$ the notation for the monopoles).
In addition we denote by $\theta_m$ the dual of the singlets $\gamma_m$.
The quiver diagram is
 \be\label{theory_D}
             \begin{tikzpicture}
		
		\def\ox{0}
		\def\oy{0}
		\def\radius{.3}
		\def\lunghBif{1.5}
		\def\yshift{.1}
	
	         \node at (\ox-5*\radius,\oy-.05) {$\theta_1,\ldots \theta_N,$};
	
		\foreach \intero in {0}   \draw[line width=.6pt]    (\ox+\intero*\lunghBif,\oy+\yshift) --  (\ox+\lunghBif+\intero*\lunghBif,\oy+\yshift) ;
		\foreach \intero in {0}   \draw[line width=.6pt]    (\ox+\intero*\lunghBif,\oy-\yshift) --  (\ox+\lunghBif+\intero*\lunghBif,\oy-\yshift) ;

		\node        at  (\ox+1.65*\lunghBif,\oy)										[circle,inner sep=2.5mm,draw=black,fill=white,thick]   {}    ;	

%
		\foreach \intero in {1}                \node	        at (\ox-\lunghBif+\intero*\lunghBif,\oy) 	[circle,inner sep=2mm,draw=black,fill=white,thick]   {$1$}    ;
						     	 	\node	at (\ox+1.1*\lunghBif,\oy) 	        [minimum height=.8cm, draw=black,fill=white,thick]   {$N+1$}   ;

	      \end{tikzpicture}
\ee
The flipping fields $\Flip^{\,\mathcal{U}}_{\,ij}$ of theory $D$ are dual to the electric meson of Theory $A$, \
so the superpotential $W_D$ becomes
\begin{eqnarray}
W_D=\m^{\pm} \sigma_\pm    + \mathcal{U}_{ij} \Flip^{\mathcal{U}}_{ij}- \sum_{m=1}^{N } \frac{(-)^m}{m}\ \theta_m\,\Tr\Big[\underbrace{ \,\Flip^{\mathcal{U}}\ldots\Flip^{\mathcal{U}}\, }_{m{\rm\ times}} \Big]\,.
\label{W_D_superP}
\end{eqnarray}
%
Both $\mathcal{U}_{ij}$ and $\Flip^{\mathcal{U}}$ belong to adjoint of $U(N+1)$, since they originate from Aharony duality.\\

The mirror of theory $D$, which we will identify with theory $C$, is now obtained by applying piecewise mirror symmetry \cite{Kapustin:1999ha}.
This procedure amounts to replace each flavor $U_i,\widetilde{U}_i$ with an SQED theory coupled to a singlet $\chi_i$,
and do the functional integration on the $U(1)$ gauge node of \eqref{theory_D}.
For each SQED theory, there is a cubic superpotential is of the form $\chi s\widetilde{s}$, if $s$ and $\tilde{s}$ denote schematically the flavors. 
We redefine the set of $\chi_i$ as follows,
\bea
\sum_{i}\chi_i=(N+1){\delta}&\quad;\quad &
\left\{
\begin{array}{ccl}
\chi_1&=&{\delta}{}-\psi_1,\\[.2cm]
\chi_2&=&{\delta}{}+\psi_2+\psi_1,\\
\ \vdots \\
\chi_{N+1}&=& {\delta}{}+\psi_{N}.
\end{array}\right.
\eea
Then the cubic superpotentials can be presented in the form
\be\label{cubic_superp}
 \sum_{i=0}^{N} s^{(i,i+1)}{\delta}\,\widetilde{s}^{(i,i+1)} +\sum_{i=1}^N \psi_i\big(  s^{(i,i+1)}\widetilde{s}^{(i,i+1)} -  s^{(i-1,i)}\widetilde{s}^{(i-1,i)}\big) 
\ee
and the resulting theory is the abelian quiver \cite{Benvenuti:2016wet}\\[.1cm]
\be\label{abelian_DC}
             \begin{tikzpicture}
		
		\def\ox{0}
		\def\oy{0}
		\def\radius{.3}
		\def\lunghBif{1.5}
		\def\yshift{.1}
		
		\foreach \intero in {0,1}   \draw[line width=.6pt]    (\ox+\intero*\lunghBif,\oy+\yshift) --  (\ox+\lunghBif+\intero*\lunghBif,\oy+\yshift) ;
		\foreach \intero in {0,1}   \draw[line width=.6pt]    (\ox+\intero*\lunghBif,\oy-\yshift) --  (\ox+\lunghBif+\intero*\lunghBif,\oy-\yshift) ;
		
		\draw[line width=.6pt,dashed]    (\ox+2*\lunghBif,\oy+\yshift) --  (\ox+0.6*\lunghBif+2*\lunghBif,\oy+\yshift) ;
		\draw[line width=.6pt,dashed]    (\ox+2*\lunghBif,\oy-\yshift) --  (\ox+0.6*\lunghBif+2*\lunghBif,\oy-\yshift) ;		
		
		\draw[line width=.6pt,dashed]    (\ox+3*\lunghBif,\oy+\yshift) --  (\ox+0.5*\lunghBif+3*\lunghBif,\oy+\yshift) ;
		\draw[line width=.6pt,dashed]    (\ox+3*\lunghBif,\oy-\yshift) --  (\ox+0.5*\lunghBif+3*\lunghBif,\oy-\yshift) ;	
	
		\draw[line width=.6pt]    (\ox+3.5*\lunghBif,\oy+\yshift) --  (\ox+4.5*\lunghBif,\oy+\yshift) ;	
		\draw[line width=.6pt]    (\ox+3.5*\lunghBif,\oy-\yshift) --  (\ox+4.5*\lunghBif,\oy-\yshift) ;	
		 \node	at (\ox+.1*\lunghBif,\oy) 	[rectangle,inner sep=2.5mm,draw=black,fill=white,thick]   {$1$}   ;
		\foreach \intero in {1,2}   	\node        at  (\ox+\intero*\lunghBif,\oy+5*\yshift) 	[circle,inner sep=2mm,draw=black,fill=white,thick]   {}    ;
		\node        at  (\ox+3.5*\lunghBif,\oy+5*\yshift) 	[circle,inner sep=2mm,draw=black,fill=white,thick]   {}    ;
		\node        at  (\ox+4.75*\lunghBif,\oy-0*\yshift) 	[circle,inner sep=2mm,draw=black,fill=white,thick]   {}    ;

		\foreach \intero in {2,3}           \node	at (\ox-\lunghBif+\intero*\lunghBif,\oy) 	[circle,inner sep=2mm,draw=black,fill=white,thick]   {$1$}    ;

		 \node	at (\ox+3.5*\lunghBif,\oy) 	[circle,inner sep=2mm,draw=black,fill=white,thick]   {$1$}    ;
		 \node	at (\ox+4.4*\lunghBif,\oy) 	[rectangle,inner sep=2.5mm,draw=black,fill=white,thick]   {$1$}   ;
 		 \node     at   (\ox-\lunghBif+3.3*\lunghBif,\oy-.75)     {$\underbrace{\rule{4.6cm}{0cm}}_{N\ nodes}$};
		 		
	      \end{tikzpicture}
\ee  
As usual, we are distinguishing the flavors $s,\tilde{s}$, with a label $(L,R)$. In particular, 
the fields $s^{(0,1)}$ and $\widetilde{s}^{(0,1)}$ are a pair of fundamental/anti-fundamental on the first link, while
the fields $s^{(i,i+1)}$ and $\widetilde{s}^{(i,i+1)}$ for $i=1,\ldots N$ are bifundamentals.
The change of variable from $\chi_{i=1,..N+1}$ to $\{{\delta},\psi_{i=1,..N}\}$ can be understood as the arrangement of $U(1)^{N+1}$ into a diagonal $U(1)$, 
parametrized by $\delta$, and the Cartan of $SU(N+1)_{\mathrm{top}}$. 
Then, we can think of $\psi_i$ as a singlet attached to the $i$-th gauge node of the quiver \eqref{abelian_DC}.  
Note that the second term of the superpotential \eqref{cubic_superp} is $\mathcal{N}=4$. 
Finally, the field ${\delta}$ has been represented with the horizontal loop on the last flavor node in \eqref{abelian_DC}. 

The mirror of theory $D$ is completed once we map the superpotential $W_D$. In order to do so we should refine the map of the operators. 
Mirror symmetry would relate the meson $\mathcal{U}_{ij}$ to the monopole matrix $\n_{\,ij}$. 
But since the meson $\mathcal{U}_{ij}$ is not traceless, there is a mismatch of representations we have to take care of. 
More precisely, we claim that the $SU(N+1)$ degrees of freedom of  $\mathcal{U}_{ij}$ are mirror to the monopole matrix $\n_{\,ij}$, which is traceless, 
while the trace $\Tr\,\mathcal{U}$ is in correspondence with $\delta$. 
The rest of the dictionary is standard: The monopole fields $\m_{\pm}$ of theory $D$ are mirror to the long meson $L_+=\prod_{i=0}^{N} s_i$ and $L_-=\prod_{i=0}^{N} \widetilde{S}_i$, and
the flipping fields $\Flip_{ij}^{\mathcal{U}}$, $\sigma_{\pm}$ and $\theta_m$ are mapped to an equivalent number of singlets, $\Flip_{ij}^{\n}$, $\Sigma'_{\pm}$, and $\Gamma'_m$.

The superpotential $\chk{W_D}$ is
\bea
\chk{W_D}&=& \sum_{i=0}^{N} s^{(i,i+1)}\delta\, \tilde{s}^{(i,i+1)} +\sum_{i=1}^N \psi_i\big(  s^{(i,i+1)}\tilde{s}^{(i,i+1)} -  s^{(i-1,i)}\tilde{s}^{(i-1,i)}\big)    \notag\\
& &+  L_{\pm}\Sigma'_{\pm} +\left(  {\delta}{}\,\Tr\,\Flip_{ij}^{\n}+ \n_{\,ij} \Flip_{ij}^{\n}\right) -  \sum_{m=1}^{N } \frac{(-)^m}{m}\ \Gamma'_m\,\Tr\Big[\underbrace{ \,\Flip_{ij}^{\n}\ldots\Flip_{ij}^{\n}\, }_{m{\rm\ times}} \Big]\,
\eea
The terms $\Gamma'_1 \Tr\Flip_{ij}^{\n}$ and $ {\delta}{}\, \Tr\Flip_{ij}^{\n} $ combine into the mass term $(\Gamma'_1+\delta)\Tr\Flip_{ij}^{\n}$. Then, both $(\Gamma'_1+\delta)$ and $\Tr\Flip_{ij}^{\n}$ can be integrated out, while
the field $\Gamma_1\equiv (\Gamma'_1-\delta)$ remains massless. After trivial redefinitions, 
\bea
\chk{W_D}&=& \sum_{i=0}^{N} s^{(i,i+1)}\Gamma_1\, \tilde{s}^{(i,i+1)} +\sum_{i=1}^N \psi_i\big(  s^{(i,i+1)}\tilde{s}^{(i,i+1)} -  s^{(i-1,i)}\tilde{s}^{(i-1,i)}\big)    \notag\\
& &+  L_{\pm}\Sigma_{\pm} + \n_{\,ij} \Flip_{ij}^{\n} -  \sum_{m=2}^{N } \frac{(-)^m}{m}\ \Gamma_m\,\Tr\Big[\underbrace{ \,\Flip_{ij}^{\n}\ldots\Flip_{ij}^{\n}\, }_{m{\rm\ times}} \Big]\,{ \equiv W_C}
\eea
The notation $\psi_i$, $s^{(i,i+1)}$, $\Flip_{ij}^{\n}$ and $\n_{\,ij}$ should be familiar from the study of theory C. 
We have found $\Flip_{ij}^{\n}\in SU(N+1)$, $\n_{\,ij}\in SU(N+1)$, and other $4(N+1)$ fields. These corresponds to bifundamentals, and fundamentals on the right and and left of \eqref{abelian_DC}, in addition to the singlets $\psi_{i=1,.N}$, the fields $\Sigma_{\pm}$, and $\Gamma_{i=1,..N}$.
Remarkably, this number is precisely the same number we determined in section \ref{nilpot_higg} from the nilpotent Higgsing.\\

{
We have not discussed how the deformation $\mathcal{L}^{FFT}$ brings  $FFT[SU(\Num)]$ down to Theory D. This would require a study of the Higgsing process on monopole fields, 
a challenge which is the behind the immediate scope of this paper. 
}

\subsection*{Operator map}
We conclude this section by recording the Chiral ring generators which we are able to map across the four dual frames $ABCD$:

\begin{itemize}
\item Theory $A$: 
\begin{itemize}
\item[$\clubsuit$]Two monopoles with $R[\Mon^\pm_A]=2-r(N+1)r$
\item[$\blacklozenge$] HB moment map, $(N+1)\times (N+1)$ traceless, with  $R[\Pi^\mathcal{Q}_A]=2r$
\item[$\spadesuit$] Flipping fields with $R[\gamma_m]=2-2rm$, $m=1,\cdots N$
\end{itemize}
\item Theory $B$: 
\begin{itemize}
\item[$\clubsuit$] Two mesons with $R[d\tilde p_{N+1}]=R[\tilde d p_{N+1}]=2-(N+1)r$
\item[$\blacklozenge$] Monopoles matrix,  $(N+1)\times (N+1)$ traceless, with $R[\Non_B]=2r$
\item[$\spadesuit$]  Dressed mesons with $R[\tilde d \Omega^i d]=2(1-r(N-i))$, $i=0,\cdots N-1$
\end{itemize}
\item Theory $C$:
\begin{itemize}
\item[$\clubsuit$] Two flipping fields with $R[\Sigma^\pm]=2-r(N+1)$
\item[$\blacklozenge$] Flipping fields,  $(N+1)\times (N+1)$  traceless, with  $R[F^\n_C]=2r$
\item[$\spadesuit$]  Flipping fields with $R[\Gamma_m]=2-2rm$, $m=1,\cdots N$
\end{itemize}
\item Theory $D$:
\begin{itemize}
\item[$\clubsuit$]Two Flipping fields with $R[\sigma^\pm]=2-(N+1)r$
\item[$\blacklozenge$] Flipping fields,  $(N+1)\times (N+1)$ traceless, with $R[F^\mathcal{U}_D]=2r$
\item[$\spadesuit$]  Flipping fields with $R[\theta_m]=2-2rm$, $m=1,\cdots N$.
\end{itemize}
\end{itemize}

\section{Partition Functions}\label{difo_gen}

In this section we study partition functions of our theories on the squashed three-sphere $S^3_b$, and we 
show that they are all equal as we move in the commutative diagram: 
\be\label{Part_func_diagrm}
             \begin{tikzpicture}
             \def\ptAx{0}
             \def\ptAy{0}
             \def\lato {2}
             \def\spazio{.3}
       
             \draw[line width=.6pt,<->]  (\ptAx+\spazio,\ptAy) -- (\ptAx+\lato-\spazio,\ptAy);
             \draw[line width=.6pt,<->]  (\ptAx,\ptAy-\spazio) -- (\ptAx,\ptAy-\lato+\spazio);
             \draw[line width=.6pt,<->]  (\ptAx+\lato,\ptAy-\spazio) -- (\ptAx+\lato,\ptAy-\lato+\spazio);
             \draw[line width=.6pt,<->]  (\ptAx+\spazio,\ptAy-\lato) -- (\ptAx+\lato-\spazio,\ptAy-\lato);

             \node at (\ptAx,\ptAy) {$A$};
             \node at (\ptAx+\lato-0.1*\spazio,\ptAy-0.05*\spazio) {$B$};
             \node at (\ptAx+\lato,\ptAy-\lato+0.1*\spazio) {$C$};
              \node at (\ptAx,\ptAy-\lato+0.1*\spazio) {$D$};

             \def\ptBx{-1}
             \def\ptBy{+1}
             \def\lato {4}
             \def\spazio{.2}
             
             \draw[line width=.6pt,<->]  (\ptBx+\spazio,\ptBy) -- (\ptBx+\lato-\spazio,\ptBy);
             \draw[line width=.6pt,<->]  (\ptBx,\ptBy-\spazio) -- (\ptBx,\ptBy-\lato+\spazio);
             \draw[line width=.6pt,<->]  (\ptBx+\lato,\ptBy-\spazio) -- (\ptBx+\lato,\ptBy-\lato+\spazio);
             \draw[line width=.6pt,<->]  (\ptBx+\spazio,\ptBy-\lato) -- (\ptBx+\lato-\spazio,\ptBy-\lato);

             \node[scale=.95] at (\ptBx-5.5*\spazio,\ptBy+0.5*\spazio) {$T[SU(\Num)]$};
             \node[scale=.95] at (\ptBx-6.8*\spazio,\ptBy-\lato-0.5*\spazio) {$FFT[SU(\Num)]$};
             \node[scale=.95] at (\ptBx+\lato+5.5*\spazio,\ptBy+0.45*\spazio) {$\chk{T[SU(\Num)]}$};
	     \node[scale=.95] at (\ptBx+\lato+7*\spazio,\ptBy-\lato-0.5*\spazio) {$FF\chk{T[SU(\Num)]}$};
 
             
             \draw[line width=.6pt,->]  (\ptBx+\spazio,0.8*\ptBy)--(\ptAx-0.75*\spazio,\ptAy+0.75*\spazio); 
             \draw[line width=.6pt,->]  (\ptBx+\lato-\spazio,0.8*\ptBy)--(\ptAx+0.5*\lato+0.75*\spazio,\ptAy+0.75*\spazio); 
             \draw[line width=.6pt,->]  (\ptBx+1*\spazio,\ptBy-\lato+0.75*\spazio)--(\ptAx-1.1*\spazio,\ptAy-0.5*\lato-1*\spazio); 
             \draw[line width=.6pt,->]  (\ptBx+\lato-1*\spazio,\ptBy-\lato+0.75*\spazio)--(\ptAx+0.5*\lato+0.7*\spazio,\ptAy-0.5*\lato-1*\spazio);

             \end{tikzpicture}        
\ee      
We follow the notation of \cite{Hama:2011ea}. We introduce the  vectors
$\vec{M}=(M_1,\ldots M_{N+1})$ and $\vec{T}=(T_1,\ldots T_{N+1})$ of real mass parameters for the flavor and topological symmetries and
 the real mass $m_A$  associated to the $U(1)_A$ symmetry. We also define
 $Q\equiv b+b^{-1}$, where $b$ is the squashing of the three-sphere. 
Then, the partition function of $T[SU(N+1)]$ can be obtained by the following set of rules:
\begin{itemize}
\item Each one of the $N$ gauge nodes, labelled by $(k)$ with $k=1,\ldots N$, 
carries a measure $dx^{(k)}=\prod_{i=1}^k {dx^{(k)}_i}\big/{k!}$ where the set $\{x_i^{(k)}\}$ represents the Coulomb Branch coordinates on the localizing locus. 
\item The contribution from vector multiplets and adjoint chirals attached to a node $(k)$ is 
\bea
Z_{\rm vec}^{(k)}&=&\prod_{i<j}^k  \frac{1}{s_b\left(\tfrac{iQ}{2}\pm(x^{(k)}_i-x^{(k)}_j) \right)} \,,\\
Z_{\rm adj}^{(k)}&=& \prod_{i,j}^k  s_b\left(m_A\pm(x^{(k)}_i-x^{(k)}_j) \right)  \,.
\eea
\item
The contribution of bifundamentals on a link $(k,k+1)$ is
\be
 {Z}_{\rm bif}^{(k,k+1)}=\prod_{i=1}^k\prod_{j=1}^{k+1} s_b\left(\tfrac{iQ}{4}-\tfrac{m_A}{2}\pm(x^{(k)}_i-x^{(k+1)}_j)\right)\,.
 \ee
%
 \end{itemize}
%
As pointed out in \cite{Jafferis:2010un}, the partition function depends holomorphically 
on the combination of the real mass parameter $m_A$, and the coefficient determining the IR R-symmetry.
Then, we will take ${\rm Im}(m_A)=-\tfrac{Q}{2}\alpha$ 
with $\alpha$ parametrizing the mixing $R=C+H+\alpha(C-H)$. In this conventions, a chiral multiplet of R-charge $r$ contributes 
with $s_b(\tfrac{iQ}{2} (1-r)- \ldots)$ to the partition function \cite{Hama:2011ea}, and
from $Z_{\rm bif}$ and $Z_{\rm adj}$ we read off
\be
R_{\rm bif}
=\frac{1-\alpha}{2}=r,\qquad
R_{\Phi}=
1+\alpha=2(1-r).
\label{R_charges_part_func}
\ee 
This is indeed the same assignment we discussed in Section~\ref{tsun_sec}.

Putting all together the partition function of the tail \eqref{quiverT} is:
\bea
\mathcal{Z}^T[N, m_A;\vec M, \vec T]&=& e^{+2\pi i T_{N+1}  \sum_{i=1}^{N+1} M_i}
\int \prod_{k=1}^N dx^{(k)} e^{2\pi i~ \xi^{(k)} \sum_{i=1}^k x^{(k)}_i} 
Z_{\rm vec}^{(k)}~Z_{\rm adj}^{(k)}~ Z_{\rm bif}^{(k,k+1)}\qquad\label{tail_part_func}
\eea
where $\xi^{(k)}=T_{k}-T_{k+1}$, and $\vec{x}^{(N+1)}\equiv\vec{M}$ a constant vector, i.e. not integrated over. 
Finally, the partition function of $T[SU(\Num)]$ is a specification of $\mathcal{Z}^T$  to the case  $\sum_{i=1}^{N+1}T_i=\sum_{i=1}^{N+1}M_i=0$,
consistent with the non-abelian global symmetry $SU(N+1)_{\rm flavor} \times SU(N+1)_{\rm top}$.

\subsection{Difference operators and dual partition functions}\label{difo}

In this section we consider the outer diagram \eqref{Part_func_diagrm}, in which  $T[SU(\Num)]$ is dual to $FFT[SU(\Num)]$ and mirror to $\chk{T[SU(\Num)]}$, and 
show that the various partition functions are all equal, as function of the global symmetry parameters: $M_p$, $T_p$ and $m_A$. 

The partition functions $\chk{T[SU(\Num)]}$ is related in a straighforward way to $\mathcal{Z}^{T}$.
The action of mirror symmetry on $T[SU(\Num)$ defines
\be
\mathcal{Z}^{\chk{T}}[N,m_A;\vec M,\vec
T]=\mathcal{Z}^T[N,-m_A;\vec T,\vec M].
\ee
In particular $\mathcal{Z}^{\chk{T}}$ is given by the same matrix integral as
$T[SU(\Num)]$, where masses $M_a$ and FI parameters
$T_a$ are swapped, and the sign of the axial mass $m_A$ inverted.
This is consistent with the fact that mirror symmetry exchanges HB and CB.

Our prescription for the partition function of ${FFT[SU(\Num)]}$ is
\be\label{presc_FFT}
\mathcal{Z}^{FFT}[N,m_A;\vec M,\vec T]=K[\vec M, -m_A]K[\vec T, m_A] \mathcal{Z}^{T}[N,-m_A;\vec M,\vec T]
\ee
 where 
\be
K[\vec x, m_A] =Z_{\rm adj}^{(N)}&=& \prod_{i,j}^N  s_b\left(m_A\pm(x_i-x_j) \right)=K[\vec x, -m_A]^{-1}  \,.
\ee
The factor $K[\vec x,\pm m_A]$ are used to introduce the contribution of flipping fields $\Flip^{\mathcal{R}}_{ij}$ and $\Flip^{\m}_{ij}$, 
for the moment map and the monopole, respectively.
The two signs of $m_A$ in $K[\vec M, -m_A]$ and $K[\vec T, m_A]$ are consistent with the fact that in
${FFT[SU(\Num)]}$ 
the HB flipping fields $\Flip^{\mathcal{R}}_{ij}$ have $R$-charge  $2r$, while the CB flipping fields $\Flip^{\m}_{ij}$ have $R$-charge $2-2r$.
Notice that the diagonal elements $i=j$ in the product $K[\vec M, -m_A]K[\vec T, m_A]$ simplify to unity. 
Therefore we can understand this product as the contribution of (singlets) adjoint fields in the $SU(N+1)_{\mathrm{flavor}}\times SU(N+1)_{top.}$.

Proving our dualities is equivalent to show that:
\bea
\mathcal{Z}^{\chk{T}}[N,m_A;\vec M,\vec T]&=&\mathcal{Z}^T[N,-m_A;\vec T,\vec M]
=\mathcal{Z}^{T}[N,m_A;\vec M,\vec T]\,,\\[.4cm]
\mathcal{Z}^{FFT}[N,m_A;\vec M,\vec T]&=&K[\vec M, -m_A] K[\vec T, m_A] \mathcal{Z}^{T}[N,-m_A;\vec M,\vec T]\nn\\[.2cm]
&=&\mathcal{Z}^{T}[N,m_A;\vec M,\vec T]\,
\eea
Our proof is based on~\cite{Bullimore:2014awa} where,
building on the results of~\cite{Gaiotto:2013bwa}, it was shown that 
$\mathcal{Z}^{T}$ is eigenfunctions of two sets
of trigonometric Ruijsenaars-Schneider (RS) Hamiltonians.
We introduce a first set of RS Hamiltonians,
\be
\mathcal{H}_r(\vec M, m_A)= \sum_{I\subset \{1,\cdots,n\}, |I|=r} \prod_{i\in I, j\not\in I} \frac{\sinh\pi b (i\tfrac{Q}{2}-m_A+M_i-M_j)}{\sinh\pi b (M_i-M_j)}   \prod_{i\in I} e^{i b\partial_{M_i}}\,.
\ee
(the other is obtained by exchanging $b\to\frac{1}{b}$). Then,
\be
\mathcal{H}_r(\vec M, m_A)\mathcal{Z}^{T}[N,m_A;\vec M,\vec T]=\chi_r(\vec T)\mathcal{Z}^{T}[N,m_A;\vec M,\vec T]
\ee
where $r=1,\cdots,N$, and $\chi_r(\vec T)$ are eigenvalues. 
Due to a peculiar property of the RS system, the same eigenfunction $\mathcal{Z}^{T}$
satisfies also the so-called $p$-$q$ dual equation: 
\be\label{dualpqRSequa}
\mathcal{H}_r(\vec T, -m_A)\mathcal{Z}^{T}[N,m_A;\vec M,\vec T]=\chi_r(\vec M)\mathcal{Z}^{T}[N,m_A;\vec M,\vec T]\,.
\ee
Upon a redefinition of parameters, the two eigenvalue equations imply the identity:
\be
\mathcal{Z}^{T}[N,m_A;\vec M,\vec T]=\mathcal{Z}^{T}[N,-m_A;\vec T,\vec M]\,.  
\ee
The same steps can be repeated for the RS Hamiltonian in which $b\to\frac{1}{b}$, thus mirror symmetry is proven~\cite{Bullimore:2014awa}.  
Quite interestingly, by considering the action of $K[\vec M, m_A]$ on the Hamiltonians
\footnote{To see this we use the following property of the double sine function 
\be
s_b(m_A\pm x)e^{i b\partial_x}s_b(-m_A\pm x)=\frac{\sinh\pi b (m_A+x+i\tfrac{Q}{2})}{\sinh\pi b (-m_A+x+i\tfrac{Q}{2})}.
\ee
}:
\be
\mathcal{H}_r(\vec M, -m_A)=K[\vec M, m_A] \mathcal{H}_r(\vec M, m_A)K[\vec M, m_A]^{-1}\,,
\ee
we can also show from \eqref{dualpqRSequa} that
\bea
\mathcal{H}_r(\vec M, -m_a)\rule{.7cm}{0pt}\mathcal{Z}^{T}[N,-m_A;\vec M,\vec T]\rule{1.3cm}{0pt}&=&\nn\\
K[\vec M, m_A]\mathcal{H}_r(\vec M, m_a)K[\vec M, m_A]^{-1}\mathcal{Z}^{T}[N,-m_A;\vec M,\vec T]&=&\nn\\
&=&\chi_r(\vec T)\mathcal{Z}^{T}[N,-m_A;\vec M,\vec T]\qquad
\label{eq:36}
\eea
Furthermore, using that $K[\vec T, m_A]$ commutes with $\mathcal{H}_r(\vec M, m_a)$ we find that the second and third terms in \eqref{eq:36} provides the additional relation
\bea
&&
\mathcal{H}_r(\vec M, m_a)\Big( K[\vec M, -m_A] K[\vec T, m_A]\mathcal{Z}^{T}[N,-m_A;\vec M,\vec T]\Big)=\nonumber\\
&&
\rule{3.5cm}{0pt}=\chi_r(\vec T)\Big( K[\vec M, -m_A]K[\vec T,m_A]\ \mathcal{Z}^{T}[N,-m_A;\vec M,\vec T]\Big)\,.
\eea
Therefore we conclude that 
 $\mathcal{Z}^{T}[N,m_A;\vec M,\vec T]$ and
 $\mathcal{Z}^{FFT}[N,m_A;\vec M,\vec T]$ satisfy the same RS
 eigenvalue equation. Of course, the same argument can be used
 for the RS Hamiltonians in which $b\rightarrow 1/b$. 
Thus we conclude that  $\mathcal{Z}^{T}[N,m_A;\vec M,\vec T]=\mathcal{Z}^{FFT}[N,m_A;\vec M,\vec T]$.

\subsection{Sequential confinement: from $T[SU(N+1)]$ to Theory A}

We now discuss the effect of the monopole deformation 
\be
\mathcal{L}^T_{\{1,\ldots,N-1\}}=\Mon^{[10\cdots00]}+\Mon^{[010\cdots00]}+\cdots+\Mon^{[00\cdots10]} \,.
\ee
on the partition function of $T[SU(N+1)]$. First, let us observe that 
on each node where the monopole potential is turned on, the symmetry $U(1)_A\times U(1)_{\mathrm{top}}$ is broken to the diagonal and consequently the FI parameters take special values related to the axial mass:
\be
\label{specia}
\xi^{(k)}=T_{k}-T_{k+1}=m_A+\tfrac{i Q}{2},
\qquad k=1,\cdots, N-1\,.
\ee
The last node is underformed, so there is no constraint on $\xi^{(N)}$.  
Following the logic of the sequential confinement, spelled out in Section~\ref{theory_A_sec},
we dualize the first gauge node, and sequentially all the nodes, by using the duality \cite{Benini:2017dud} between 
\be
U(k){\rm \ with\ }k+1{\rm\ flavors\ and\ } \cW=\Mon^+\quad \leftrightarrow 
\quad WZ{\rm\ model\ with \ }\cW=\gamma \det M.
\ee 
At the level of the partition functions this duality is obtained from the following evaluation formula,:
%
%
%
 %
%
\bea
&&
\int dx^{(k)} e^{-\pi i(\eta- i Q  )\sum_{j=1}^k x^{(k)}_j } ~Z_{\rm vec}^{(k)}~ \widetilde{Z}^{(k,k+1)}_{\rm }[\vec{M},\vec{\mu}\,]=\nonumber\\
&&
\rule{.8cm}{0pt}
 (-)^{k}e^{-\pi i\sum_{a=1}^{k+1}(\eta-i Q +2 \mu_a)M_a}  s_b\left(\tfrac{iQ}{2}-\eta\right) \prod_{i,j=1}^{k+1} s_b\left(\tfrac{iQ}{2}-\mu_i-\mu_j-M_i+M_j)\right)\qquad
  \label{sara_formula_1}
\eea
with the definition
\bea
 \widetilde{Z}^{(N_c,N_f)}_{\rm }[\vec{M},\vec{\mu}\,]&=&\prod_{i=1}^{N_c}\prod_{j=1}^{N_f} s_b\left(\tfrac{iQ}{2}-\mu_j\pm(x^{(N_c)}_i-M_j)\right)\,
 \label{more_gen_Zbif}
\eea
and the constraint from the monopole superpotential
\be
\eta=iQ-2\sum_{a=1}^{k+1}\mu_a\ .
\ee
%
%
%
We will actually need the identity \eqref{sara_formula_1} specialised to the case in which in  the electric theory the fundamental chirals couple to the adjoint breaking the $SU(k+1)\times SU(k+1)$ global symmetry to the diagonal and consequently the parameters $\mu_a$ are specialised to $\mu_a=\frac{i Q}{4}+\tfrac{m_A}{2}$ for $a=1,\cdots, k+1$.  
The constraint now reads, 
\be
\eta=i Q-(k+1)\left(m_A+\tfrac{iQ}{2}\right)
\ee
and
%
we find
\be\boxed{
\begin{array}{lll}
&&\\
&&
\displaystyle \int  dx^{(k)}  e^{\pi i (k+1)(m_A+{iQ}/{2})\sum_{j=1}^k x^{(k)}_j}~ Z_{\rm vec}^{(k)} ~Z_{\rm bif}^{(k,k+1)}= 
\\[.5cm]
&&
\displaystyle
\rule{.8cm}{0pt}(-)^k 
  e^{\pi i k(m_A+{iQ}/{2})\sum_{a=1}^{k+1} M_a}  s_b\left(-\tfrac{iQ}{2}+(k+1)(m_A+\tfrac{i Q}{2})\right) \left(Z_{\rm adj}^{(k+1)}\right)^{-1}\,
   \\
 &&
 \end{array}
 } 
\label{mofor}
\ee
where we identified  $\prod_{a,b=1}^{k+1} s_b(-m_A+y_a-y_b)= \left(Z_{\rm adj}^{(k+1)}\right)^{-1}$.
At this point, we can apply this identity to $\mathcal{Z}^{T}[N,m_A;\vec M,\vec T]$, 
with $\vec T$ specialised as in eq. (\ref{specia}), 
starting from the first node, where the adjoint is a (gauge) singlet, and sequentially by promoting
each time the real mass parameters to dynamical variables, i.e. $M_i\rightarrow x_i$.
Consider the first few dualizations as a warm-up, we will highlight some crucial simplifications. 
Focusing on the integrands, the partition function reads
\bea
&&\int dx^{(1)} e^{2\pi i (m_A+{iQ}/{2}) x^{(1)}_1}s_b(m_A) Z_{\rm bif}^{(1,2)}  \times \nonumber\\
&& 
\int dx^{(2)} e^{2\pi i (m_A+{iQ}/{2}) (x^{(2)}_1+x^{(2)}_2)} Z_{\rm vec}^{(2)} Z_{\rm adj}^{(2)} Z_{\rm bif}^{(2,3)}  \ldots  \qquad
\eea
which becomes
\bea
&\sim& s_b(m_A)\int dx^{(2)} e^{3\pi i (m_A+{iQ}/{2}) (x^{(2)}_1+x^{(2)}_2)}  Z_{\rm vec}^{(2)} Z_{\rm bif}^{(2,3)} \times\nonumber\\
& & 
\int dx^{(3)} e^{2\pi i(m_A+{iQ}/{2}) (x^{(3)}_1+x^{(3)}_2+x^{(3)}_3)} Z_{\rm vec}^{(3)} Z_{\rm adj}^{(3)} Z_{\rm bif}^{(3,4)}  \ldots  \qquad
\eea
The effect of the confinement of the $U(1)$ node has been to cancel the adjoint on the $U(2)$ node and shift the FI. 
Both these modifications are such that we can apply \eqref{mofor} to the $U(2)$ node. This procedure goes on sequentially.  
%
%
%
%
%
%

After confining all nodes but the last one we obtain:
\bea
\nonumber
&
&  e^{+2\pi i T_{N+1}  \sum_{i=1}^{N+1} M_i} \prod_{l=0}^{N-1} s_b\left(-\tfrac{iQ}{2}+(l+1)(m_A+\tfrac{i Q}{2})\right) \times \\
&&
\int dx^{(N)} e^{2\pi i \left[ \frac{N-1}{2}(m_A+iQ/2) +\xi^{(N)} \right] \sum_{j=1}^{N} x^{(N)}_j }\,Z_{\rm vec}^{(N)} Z_{\rm bif}^{(N,N+1)} \qquad
\eea
On the first line we recognize the contribution of the fields $\gamma_l$. Indeed, as explained
around eq.\eqref{R_charges_part_func},  we can read out the R-charges  by looking at the arguments of the $s_b$ functions and we find:
$(1-R_{\gamma_l})=-1+(1-\alpha)(l+1)$, from which follows the solution $R_{\gamma_l}=2-2(l+1)r$. 
This is the same assignment of R-charges we read off from the superpotential $W_A=\sum_{m=1}^N \frac{(-1)^m}{m}\gamma_m \Tr((\Pi^\mathcal{Q})^{m})$,
if we identify the indexes as $l+1=m$.\ 

The partition function of theory $A$ is finally obtained by making explicit the values of $T_1,\ldots T_N$, using the constraint $\sum_{i=1}^{N} T_i=0$. Then
\be\label{value_Ti}
T_i&=&\frac{(N+1) -2i}{2}\left(m_A+\frac{i Q}{2}\right),\qquad i=1,\ldots N
\ee
and
\be
T_{N+1}=T_N-\xi^{(N)}=-\frac{(N-1)}{2}\left(m_A+\frac{i Q}{2}\right)-\xi^{(N)}\,.
\ee

The result is
\bea
\nonumber
\mathcal{Z}^A[N,m_A;\vec{M},T_{N+1}]&=&  e^{+2\pi i T_{N+1} \sum_{i=1}^{N+1} M_i} \prod_{l=0}^{N-1} s_b\left(-\tfrac{iQ}{2}+(l+1)(m_A+\tfrac{i Q}{2})\right) \times \\
&&
\rule{.5cm}{0pt}\int dx^{(N)} e^{-2\pi i T_{N+1} \sum_{j=1}^{N} x^{(N)}_j }\,Z_{\rm vec}^{(N)} Z_{\rm bif}^{(N,N+1)}\,.
\label{part_func_A}
\eea

\subsection{Nilpotent mass deformation: from $T[SU(\Num)]^V$ to Theory B}

In $\chk{T[SU(\Num)]}$ the parameters $T_i$ become real masses for the flavor symmetry and $m_A$ changes sign, according to mirror symmetry. 
The identity between partition functions is indeed
\be\label{mirror_relation_partf}
\mathcal{Z}^{T}[N,m_A;\vec M,\vec T]=\mathcal{Z}^T[N,-m_A;\vec T,\vec M]\equiv \mathcal{Z}^{\chk{T}}[N,m_A;\vec M,\vec T]\,
\ee
The new FI parameters entering the mirror partition function are $\zeta^{(a)}=M_a-M_{a+1}$. 
The values of the $T_i$ we fixed in \eqref{value_Ti} lead to a telescopic cancellation of the one-loop contributions of $N-1$ flavors. In more details,
\bea
\nonumber
&&\prod_{i=1}^N\prod_{j=1}^{N+1} s_b\left(\tfrac{iQ}{4}+\tfrac{m_A}{2}\pm(x^{(N)}_i-T_j)\right)=
\\\nonumber
&&
\prod_{i=1}^N  s_b\left(\tfrac{iQ}{4}+\tfrac{m_A}{2}\pm(x^{(N)}_i-T_{N+1})\right) 
\prod_{j=1}^{N} s_b\left(\tfrac{iQ}{4}+\tfrac{m_A}{2}\pm\left(x^{(N)}_i- \tfrac{(N+1) -2j}{2}\left(m_A+\tfrac{i Q}{2}\right)\right)\right)=\\\nonumber
&&
\prod_{i=1}^N s_b\left(\tfrac{iQ}{4}+\tfrac{m_A}{2}\pm(x^{(N)}_i-T_{N+1})\right)
\qquad  \frac{ s_b\left(x^{(N)}_i+\tfrac{N}{2}\left(m_A+\tfrac{i Q}{2}\right)\right) }{ s_b\left(x^{(N)}_i- \tfrac{N}{2}\left(m_A+\tfrac{i Q}{2}\right)\right)}
 \eea
where we highlighted the cancellations in the last line. 
The partition function of theory $B$ is then
\bea
\mathcal{Z}^B[N,m_A;\vec{M},T_{N+1}] &=& e^{2\pi i M_{N+1}  T_{N+1} } 
\int \prod_{k=1}^{N-1} dx^{(k)}e^{2\pi i \zeta^{(k)} \sum_{a=1}^k x^{(k)}_a} Z_{\rm vec}^{(k)}~Z_{\rm adj}^{(k)}~ Z_{\rm bif}^{(k,k+1)}\nn\\
&&
\times \int dx^{(N)}e^{2\pi i \zeta^{(N)}\sum_{a=1}^N x^{(N)}_a} Z_{\rm vec}^{(N)}~Z_{\rm adj}^{(N)}~ \widetilde{Z}^{(N,1)}[T_{N+1},\tfrac{iQ}{4}-\tfrac{m_A}{2}]~Z^d\nn\qquad\\
\eea
where we used $\sum_{i=1}^N T_i=0$ in the prefactor, and defined
\bea
Z^d&=&\prod_{i=1}^N s_b\left(\tfrac{N}{2}\left(m_A+\tfrac{i Q}{2}\right)\pm x^{(N)}_i\right)\qquad
\eea
The contributions $Z^d$ corresponds to the fundamentals fields $d$ and $\tilde d$. Looking at the coefficient of $\tfrac{i Q}{2}$ we see that $1-R_d=\tfrac{N}{2}(1-\alpha)$ so $R_d=1- N r$ as expected.
The contribution $\widetilde{Z}^{(N,1)}$ originates from two fundamentals chirals with R-charge $(1-r)$, which are still part of the tail in the quiver diagram.
 
\subsection{Partition Functions on the A-to-D side}
We obtain theory $D$ from theory $A$ by applying Aharony duality.
As reviewed in Appendix~\ref{App_Integrals}, Aharony duality 
is implemented by the following integral identity,
\bea
&&
 \int dx^{(N_c)} e^{-\pi i \lambda \sum_j x^{(N_c)}_j} Z_{\rm vec}^{(N_c)} \widetilde{Z}^{(N_c,N_f)}[\vec{M},\vec{\mu}\,]= \nn\\
&&
\rule{0.5cm}{0pt}
{ e^{-i\pi \lambda \sum_{a=1}^{N_f } M_a } } s_b\left(\tfrac{iQ}{2}(N_c-N_f)  +|\vec{\mu}| \pm \tfrac{\lambda}{2}\right) \prod_{a,b=1}^{N_f}s_b\left(\tfrac{iQ}{2}-\mu_a-\mu_b-M_a+M_b \right) \nn\\
&&
\rule{1cm}{0pt}
\times\int dx^{(N_f-N_c)} e^{-\pi i \lambda \sum_j x^{(N_f-N_c)}_j} Z_{\rm vec}^{(N_f-N_c)} \widetilde{Z}^{(N_f-N_c,N_f)}[-\vec{M},\tfrac{iQ}{2}-\vec{\mu}\,]
\qquad
\label{aharec}
\eea
with $\widetilde{Z}^{(N_c,N_f)}$ defined in \eqref{more_gen_Zbif}. 

For $N_c=N$, $N_f=N+1$ and $\lambda=2T_{N+1}$ , the l.h.s of \eqref{aharec} coincides with the integrand of $\mathcal{Z}^A$.
However, since theory $A$ has a non trivial superpotential, which breaks $SU(N_f)\times SU(N_f)$ to $SU(N_f)\times U(1)_A$, 
we need to consider the identification $\mu_a=\frac{i Q}{4}+\tfrac{m_A}{2}$ for $a=1,\cdots, N_f$. 
Finally, we obtain $\mathcal{Z}^D$ upon including the prefactor associated to the dual of the fields $\gamma_{l\ge 0}$ of theory $A$. 
We find:
\bea
&&
\mathcal{Z}^D[N,m_A;\vec{M},{T_{N+1}}]=\nn\\
&& 
\rule{0.5cm}{0pt} 
 s_b\left(-\tfrac{iQ}{2}+\tfrac{N+1}{2}( m_A+\tfrac{iQ}{2}) \pm\xi  \right) \prod_{a,b=1}^{N+1}s_b\big(-m_A+M_a-M_b \big)\nn\\
&&  
\rule{1cm}{0pt}
\times\prod_{l=0}^{N-1} s_b\left(-\tfrac{iQ}{2}+(l+1)(m_A+\tfrac{i Q}{2})\right) \int  dx  e^{-2\pi i T_{N+1} x }\,  \widetilde{Z}^{(1,N+1)}[-\vec{M},\tfrac{iQ}{4}-\tfrac{m_A}{2}\,]\qquad\nn\\
\label{part_func_theoryD}
\eea
Some comments on $\mathcal{Z}^D$ are in order. There is a cancellation of contact terms when using \eqref{aharec} on the integrand of theory $A$. 
This is so because the FI of theory $A$, compared to $T[SU(\Num)]$, has been reduced to $T_{N+1}$ during the sequential confinement. 

In the notation of Section \ref{ABCD_sec},
we recognize in the first line of $\mathcal{Z}^D$ the contribution of the two singlets, $\sigma_{\pm}$, and that of $(N+1)^2$ singlets $\Flip^{\mathcal{U}}_D$. 
The fields $\sigma_{\pm}$ are flipping fields for the monopoles, and the fields $\Flip^{\mathcal{U}}_D$ flip the meson. 
From the arguments of the $s_b$ we read out $(1-R_{\sigma_\pm})=-1+\tfrac{N+1}{2}(1-\alpha)$, or $R_{\sigma_{\pm}}=2-(N+1)r$. 
We then find that $(1-R_{\Flip^{\mathcal{U}}_D })=\alpha$, or $R_{\Flip^{\mathcal{U}}_D }=1-\alpha=2r$.
Notice also that the contribution of  $\theta_1$ (the coefficient with $l=0$ in the product) cancels the diagonal contributions of the singlets $\Flip^{\mathcal{U}}_D$, effectively enforcing the tracelessness of $\Flip^{\mathcal{U}}_D$.

\subsection{$T[SU(2)]$} \label{sec_TS2_full}

The case of $T[SU(2)]$ is simple enough to compute the partition function explicitly. In this case,
our monopole deformation is empty, thus the partition function of theory $A$  is directly that of $T[SU(2)]$, with the specification $T_1=0$.
The presence of flipping fields, and the non trivial mapping of parameters across the commutative diagram, 
makes the equalities of partition functions a nice exercise to go through. In these computations we will keep $M_1\neq M_2$. 

The integrand of the $T[SU(2)]$ partition function can be quickly evaluated by residue integration \cite{Pasquetti:2011fj}.
The full partition function will also include a factor of $Z^{(1)}_{adj}=s_b(m_A)$, and an exponential prefactor. In our conventions, $\xi=T_1-T_2$,
\bea
\mathcal{Z}^T[1,m_A,\vec{M},\vec{T}]= e^{2\pi i  T_2(M_1+M_2)} s_b(m_A)\int_{-\infty}^{+\infty} dx\, 
e^{2\pi i x \xi}\, \prod_{i=1}^2 \frac{ s_b(+\tfrac{iQ}{4}-\tfrac{m_A}{2} + (x-M_i))}{ s_b( -\tfrac{iQ}{4}+\tfrac{m_A}{2}+(x-M_i)}\nn\\
\label{integralZT12}
\eea
We pick poles of the two $s_b$ functions at the numerator, and assign to the set of poles labelled by $M_i$, the series $\mathscr{S}_i$ defined as,
\bea
&&
\mathscr{S}_i = e^{2\pi i \xi M_i} e^{i\pi \xi (iQ/2+m_A)}\frac{ s_b(\tfrac{iQ}{2}+(-)^i(M_2-M_1)) }{ s_b(m_A) s_b(m_A+(-)^i(M_2-M_1)) } \times \nn\\[.2cm]
&&
\rule{1cm}{0pt}~_2\varphi_1^{(q)}\left[ q^{ \frac{1}{2} }e^{2\pi b m_A}, q^{ \frac{1}{2} }e^{2\pi b ((-)^{i} (M_2-M_1)+m_A)};  qe^{(-)^{i} 2\pi b(M_2-M_1)}; q^{\frac{1}{2}}e^{-2\pi b(\xi+m_A)} \right]
\times\nn\\
&&
\rule{1cm}{0pt}~_2\varphi_1^{(\tilde{q})}\left[ \tilde{q}^{ \frac{1}{2} }e^{\frac{2\pi}{b} m_A}, \tilde{q}^{ \frac{1}{2} }e^{\frac{2\pi}{b} ((-)^{i} (M_2-M_1)+m_A)};  \tilde{q}e^{(-)^{i} \frac{2\pi}{b}(M_2-M_1)}; \tilde{q}^{\frac{1}{2}}e^{-\frac{2\pi}{b}(\xi+m_A)} \right]
\label{serie_tsu2}
\eea
Recall the definitions $q=e^{2\pi i bQ}$ and $\tilde{q}=e^{2\pi iQ/b}$. 
The hypergeometric function$~_2\varphi_1^{(q)}$ admits the series representation
\bea\label{def_2phi1}
~_2\varphi_1^{(q)}\left[a,b;c;z\right]=\sum_{n} \frac{ (a;q)_n (b;q)_n}{(q;q)_n (c;q)_n} z^n\,, \qquad |q|< 1.
\eea
If $|q|<1$ and $|\tilde{q}|>1$ we use the relation
\be
~_2{\varphi}_1^{(\tilde{q})}\left[a,b;c;z\right]=~_2\varphi_1^{(1/\tilde{q})}\left[ a^{-1},b^{-1};c^{-1}; abz/(\tilde{q}c)\right]\,,
\ee
where the r.h.s can be expanded out as in \eqref{def_2phi1}. 
The partition function is invariant under $b\leftrightarrow 1/b$ and can be written as the sum 
$\sum_{i=1}^2 e^{2\pi i  T_2(M_1+M_2)} s_b(m_A)\, \mathscr{S}_i$.
We will work with its factorized expression \cite{Nieri:2015yia}, namely 
\be\label{factorization_mat}
\mathcal{Z}^T[1,m_A,\vec{M},\vec{T}]= e^{2\pi i \mathcal{P}} \left[ \begin{array}{l} \mathcal{B}_1^{(\tilde{q})}, 
\mathcal{B}_2^{(\tilde{q})} \end{array}\right] \left[ \begin{array}{cc} 1 & 0 \\ 0 & e^{2\pi i (M_2-M_1)(\xi+m_A-iQ/2)} \end{array}\right]
\left[ \begin{array}{c} \mathcal{B}_1^{({q})}\\ \mathcal{B}_2^{({q})} \end{array}\right]
\ee
In \eqref{factorization_mat} we isolated the following exponential prefactor
\be\label{prefaTSU2}
\mathcal{P}=(T_1M_1+T_2M_2)+\tfrac{(T_1-T_2)(m_A+iQ/2) +( M_1-M_2)(m_A-iQ/2)}{2}+\tfrac{(m_A^2+Q^2/4)}{4}
\ee
and defined the holomorphic blocks $\mathcal{B}_{i=1,2}$ associated to the series $\mathscr{S}_{i=1,2}$:\footnote{Compared to \cite{Nieri:2015yia} 
we do not introduce $\Theta$-functions to factorize exponential terms, but we use the `factorization' matrix \eqref{factorization_mat}.} 
\bea
&&
\mathcal{B}_2^{(q)}[\vec{M},\vec{T}]=\mathcal{B}_1^{(q)}[-\vec{M},\vec{T}],\\[.2cm]
&&
\mathcal{B}_1^{(q)}[\vec{M},\vec{T}]=\frac{ (qe^{2\pi b (M_1-M_2)},q)_{\infty} }{ (q^{\frac{1}{2}}e^{2\pi b( M_1-M_2+m_A)},q)_{\infty} } \times\nn\\
&&
\rule{.5cm}{0pt}_2\varphi_1^{(q)}\left[ {q^{\frac{1}{2}}e ^{2\pi b m_A },q^{\frac{1}{2}}e^{2\pi b (M_1-M_2+m_A)};  qe^{2\pi b (M_1-M_2)} }, q^{\frac{1}{2}}e^{2\pi b (T_2-T_1-m_A)}\right]\qquad
\eea
The holomorphic blocks of theory $A$, $\mathcal{B}^{A}_{i=1,2}$, can be defined from $\mathcal{B}_{i=1,2}$ by setting $T_1=0$.  
Whenever needed we will understand $b\rightarrow b^{-1}$ in the conjugate blocks $\mathcal{B}_{i=1,2}^{(\tilde{q})}$.

\subsubsection*{Aharony Duality} 
The partition function of theory $D$ is given by \eqref{part_func_theoryD},
\be
\mathcal{Z}^{D}[1,m_A;\vec{M},-\xi]&=&
s_b\left( m_A \pm\xi  \right) \prod_{a,b=1}^{2}s_b\big(-m_A+M_a-M_b \big)\times\nn\\
&&  
\rule{.5cm}{0pt}
s_b\left(m_A \right) \int_{-\infty}^{+\infty} dx\, e^{2\pi i x \xi}\, \prod_{i=1}^2 \frac{ s_b(+\tfrac{iQ}{4}+\tfrac{m_A}{2} + (x+M_i))}{ s_b( -\tfrac{iQ}{4}-\tfrac{m_A}{2}+(x+M_i)}\qquad
\label{ZDTSU2}
\ee
%
%
The FI parameter $\xi$ should be fixed to be $-T_2$, but we are keeping it generic for comparison with $Z^{FFT}$ in the next section. 
The integrand \eqref{ZDTSU2} can be evaluated by residue integration, as in \eqref{serie_tsu2}. The modifications are minors so we will not repeat them. 
Instead, after writing the partition function in the factorized form, we show how the blocks map into each other when going from theory $A$ to theory $D$. 
The factorization of the partition function is
\be
\mathcal{Z}^{D}[1,m_A;\vec{M},-\xi]=
e^{2\pi i \mathcal{P}} \left[ \begin{array}{l} \mathcal{B}_1^{D(\tilde{q})}, \mathcal{B}_2^{D(\tilde{q})} \end{array}\right] \left[ \begin{array}{cc} e^{2\pi i (M_2-M_1)(\xi+m_A-iQ/2)} & 0 \\ 0 & 1 \end{array}\right]
\left[ \begin{array}{c} \mathcal{B}_1^{D({q})}\\ \mathcal{B}_2^{D({q})} \end{array}\right]\qquad
\ee
where $\mathcal{P}$ is the same prefactor \eqref{prefaTSU2} and the holomorphic blocks are 
\bea
&&
\mathcal{B}_2^{D(q)}[\vec{M},\vec{T}]=\mathcal{B}_1^{D(q)}[-\vec{M},\vec{T}],\\[.2cm]
&&
\mathcal{B}_1^{D({q})}[\vec{M},\vec{T}]= \frac{ (q e^{-2\pi b(M_1-M_2)},q)_{\infty} }{ (q^{\frac{1}{2}} e^{-2\pi b (M_1-M_2-m_A) },q)_{\infty} } 
\frac{ ( q^{ \frac{1}{2} } e^{-2\pi b(\xi-m_A) },q) }{ ( q^{ \frac{1}{2} } e^{-2\pi b(\xi+m_A) },q)} \times\nn\\
&&
\rule{0.5cm}{0pt}
_2\varphi_1^{(q)}\left[ {q^{\frac{1}{2}}e ^{-2\pi b m_A },q^{\frac{1}{2}}e^{2\pi b (M_2-M_1-m_A)};  qe^{2\pi b (M_2-M_1)} }, q^{\frac{1}{2}}e^{2\pi b (-\xi+m_A)}\right] \qquad
\eea
By using the second of the Heine's identities \cite{Gasper}, 
\be\label{Heine3}
~_2\varphi_1^{(q)}[ a,b;c;z]= \frac{ (ab z/c)_{\infty} }{ (z)_{\infty} } ~_2\varphi^{(q)}_1[ c/a,c/b; c; abz/c]
\ee
and its $\tilde{q}$ analog, which in this case coincides with \eqref{Heine3}, we find
\be\label{changeADblocks}
\left[ \begin{array}{c} \mathcal{B}_1^{D({q})}\\ \mathcal{B}_2^{D({q})} \end{array}\right]=\left[\begin{array}{cc} 0 & 1 \\ 1 & 0 \end{array}\right]
\left[ \begin{array}{c} \mathcal{B}_1^{A({q})}\\ \mathcal{B}_2^{A({q})} \end{array}\right]
\ee
Thus, we have shown the equality
\be\label{equaDS2}
\mathcal{Z}^{D}[1,m_A;\vec{M},T_2]=\mathcal{Z}^{A}[1,m_A;\vec{M},T_2].
\ee

\subsubsection*{Flip-Flip duality}

In order to compute $\mathcal{Z}^{FFT}$ we follow our prescription \eqref{presc_FFT}. Considering $Z^T[1,-m_A;\ldots]$, with $Z^T$ given in \eqref{integralZT12}, we obtain 
\bea
\mathcal{Z}^{FFT} [1,m_A,\vec{M},\vec{T}]&=&K[\vec M, -m_A]K[\vec T, m_A]\ s_b(-m_A)\times\\
&&\quad e^{2\pi i  T_2(M_1+M_2)} \int_{-\infty}^{+\infty} dx\, e^{2\pi i x \xi}\, \prod_{i=1}^2 \frac{ s_b(+\tfrac{iQ}{4}+\tfrac{m_A}{2} + (x-M_i))}{ s_b( -\tfrac{iQ}{4}-\tfrac{m_A}{2}+(x-M_i)}\nn
\eea
Recall that $K[\vec x, m_A] = \prod_{i,j}  s_b\left(m_A\pm(x_i-x_j) \right)$, therefore
\bea\label{FFTS2}
\mathcal{Z}^{FFT} [1,m_A,\vec{M},\vec{T}]&=&s_b\left( m_A \pm\xi  \right) s_b\big(-m_A\pm(M_1-M_2) \big)s_b(-m_A)\times\\
&&\quad e^{2\pi i  T_2(M_1+M_2)} \int_{-\infty}^{+\infty} dx\, e^{2\pi i x \xi}\, \prod_{i=1}^2 \frac{ s_b(+\tfrac{iQ}{4}+\tfrac{m_A}{2} + (x-M_i))}{ s_b( -\tfrac{iQ}{4}-\tfrac{m_A}{2}+(x-M_i)} \nn
\eea
Very similarly to the $TSU(2)$ computation, we evaluate the integrand in \eqref{FFTS2} and factorize the result into
\be
\mathcal{Z}^{FFT} [1,m_A,\vec{M},\vec{T}] = 
e^{2\pi i \mathcal{P}} \left[ \begin{array}{l} \mathcal{B}_1^{FF(\tilde{q})}, \mathcal{B}_2^{FF(\tilde{q})} \end{array}\right] \left[ \begin{array}{cc} 1 & 0 \\ 0 & e^{2\pi i (M_2-M_1)(\xi+m_A-iQ/2)} \end{array}\right]
\left[ \begin{array}{c} \mathcal{B}_1^{FFT({q})}\\ \mathcal{B}_2^{FFT({q})} \end{array}\right]\nn
\ee
where, by using again \eqref{Heine3} on the holomorphic blocks $\mathcal{B}_{i=1,2}^{FF}$, we find
\be
\left[ \begin{array}{c} \mathcal{B}_1^{FF({q})}\\ \mathcal{B}_2^{FF({q})} \end{array}\right]=\left[\begin{array}{cc} 1 & 0 \\ 0 & 1 \end{array}\right]
\left[ \begin{array}{c} \mathcal{B}_1^{({q})}\\ \mathcal{B}_2^{({q})} \end{array}\right]
\ee
Thus we have shown that
\be\label{equaFFTS2}
\mathcal{Z}^{FFT} [1,m_A,\vec{M},\vec{T}] =\mathcal{Z}^{T} [1,m_A,\vec{M},\vec{T}]
\ee
Notice that our definition of $\mathcal{Z}^{FFT}$ in \eqref{presc_FFT}, which was strongly motivated by the use of difference operators, has correctly captured possible field theory contact terms. 

In Section \ref{flipflip_sec} we used a field theory argument to show that Aharony duality applied to $T[SU(2)]$ is related to Flip-Flip duality. This is consistent
with the observation that \eqref{equaFFTS2} and \eqref{equaDS2} follow from the same Heine's identity \eqref{Heine3}, i.e. they are not independent. 
However, we did not obtain Flip-Flip duality directly, and we insisted on some additional manipulations. 
These manipulations will also be visible at the level of the partition function: 
Consider the action of Aharony duality on $T[SU(2)]$ by implementing \eqref{aharec} on the integrand of $Z^T$. We denote this by $\mathcal{A}\circ Z^T$. 
Then, we find the relation
\be\label{final_FFTS2}
\mathcal{Z}^{FFT} [1,m_A,-\vec{M},\vec{T}]=e^{-2\pi T_1 (M_1+M_2)}\mathcal{A}\circ Z^T[1,m_A,\vec{M},\vec{T}]
\ee
The contribution of the $SU(2)_{\mathrm{flavor}}\times SU(2)_{\mathrm{top}}$ flipping fields in $FFT[SU(2)]$ comes out as follows:  On the r.h.s, of \eqref{final_FFTS2} we find
\be
s_b(m_A) \times \Big[ s_b(\pm \xi +m_A) \prod_{a,b=1}^2 s_b(-m_A+M_a-M_b)\Big]
\ee
where the terms in parenthesis $[\ldots]$ are introduced by the Aharony duality. Then,
one of the diagonal contributions, i.e. $a=b=1$ or $a=b=2$, simplifies with the original $s_b(m_A)$ of $T[SU(2)]$, and we recover the same prefactors as in \eqref{FFTS2}. 
Furthermore, when the constraint $M_1+M_2=0$ is imposed, the relation \eqref{final_FFTS2} implies the equalities
\be
\mathcal{Z}^{FFT} [1,m_A]=\mathcal{A}\circ Z^T[1,m_A]=\mathcal{Z}^{D}[1,m_A;T_2-T_1]\ ,
\ee
as expected from the field theory argument. 

\subsubsection*{Mirror Symmetry} 
Explicit computations about the partition functions of $T[SU(2)]$ and $\chk{T}[SU(2)]$ 
have been done in \cite{Nieri:2015yia}. For sake of completeness, we repeat them in our notation to show consistency. 
We shall refer directly to theory $A$ and theory $B$, since the monopole deformation is empty.

We write $\zeta=M_1-M_2$ and
\be
\mathcal{Z}^{\chk{T}}[1,m_A;\vec{M},\vec{T}]=e^{+2\pi i M_2(T_1+T_2) } s_b(-m_A)\int_{-\infty}^{+\infty} dx e^{-2\pi i \zeta x}\prod_{i=1}^2 \frac{ s_b(\tfrac{iQ}{4} +\tfrac{m_A}{2} +(x+T_i) ) }{  s_b(-\tfrac{iQ}{4} -\tfrac{m_A}{2} +(x+T_i) )}\nn\\
\ee 
Notice the change of variables $x\rightarrow -x$ compared to \eqref{mirror_relation_partf}. In its factorized form we can then extract the same prefactor $\mathcal{P}$, given in \eqref{prefaTSU2}, and obtain,
\be
\mathcal{Z}^{\chk{T}}[1,m_A;\vec{M},\vec{T}]=
e^{2\pi i \mathcal{P}} \left[ \begin{array}{l} \mathcal{B}_1^{B(\tilde{q})}, \mathcal{B}_2^{B(\tilde{q})} \end{array}\right] \left[ \begin{array}{cc} 1 & 0 \\ 0 & e^{2\pi i (M_2-M_1)\xi}e^{-i\pi(m_A^2+Q^2/4)} \end{array}\right]
\left[ \begin{array}{c} \mathcal{B}_1^{B({q})}\\ \mathcal{B}_2^{B({q})} \end{array}\right]\nn\\
\ee
where the blocks are
\bea
&&
\mathcal{B}_1^{B(q)}[\vec{M},\vec{T}]=\frac{(qe^{-2\pi b(T_1-T_2)},q)_{\infty}}{ (q^{\frac{1}{2}} e^{-2\pi b(T_1-T_2+m_A)},q)_{\infty}} \times\nn\\
&&
\rule{0.5cm}{0pt}_2\varphi_1^{(q)}\left[ {q^{\frac{1}{2}}e ^{-2\pi b m_A },q^{\frac{1}{2}}e^{2\pi b (T_2-T_1-m_A)};  qe^{2\pi b (T_2-T_1)} }, q^{\frac{1}{2}}e^{2\pi b (M_1-M_2+m_A)}\right] \qquad\\[.3cm]
&&
\mathcal{B}_2^{B(q)}[\vec{M},\vec{T}]= \frac{(e^{-2\pi b(T_1-T_2)},q)^{-1}_{\infty}}{ (q^{\frac{1}{2}} e^{-2\pi b(T_1-T_2-m_A)},q)^{-1}_{\infty}} \times\nn\\
&&
\rule{0.5cm}{0pt}
_2\varphi_1^{(q)}\left[ {q^{\frac{1}{2}}e ^{-2\pi b m_A },q^{\frac{1}{2}}e^{2\pi b (T_1-T_2-m_A)};  qe^{2\pi b (T_1-T_2)} }, q^{\frac{1}{2}}e^{2\pi b (M_1-M_2+m_A)}\right] \qquad
\eea
Given the form of the integrand we readily find $\mathcal{B}_1^{B(q)}[\vec{M},\vec{T}]=\mathcal{B}_1^{A(q)}[-\vec{T},-\vec{M}]$. 
The map between blocks under mirror symmetry is more involved than \eqref{changeADblocks}. It can be derived from the use of the first Heine's identity
\be
~_2\varphi_1^{(q)}[ a,b;c;z]= \frac{ (b)_{\infty}(az)_{\infty} }{ (c)_{\infty} (z)_{\infty} } ~_2\varphi^{(q)}_1[ c/b,z,az;b]
\ee
and the analitic continuation formulas \cite{Nieri:2015yia}. The result is 
\be\label{fromAB_block}
\left[\begin{array}{c} \mathcal{B}^{B(q)}_1 \\  \mathcal{B}^{B(q)}_2 \end{array}\right]= 
\left[ \begin{array}{cc} 1 & 0 \\ c^{(q)}_{21} & c^{(q)}_{22} \end{array}\right] 
\left[\begin{array}{c} \mathcal{B}^{A(q)}_1 \\  \mathcal{B}^{A(q)}_2 \end{array}\right]\qquad 
\ee
with 
\bea
c_{22}^{(q)}&=& 
\frac{ (q^{\frac{1}{2} } e^{2\pi b(M_1-M_2-m_A)},q)_{\infty}  (q^{\frac{1}{2} } e^{-2\pi b(M_1-M_2-m_A)},q)_{\infty}  }{
(e^{2\pi b(M_1-M_2)},q)_{\infty} (qe^{-2\pi b(M_1-M_2)},q)_{\infty} }\\[.3cm]
c_{21}^{(q)}&=&
\frac{ e^{-2\pi b(T_1-T_2)} ( e^{2\pi b(T_1-T_2+M_1-M_2)} ,q)_{\infty} (q e^{-2\pi b(T_1-T_2+M_1-M_2)},q)_{\infty} ( q^{\frac{1}{2}} e^{\pm2\pi b m_A} ,q)_{\infty} }{
( e^{-2b\pi (T_1-T_2)},q)_{\infty} (qe^{2b\pi (T_1-T_2)},q)_{\infty} (e^{2\pi b (M_1-M_2)},q)_{\infty} (qe^{-2\pi b (M_1-M_2)},q)_{\infty}
}\nn\\
\eea
For the conjugate blocks we get 
\be\label{fromAB_blocktilde}
\left[\begin{array}{c} \mathcal{B}^{B(\tilde{q})}_1 \\  \mathcal{B}^{B(\tilde{q})}_2 \end{array}\right]= 
\left[ \begin{array}{cc} 1 &\  -c_{22}^{(\tilde{q})}/c_{21}^{(\tilde{q})} \\ 0 & c^{(\tilde{q})}_{22} \end{array}\right] 
\left[\begin{array}{c} \mathcal{B}^{A(\tilde{q})}_1 \\  \mathcal{B}^{A(\tilde{q})}_2 \end{array}\right]\qquad 
\ee
The connection matrix in \eqref{fromAB_blocktilde} is essentially the inverse of \eqref{fromAB_block}, transposed. 
\section{Spectral dualities}\label{spd}
In this section we connect the dualities discussed in the first part
of the paper to a class of $3d$ dualities which we call spectral
dualities since they have their origin in $5d$ spectral
dualities, or  fiber-base duality in topological string.

In the introduction we claimed that $3d$ spectral dual pairs can be regarded as
$3d$ $\mathcal{N}=2$ theories living on a codimension-two
defect which is coupled to a (trivial) $5d$ $\mathcal{N}=1$ theory. 
The starting point of this construction is a toric CY three-fold $\mathfrak{X}$ which engineers a 
$5d$ $\mathcal{N}=1$ linear quiver theory.
A $3d-5d$  coupled system can be  obtained via Higgsing, by tuning the K\"ahler parameters of the CY $\mathfrak{X}$ in a specific way.
The resulting CY will be denoted by
$\mathcal{X}$\footnote{We denote by $\mathfrak{X}$ or $\mathcal{X}$ the toric
  graphs together with the values of complexified K\"ahler parameters
  of the corresponding CY manifold. In particular $\mathfrak{X}$ has generic K\"ahler parameters whereas in $\mathcal{X}$ they are tuned to special quantised values.}
  and since we will be considering a {\it complete} Higgsing it will correspond  to the $3d$ 
  theory $\mathcal{T}_{\mathcal{X}}$ coupled to $5d$ free hypermultiplets.
  In particular the  topological string partition function we started with reduces to the partition function of our
$3d$ theory $\mathcal{T}_{\mathcal{X}}$. From the original fiber-base duality of the CY, 
we can then infer the existence of $3d$ dualities, which we will discuss in the next section. 

More precisely, we have found the following relation between the holomorphic
block ($D^2\times S^1$ partition function) ${\mathcal
  B}^{\alpha_0}_{\mathcal{T}_{\mathcal{X}}}$ evaluated on a contour
$\alpha_0$ and the partition function of the Higgsed topological
string on  $\mathcal{X}$:
\begin{equation}
  {\mathcal B}^{\alpha_0}_{\mathcal{T}_{\mathcal{X}}}=\mathcal{E}_{\mathcal{T}_{\mathcal{X}}}
  Z^{\alpha_0}_{\mathrm{cl}, \mathcal{T}_{\mathcal{X}}}Z^{\alpha_0}_{1 \mathrm{-loop}, \mathcal{T}_{\mathcal{X}}}
  Z^{\alpha_0}_{\mathrm{vort}, \mathcal{T}_{\mathcal{X}}}= G Z^{\mathcal{X}}_{1 \mathrm{-loop, top}}Z^{\mathcal{X}}_{\mathrm{vort}, \mathrm{top}}\,.\label{eq:3}
\end{equation}
We have separated the topological string partition function on the
r.h.s.\ of Eq.~(\ref{eq:3}) into two pieces, $Z^{\mathcal{X}}_{1
  \mathrm{-loop, top}}$ and $Z^{\mathcal{X}}_{\mathrm{vort},
  \mathrm{top}}$ which coincide with the vortex part of the $3d$
partition function.  $Z^{\mathcal{X}}_{1 \mathrm{-loop, top}}$ is
independent of the $3d$ FI parameters, and hence of the corresponding
K\"ahler parameters of the CY, while $Z^{\mathcal{X}}_{\mathrm{vort},
  \mathrm{top}}$ does depend on them. Finally $G$ denotes a possible fiber-base invariant
prefactor and $\mathcal{E}_{\mathcal{T}}$ a contact term.
  
\begin{figure}[h]
  \centering
  \begin{tabular}{c}
    \includegraphics{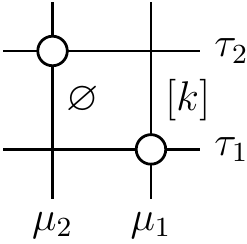}\\
    a)
  \end{tabular} \qquad   \begin{tabular}{c}
    \includegraphics{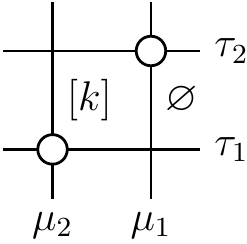}\\
    b)
  \end{tabular}
  \caption{Two toric CY diagrams corresponding to two vacua of the
    $3d$ $FT[SU(2)]$ theory. Notice that the spectral parameters of
    external legs are the same in both cases.}
  \label{fig:contours-cy}
\end{figure}

  The choice of contour
$\alpha_0$ on which the holomorphic block is evaluated corresponds to a particular
way of tuning the K\"ahler parameters to implement the Higgsing.
 More concretely, the different contours
  correspond to Higgsed toric CY's, in which the spectral parameters
  in the \emph{external} legs of the toric diagram are fixed, while
  the internal ones can vary. For example, two Higgsed CY's corresponding to
  two contours (or vacua) of the $FT[SU(2)]$ theory are shown in
  Fig.~\ref{fig:contours-cy} (see sec.~\ref{sec:spectr-dual-from} for
  notations).

In the following we present our two main spectral dual pairs: 

\begin{itemize}
\item[1)]
${FT[SU(\Num)]}\leftrightarrow\chk{FT[SU(\Num)]}$, which is obtained from the $T[SU(\Num)]$ commutative diagram, upon
flipping the edge $FFT[SU(\Num)]\leftrightarrow\chk{T[SU(\Num)]}$.

\item[2)]
$\mathcal{T}\leftrightarrow \mathcal{T}'$, which is obtained from
the $D\leftrightarrow B$ duality in the $ABCD$ framework.
\end{itemize}

After discussing the field theory evidence of these dualities we will
see how the holomorphic blocks  of each theory can be obtained via Higgsing from a
topological string partition function and we will then explicitly see
how the spectral duality descends from the fiber-base duality.

\subsection{$FT[SU(\Num)]$ and its spectral dual }\label{FT_spectral_sec}

Our starting point is the duality $FFT[SU(\Num)] \leftrightarrow \chk{T[SU(\Num)]}$, on the SW-NE diagonal of the diagram \ref{tffweb}. 
Recall that  $FFT[SU(\Num)]$  has two sets of singlets $F^{\mathcal{R}}_{ij}$ and $F^{{\rm m}}_{ij}$ which flip the HB and CB moment maps: 
\begin{equation}
\cW_{FFT[SU(\Num)]}=W^T[ {\Theta}, \mathbb{R} ] +\Pi^{\mathcal{R}}_{ij}
F^{\mathcal{R}}_{ij} +{\rm m}_{ij}  F^{{\rm m}}_{ij} \,.
\end{equation}

We now add another set of $(N+1)^2$ singlets $F^{T}_{ij}$ and deform the  $FFT[SU(\Num)]$ theory by the  superpotential
$\delta \cW= F^T_{ij} F^{\rm m}_{ij}$. We are basically flipping twice
the Coulomb branch of $T[SU(\Num)]$ and since $flip^2=1$, as it is easy
to see by using the equations of motion, we find a new theory, which we
call $FT[SU(\Num)]$, where only the Higgs branch moment map is flipped:
\begin{equation}
\cW_{FT[SU(\Num)]}=W^T[ {\Theta}, \mathbb{R} ] +\Pi^{\mathcal{R}}_{ij}  F^{\mathcal{R}}_{ij} \,.
\end{equation}

On the dual side $\chk{FT[SU(\Num)]}$, we proceed similarly. We add new $(N+1)^2$, singlets which we call
$ F^{\rm \mathcal{P}}_{ij} $, and the superpotential deformation 
$\delta \cW= F^{\rm \mathcal{P}}_{ij} \Pi^{\mathcal{P}}_{ij}$. This deformation is dual to that for $FT[SU(\Num)]$, 
since in the commutative diagram the singlets $F^{\rm m}_{ij}$ are mapped into the mesons moment map $\Pi^{\mathcal{P}}_{ij}$.
The resulting
$\chk{FT[SU(\Num)]}$ theory has
\begin{equation}
\cW_{\chk{FT[SU(\Num)]}}=W^T[ {\Omega}, \mathbb{P} ] +\Pi^{\mathcal{P}}_{ij} F^{\mathcal{P}}_{ij} \,.
\end{equation}

If we assign  $R$-charge $r$ to the quarks, on the side of $FT[SU(\Num)]$ we find a monopole matrix with $R[{\rm m}_{ij}]=2-2r$ on the CB, 
and $R[F^{\mathcal{R}}_{ij}]=2-2r.$
On the  side of $\chk{FT[SU(\Num)]}$ we again assign $R$-charge $r$ to the quarks so that again we will find a monopole
matrix with $R[\mathcal{N}_{ij}]=2-2r$, and $R[ F^{\mathcal{P}}_{ij} ]=2-2r$. The operator map will be:
\be
\label{spm1}
 F^{\mathcal{R}}_{ij} \leftrightarrow \mathcal{N}_{ij} \,,\qquad \qquad   {\rm m}_{ij}  \leftrightarrow F^{\mathcal{P}}_{ij}\,.
\ee
The first evidence of this duality was obtained in \cite{Zenkevich:2017ylb} using difference operators acting on the holomorphic blocks. 
The argument is similar to our discussion in section \ref{difo}.

The partition function of the ${FT[SU(\Num)]}$ theory is simply obtained by multiplying the partition function of  ${T[SU(\Num)]}$  
by the contribution  of the  flipping singlets which transform in the adjoint of the $SU(N+1)$  flavor symmetry:
\be
\mathcal{Z}^{FT}[N,m_A;\vec M,\vec T]\equiv K'[\vec M, m_A] \mathcal{Z}^{T}[N,m_A;\vec M,\vec T]\,,
\ee
where the prime indicates that we removed the trace part from the singlet contribution.

Considering the map of operators in \eqref{spm1} we see that flavor and topological fugacities 
will be swapped in the partition function of the dual theory, but the sign of $m_A$ will not change, consistently with our R-charge assignment. We have:
\be
\mathcal{Z}^{\chk{FT}}[N,m_A;\vec M,\vec T]\equiv\mathcal{Z}^{FT}[N,m_A;\vec T,\vec M]\,.
\ee
Proving our spectral duality at the level of partition functions requires to prove the following identity:
\be
\mathcal{Z}^{\chk{FT}}[N,m_A;\vec M,\vec T]=\mathcal{Z}^{FT}[N,m_A;\vec T,\vec M]=\mathcal{Z}^{FT}[N,m_A;\vec M,\vec T]\,.
\ee
But this is immediate if we consider the identity for the duality between ${FFT[SU(\Num)]}$ and $\chk{T[SU(\Num)]}$\footnote{In the ${FFT[SU(\Num)]}$ partition function we can equivalently use $K$ or $K'$ since the trace part will cancel out between the two set of singlets.}: 
\begin{eqnarray}
\mathcal{Z}^{{FFT}}[N,m_A;\vec M,\vec T]&=&K'[\vec T, m_A] K'[\vec M, -m_A] \mathcal{Z}^{T}[N,-m_A;\vec M,\vec T]=\mathcal{Z}^{T}[N,-m_A;\vec T,\vec M]\,.\nonumber \\
\end{eqnarray}
The additional flipping, which lead us to the spectral dual pair, is trivially implemented by moving the contribution of the singlets from the left to the right:
\be
 K[\vec M, -m_A] \mathcal{Z}^{T}[N,-m_A;\vec M,\vec T]= K[\vec T, -m_A] \mathcal{Z}^{T}[N,-m_A;\vec T,\vec M]\,,
\ee
which up to $m_A\to -m_A$ is the identity we were looking for.

\begin{figure}[h]
  \centering
  \includegraphics{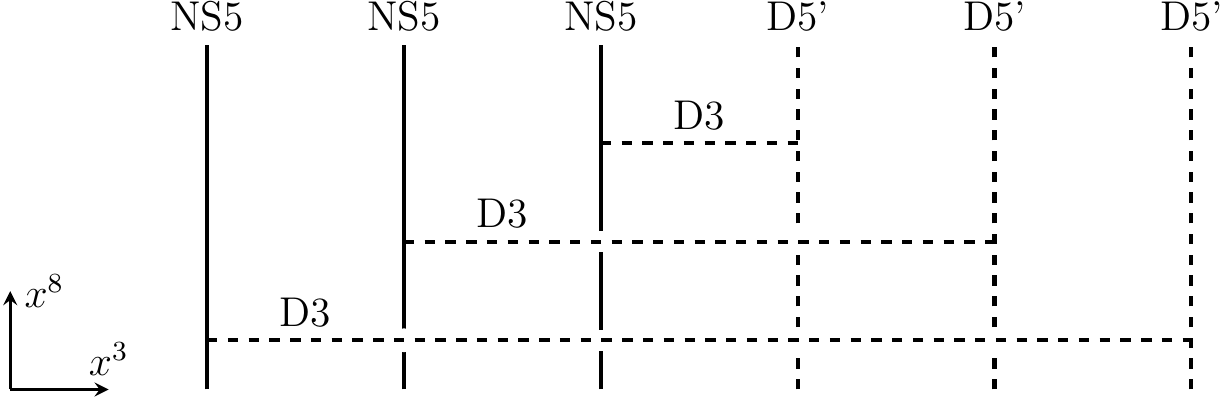}
  \caption{The brane setup giving rise to the $3d$ $FT[SU(3)]$ gauge
    theory. Notice that the NS5 and D5' branes form a $(p,q)$-brane
    web in the directions $(4,9)$ (not shown) and coincide in the
    directions $(7,8)$ ($x^8$ is vertical in the picture).}
  \label{fig:branes-flipped}
\end{figure}

\begin{figure}[h]
  \centering
  \includegraphics{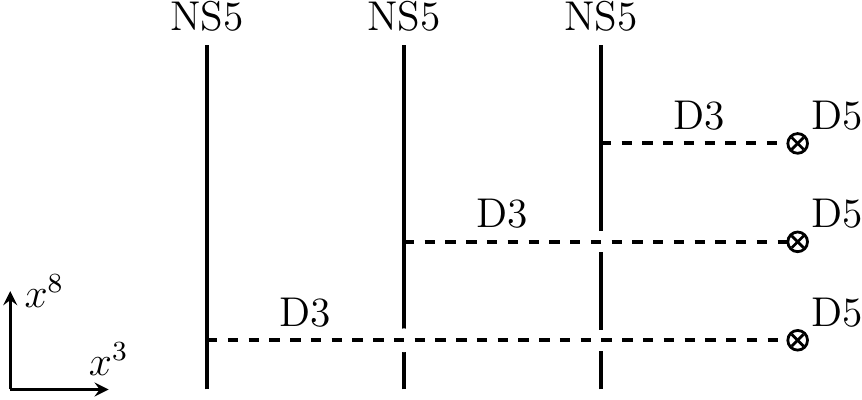}
  \caption{The brane setup giving rise to the $3d$ $T[SU(3)]$ gauge
    theory. The NS5 and D5 branes are perpendicular in all
    non-spacetime directions.}
  \label{fig:branes-unflipped}
\end{figure}

It is interesting to observe that $FT[SU(N+1)]$ and its spectral dual,
similarly to $T[SU(N+1)]$ and its mirror dual, describe the low
energy theory on a stack of D3 branes suspended between NS5 and D5
branes.  Crucially, however, the IIB brane setup for $FT[SU(N+1)]$
involves D5 branes spanning the $012478$ directions as shown in
Tab.~\ref{tab:2} (see also Fig.~\ref{fig:branes-flipped}), so we call
them D5' to distinguish them from the ones relevant for
$T[SU(N+1)]$ in Fig.~\ref{fig:branes-unflipped}.
\begin{table}[h]
  \centering
  \begin{tabular}{c|c|c|c|c|c|c|c|c|c|c}
      & $0$ & $1$ & $2$ & $3$ & $4$ & $5$ & $6$ & $7$ & $8$ & $9$\\
      \hline
      NS5 & $-$ & $-$ & $-$ &  &  &  &  & $-$ & $-$ & $-$\\
      D5' & $-$ & $-$ & $-$ &  & $-$ &  &  & $-$ & $-$  & \\
      D3 & $-$ & $-$ & $-$ & $-$ &  &  &  &  &  &  
\end{tabular}
\caption{The brane setup giving rise to the $3d$ $FT[SU(N+1)]$ gauge theory.}
  \label{tab:2}
\end{table}

The difference between the two set-up is a ``brane flip'' --- the D5'
and D5 branes are transformed into each other under the exchange of
directions $56 \leftrightarrow 78$. The set-up in Tab.~\ref{tab:2} and
Fig.~\ref{fig:branes-flipped}, which preserves $\mathcal{N}=2$
supersymmetry, is also invariant under the action of Type IIB
$S$-duality which turns the NS5 branes into D5' branes leaving the D3
branes invariant and explains the spectral self-duality of
$FT[SU(N+1)]$.  Notice also that in the brane-realisation the
$(N+1)^2$ singlets fields which flip the mesons  correspond
to the degrees of freedom of the D3 branes moving in directions $78$
between two D5' branes (one hyper for each D3 segment) and between a
D5' and an NS5 \cite{Giveon:1998sr}.

At this point it is tempting to speculate that performing also the
flip of the CB moment map to obtain $FFT[SU(N+1)]$ corresponds to
rotating also the NS5 into NS5'. This would give a new $\mathcal{N}=4$
set-up with NS5' and D5' equivalent to the one in Tab.~\ref{tabe:1} as
consistent with the duality $FFT[SU(N+1)]\leftrightarrow T[SU(N+1)]$.

\begin{figure}[h]
 \centering
 \begin{tabular}{cc}
   \includegraphics{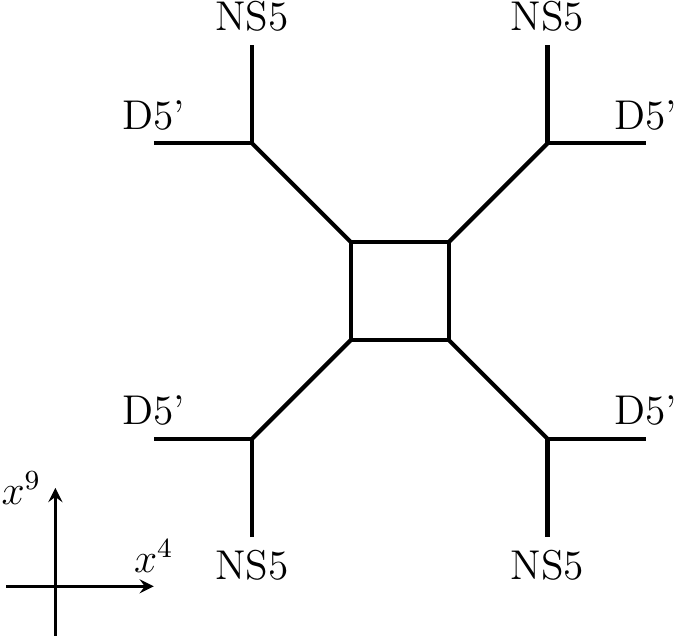}&
 \includegraphics{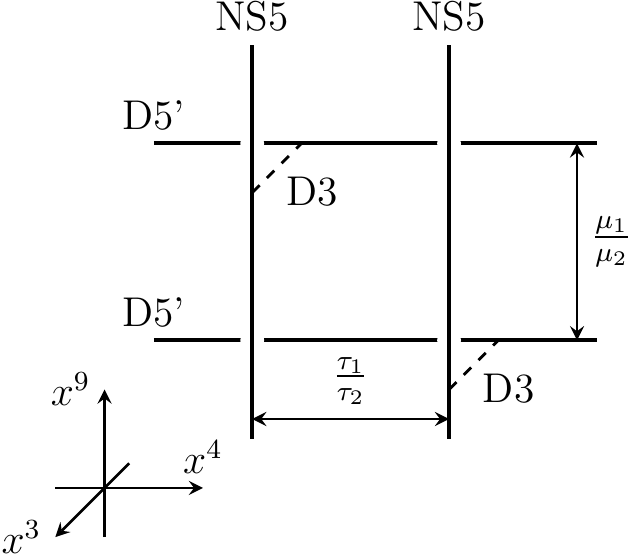}\\
   a)& b)
 \end{tabular}
 \caption{a) The $(p,q)$ five-brane web formed by pairs of
   intersecting D5' and NS5 branes in the $49$ plane, corresponding to
   the $5d$ $\mathcal{N}=1$ $SU(2)$ gauge theory with four fundamental
   hypermultiplets. b) The Higgs branch of the $5d$ theory corresponds
   to the configuration of five-branes separated along the $x^3$
   direction.  Here we consider the case where two D3 branes (here
   depicted as dashed lines) are stretching between the five-branes.}
 \label{fig:100}
\end{figure}

Coming back to the $\mathcal{N}=2$ setup for $FT[SU(N+1)]$ now an
interesting possibility arises. Consider the set-up in
Tab.~\ref{tab:2} but \emph{without} D3 branes.  We assume for a moment
that all the five-branes sit at the same point in $x^3$ direction. The
NS5 and D5' branes will form a $(p,q)$ web in the $49$ plane, as shown
in Fig.~\ref{fig:100}~a) for the simplest example of $N=1$.

The worldvolume theory on the five-branes is the $5d$ $\mathcal{N}=1$
gauge theory living in the $01278$ space. The positions of the
five-branes in the $49$ plane correspond to Coulomb moduli, couplings
and masses of the gauge theory.  In particular for the ``square''
$(p,q)$-web formed by $(N+1)$ NS5 and $(N+1)$ D5' the worldvolume
theory is the $U(N+1)^N$ $5d$ linear quiver theory with $(N+1)$
fundamental hypermultiplets at each end.

If we now go to the Higgs branch of this $5d$ theory where the NS5 and
D5' branes are separated in the $x^3$ direction we can stretch D3
branes between them as in Fig.~\ref{fig:100}~b), arriving precisely at
the setup of Tab.~\ref{tab:2}. Hence we explicitly realize the
$FT[SU(N+1)]$ theory as a defect theory appearing in the Higgs branch of the $5d$ theory.
 This realisation of $FT[SU(N+1)]$ as a defect theory has been discussed extensively in Section 3 of
\cite{Zenkevich:2017ylb}, here we summarize the salient points.

First of all there it was explicitly shown how the Higgsing
prescription can be implemented starting from the topological string
partition function for the toric CY $\mathfrak{S}$ with square toric
graph (with $(N+1)$ vertical and  $(N+1)$ horizontal legs).
  In particular the partition function of $FT[SU(N+1)]$ on $D^2\times
S^1$, evaluated on a reference contour $\alpha_0$ $ {\mathcal
  B}^{\alpha_0}_{FT[SU(N+1)]}(\vec \mu,\vec \tau, t)$ where the
parameters $\mu_i$, $\tau_j$, $t$ are exponentiated versions of $M_i$, $T_j$, $m_A$, is obtained from
$Z^{\mathcal{S}}_{\mathrm top}$ when the \emph{complete} Higgsing
pattern (eqs. (3.8) and (3.9) in  \cite{Zenkevich:2017ylb}) $\mathfrak{S}\to \mathcal{S}$ is implemented: \begin{equation} 
{\mathcal B}^{\alpha_0}_{FT[SU(N+1)]}(\vec \mu,\vec \tau, t)= G Z^{\mathcal{S}}_{1 \mathrm{-loop, top}}Z^{\mathcal{S}}_{\mathrm{vort}, \mathrm{top}}(\vec \mu,\vec \tau, t)\,,
\end{equation}
where $G$ is a fiber-base invariant factor. The parameters $\mu_i$, $\tau_j$ are identified with K\"ahler
parameters while the exponentiated axial mass $t$ is identified with
one of the equivariant $\Omega$-background parameters on
$\mathbb{R}^4_{q,t}\times S^1$.

Then in \cite{Zenkevich:2017ylb} it was observed that the topological
string partition function is invariant under fiber-base duality (which
in the case of the \emph{square} diagram $\mathfrak{S}$ is
\emph{self}-duality) even after Higgsing:
\begin{equation}
Z^{\mathcal{S}}_{1 \mathrm{-loop, top}}Z^{\mathcal{S}}_{\mathrm{vort}, \mathrm{top}}(\vec \mu,\vec \tau, t)=
Z^{\mathcal{S}'}_{1 \mathrm{-loop, top}}Z^{\mathcal{S}'}_{\mathrm{vort}, \mathrm{top}}(\vec \mu,\vec \tau, t)\,,
\end{equation}
which implies the $3d$ spectral self-duality of the $3d$ blocks:
\begin{equation}
{\mathcal B}^{\alpha_0}_{FT[SU(N+1)]}(\vec \mu,\vec \tau, t)={\mathcal
  B}^{\alpha_0}_{FT[SU(N+1)]}(\vec \tau,\vec \mu, t)={\mathcal
  B}^{\alpha_0}_{\chk{FT[SU(N+1)]}}(\vec \mu,\vec \tau, t)\,.\label{eq:4}
\end{equation}
Fiber-base duality exchanges the K\"ahler parameters of the base with
that of the fiber, thus it exchanges $\mu_i$ and $\tau_j$, but $t$ is
left untouched since it is the parameter of the $\Omega$-background
(or, equivalently, the refinement parameter of refined topological
string).

\subsection{A new spectral dual pair}
\label{sec:new-spectral-dual}
The reasoning that led us to state the spectral duality between
$FT[SU(\Num)]$ and $\chk{FT[SU(\Num)]}$ can be used on theory $D$ and
theory $B$ to obtain a daughter spectral duality. Recall that theory $D$ is SQED with $(N+1)$ flavors, 
$U_i$ and $\widetilde{U}_i$, mesonic and monopole flipping fields, and superpotential
\begin{eqnarray}
W_D=\m^{\pm} \sigma_\pm    + \mathcal{U}_{ij} (\Flip^{\mathcal{U}}_D)_
{ij}- \sum_{m=1}^{N } \frac{(-)^m}{m}\ \theta_m\,\Tr\Big[\underbrace{ \,\Flip^{\mathcal{U}}_D\ldots\Flip^{\mathcal{U}}_D
\, }_{m{\rm\ times}} \Big]\,.
\label{W_D_superP}
\end{eqnarray}
The flipping fields $\sigma_\pm$ and $\Flip^{\mathcal{U}}_{ij}$ originated from Aharony duality on theory $A$. 

To arrive at theory  $\mathcal{T}$  we flip the singlets $\sigma_\pm$ and $\theta_{m\ge 2}$, since $flip^2=1$ we arrive at:
\begin{eqnarray}
\mathcal{W}_{\mathcal{T}}=\phi (\Flip^{\mathcal{U}}_D)_{ii} + \mathcal{U}_{ij} (\Flip^{\mathcal{U}}_D)_{ij} \,,
\end{eqnarray}
where we redefined $\theta_1=\phi$, for simplicity. 
We then can use  the F-terms of $\phi$ and consider traceless flipping fields. 

Theory $\mathcal{T}'$ is obtained from theory $B$ upon repeating the same two operations that define theory $\mathcal{T}$. 
From the operator map given in section \ref{ABCD_sec},  we see that
the fields $\theta_{m\ge2}$ correspond to dressed mesons of theory $B$:
\bea
\begin{array}{clc}%
\theta_2&\leftrightarrow&\tilde{d}\underbrace{\,\checkPhi_N \cdots\checkPhi_N }_{N-2{\rm\ times} } {d}\\
\vdots\\[.2cm]
\theta_N&\leftrightarrow&\tilde{d} {d}
\end{array}
\eea
while  the monopoles $\sigma^\pm$ are mapped to the  two mesons $\tilde d p$ and $d \tilde p$.
So we have:
\be
\mathcal{W}_{\mathcal{T}'}=W_B+ \Flip_+ \tilde d~ p+ \Flip_- d ~\tilde p+\sum_{k=0}^{N-2} \Flip_k  ~\tilde{d} ~\checkPhi_N^k ~ d
\ee

The equality of the partition functions  $\mathcal{Z}_{\mathcal{T}}=\mathcal{Z}_{\mathcal{T}'}$  follows from the equality $\mathcal{Z}^D=\mathcal{Z}^B$ simply  by reshuffling the flipping fields and we obtain:\footnote{For later convenience 
we have changed the sign of  $\xi$ in $\mathcal{Z}_{\mathcal{T}}$. On the dual side $\mathcal{Z}_{\mathcal{T}'}$ we
on top replacing $\xi\to -\xi$, we  also change  the signs of the integration variables $x_i^{(k)}\to-x_i^{(k)} $.}

\bea
\nonumber
Z_{\mathcal{T}}&=&s_b(m_A)\prod_{a,b=1}^{N+1}s_b\big(-m_A+M_a-M_b \big)
\int dx \, e^{-2\pi i \xi x} \,
\prod_{a=1}^{N+1}   s_b\big(\tfrac{iQ}{4}+\tfrac{m_A}{2} \pm(x_j-M_a)\big)=\nonumber\\
&&=  e^{-2\pi i M_{N+1}  \xi } s_b\left(\tfrac{iQ}{2}-\tfrac{N+1}{2}( m_A+\tfrac{iQ}{2}) \pm\xi  \right) \prod_{l=1}^{N-1} s_b\left(\tfrac{iQ}{2}-(l+1)(m_A+\tfrac{i Q}{2})\right)\nn \\
&&
\int \prod_{k=1}^{N-1} dx^{(k)}e^{-2\pi i \zeta^{(k)} \sum_{a=1}^k x^{(k)}_a} Z_{\rm vec}^{(k)}~Z_{\rm adj}^{(k)}~ Z_{\rm bif}^{(k,k+1)}
 \int dx^{(N)}e^{-2\pi i \zeta^{(N)}\sum_{a=1}^N x^{(N)}_a} Z_{\rm vec}^{(N)}~Z_{\rm adj}^{(N)}~\nn\\
&&
\rule{1.3cm}{0pt}
\times\prod_{i=1}^{N} s_b\left(\tfrac{iQ}{4}+\tfrac{m_A}{2}\pm(x^{(N)}_i-\xi)\right)\,\prod_{i=1}^N s_b\left(\tfrac{N}{2}\left(m_A+\tfrac{i Q}{2}\right)\pm x^{(N)}_i\right)=
\mathcal{Z}_{\mathcal{T}'} 
\eea
In the first line we can notice the cancellation of the trace-part of the flipping fields with the singlet $\phi$.

\subsection*{Holomorphic blocks}
In this section realise theory ${\mathcal{T}}$ and ${\mathcal{T}'}$ as defect
theories via Higgsing. First of all we need the holomorphic blocks,
i.e.\ $D^2\times S^1$ partition functions evaluated on a reference
contour.

The block integrands $\Upsilon_{\mathcal{T}}$ and
$\Upsilon_{\mathcal{T}'}$ can be easily obtained by taking the
``square-root'' of the $S^3_b$ integrand as observed
in~\cite{Beem:2012mb}, and reviewed in~\cite{Pasquetti:2016dyl}. Their explicit expression can be found
 in Eqs.~(\ref{upst}) and~(\ref{upstp}) in the Appendix.

We then evaluate the block integrands on a basis of contours $\Gamma_\alpha$ with  $\alpha=1,\cdots N+1$ which are in one-to-one correspondence withe the $(N+1)$ SUSY vacua of the theory. Similarly we will evaluate the block integrand for the spectral dual theory on a basis of contour to obtain the blocks  ${\mathcal  B}^{\beta}_{\mathcal{T}'}$:
\be
{\mathcal  B}^{\alpha}_{\mathcal{T}}=\oint_{\Gamma_\alpha} \Upsilon_{\mathcal{T}}\,, \qquad \qquad {\mathcal  B}^{\beta}_{\mathcal{T}'}=\oint_{\Gamma_\beta} \Upsilon_{\mathcal{T}'}\,.
\ee
Testing the the spectral duality at the level of the blocks requires
to establish a  map between each element of the basis of
theory ${\mathcal{T}}$ and ${\mathcal{T}'}$.  In terms of field theory objects,
the matrix elements of this map are partition functions of $2d$
theories living on the interface between theory ${\mathcal{T}}$ in
vacuum $\alpha$ and ${\mathcal{T}'}$ in vacuum $\beta$. Geometrically
the interface is a torus $\partial(D_2 \times S^1) = T^2$, the
equivariant parameter $q$ of the $D_2 \times S^1$ background plays the
role of the complex structure of the boundary torus and the $2d$
partition function is a version of elliptic index, hence expressed in
terms of Jacobi theta-functions $\theta_q$. However we will not be
concerned with evaluating fully the matrix of transition coefficients.

We limit ourselves to the evaluation of the blocks of ${\mathcal{T}}$ on a \emph{reference} contour $\Gamma_{\alpha_0}$. On the dual side we are able to identify the corresponding contour which we also call $\Gamma_{\alpha_0}$.
The details of the calculations can be found in the appendix, here we give the final result:
\be
{\mathcal   B}^{\alpha_0}_{\mathcal{T}}=\oint_{\Gamma_{\alpha_0}} \Upsilon_{\mathcal{T}}=\mathcal{E}_{\mathcal{T}}
  Z^{\alpha_0}_{\mathrm{cl}, \mathcal{T}}Z^{\alpha_0}_{1 \mathrm{-loop}, \mathcal{T}}
  Z^{\alpha_0}_{\mathrm{vort}, \mathcal{T}}\,.
\ee
The explicit forms of $ Z^{\alpha_0}_{\mathrm{cl}, \mathcal{T}}Z^{\alpha_0}_{1 \mathrm{-loop}, \mathcal{T}}$  and 
$ Z^{\alpha_0}_{\mathrm{vort}, \mathcal{T}}$ are given in   in Eqs.~(\ref{colt}), (\ref{eq:39}). On the dual side we have:
\be
{\mathcal  B}^{\alpha_0}_{\mathcal{T}'}=\oint_{\Gamma_{\alpha_0}}\Upsilon_{\mathcal{T}'}=  \mathcal{E}_{\mathcal{T}'}
  Z^{\alpha_0}_{\mathrm{cl}, \mathcal{T}'}Z^{\alpha_0}_{1 \mathrm{-loop}, \mathcal{T}'}
  Z^{\alpha_0}_{\mathrm{vort}, \mathcal{T}}\,.
\ee
The explicit forms of $  Z^{\alpha_0}_{\mathrm{cl}, \mathcal{T}'}Z^{\alpha_0}_{1 \mathrm{-loop}, \mathcal{T}'}$ and 
$ Z^{\alpha_0}_{\mathrm{vort}, \mathcal{T}'}$ are given in  Eqs.~(\ref{coltp}), (\ref{eq:33}).

\subsection{Spectral duality from Fiber-Base}
\label{sec:spectr-dual-from}
In this section we explain how the $3d$ spectral duality between
theories $\mathcal{T}$ and $\mathcal{T}'$ follows from fiber-base
duality of refined topological string.  First of all we need to
establish the Higgsing prescription which allows us to obtain
${\mathcal B}^{\alpha}_{\mathcal{T}}$ and ${\mathcal
  B}^{\alpha}_{\mathcal{T}'}$ from refined topological string
partition functions with tuned K\"ahler parameters.

Refined topological strings provide a deformation of the topological
A-model partition function on toric CY threefolds. Apart from the
exponentiated string coupling $q = e^{-g_s}$ the deformation depends
on an additional parameter $t$, so that for $t=q$ the conventional
partition function is recovered. The rules for computing partition
were introduced in \cite{Iqbal:2007ii}. Here we briefly recall that
the main ingredient is the trivalent \emph{refined vertex}
\begin{multline}
  \label{eq:23}
  C_{ABC}(t,q) =\quad 
    \includegraphics[valign=c]{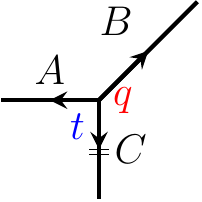} \quad = q^{\frac{||B||^2 + ||C||^2}{2}}
  t^{-\frac{||B^{\mathrm{T}}||^2 + ||C^{\mathrm{T}}||^2}{2}} M^{(q,t)}_C
  \left(t^{-\rho}\right) \times\\
  \times\sum_D \left( \frac{q}{t}
  \right)^{\frac{|D|+|A|-|B|}{2}}  \chi_{A^{\mathrm{T}}/D}
  \left(q^{-C} t^{-\rho}\right) \chi_{B/D}\left(t^{-C^{\mathrm{T}}} q^{-\rho} \right), 
\end{multline}
associated to a vertex of the toric diagram, i.e.\ to a $\mathbb{C}^3$
patch. $A$, $B$ and $C$ are Young diagrams assigned to the
intermediate legs of the toric diagram. In a generic toric diagram obtained gluing trivalent vertices,
 each intermediate leg is
geometrically a compact 2-cycle $\mathbb{P}^1$, to which corresponds a
K\"ahler parameter $k = \int_{\mathbb{P}^1} \omega$, where $\omega$ is
the K\"ahler form on the CY $\mathcal{X}$. $k$, together with the integral of the
$B$-field $b = \int_{\mathbb{P}^1} B$ defines the exponentiated
complexified K\"ahler parameter $Q = e^{-b+ik}$. The partition
function is given by the sum of the product of refined topological
vertices with additional weights of the form $Q^{|A|}$ for each
intermediate leg. The sum is carried over all Young diagrams on the
intermediate legs with empty diagrams assigned to the external legs.

\begin{figure}[h]
  \centering
  \includegraphics{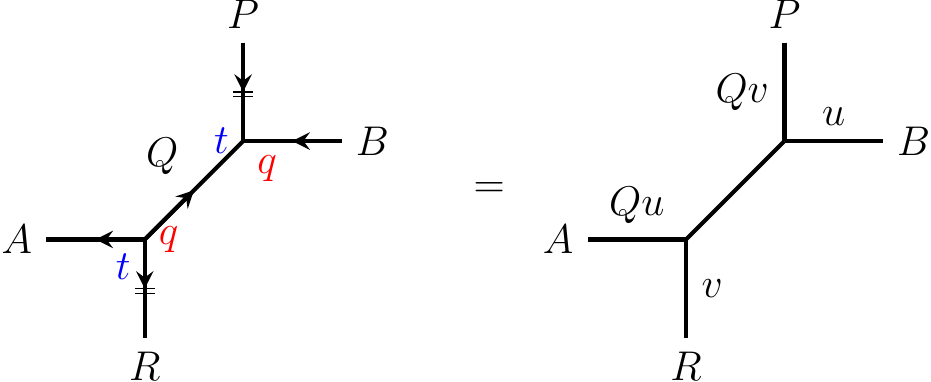}
  \caption{Resolved conifold geometry in refined topological
    strings. The double ticks denote the preferred direction, and $t$
    and $q$ indicate the respective legs of the refined topological
    vertices. $Q$ is the exponentiated complexified K\"ahler parameter
    of the base $\mathbb{P}^1$ (drawn as an intermediate diagonal
    edge). $A$, $B$, $P$ and $R$ are Young diagrams 
    associated with the outer legs. The right picture is the
    simplification of the left one with spectral parameters on the
    legs playing the roles of K\"ahler parameters.}
  \label{fig:110}
\end{figure}

It will be more convenient for us to use \emph{spectral} parameters,
assigned to all the legs of the diagram, instead of \emph{K\"ahler}
parameters associated only with the intermediate
edges. Fig.~\ref{fig:110} explains the identification for the basic
example we will need in our setup, the resolved conifold geometry. The
piece of the partition function corresponding to the resolved conifold
from Fig.~\ref{fig:110} is given by
\begin{multline}
  \label{eq:22}
     Z_{\mathrm{conifold}} \left( \left.
  \begin{smallmatrix}
     & P & \\
     A&  & B\\
     & R &
  \end{smallmatrix}
\right| Q, q,t \right) = \sum_C (-Q)^{|C|} C_{ACR}(t,q)
C_{B^{\mathrm{T}} C^{\mathrm{T}} P^{\mathrm{T}}}(q,t) =\\
= Z \left( \left.
  \begin{smallmatrix}
     & \varnothing & \\
     \varnothing&  & \varnothing\\
     & \varnothing &
  \end{smallmatrix}
\right| Q, q,t \right)  q^{\frac{||R||^2 - ||P||^2}{2}}
t^{\frac{||P^{\mathrm{T}}||^2 - ||R^{\mathrm{T}}||^2}{2}} \left( \frac{q}{t}
\right)^{\frac{|A| - |B|}{2}}
M_R^{(q,t)}(t^{-\rho}) M_{P^{\mathrm{T}}}^{(t,q)} (q^{-\rho})
G_{RP}^{(q,t)} \left( \sqrt{\frac{q}{t}} Q \right) \times\\
\times \sum_C (-Q)^{|C|}
\chi_{A^{\mathrm{T}}/C^{\mathrm{T}}} \left(p_n (t^{-\rho} q^{-R}) - p_n \left(
    \sqrt{\frac{q}{t}} Q t^{-\rho} q^{-P}  \right)\right) \times\\
\times \chi_{B/C}
\left( p_n (q^{-\rho} t^{-P^{\mathrm{T}}}) - p_n \left(
    \sqrt{\frac{t}{q}} Q q^{-\rho} t^{-R^{\mathrm{T}}} \right) \right)
\end{multline}

In what follows we normalize $Z_{\mathrm{conifold}}$ so that it is an
identity when all the external legs are empty, i.e.\ we divide by
$Z \left( \left.
  \begin{smallmatrix}
     & \varnothing & \\
     \varnothing&  & \varnothing\\
     & \varnothing &
  \end{smallmatrix}
\right| Q, q,t \right)$.

\begin{figure}[h]
  \centering
  \begin{tabular}{c}
    \includegraphics{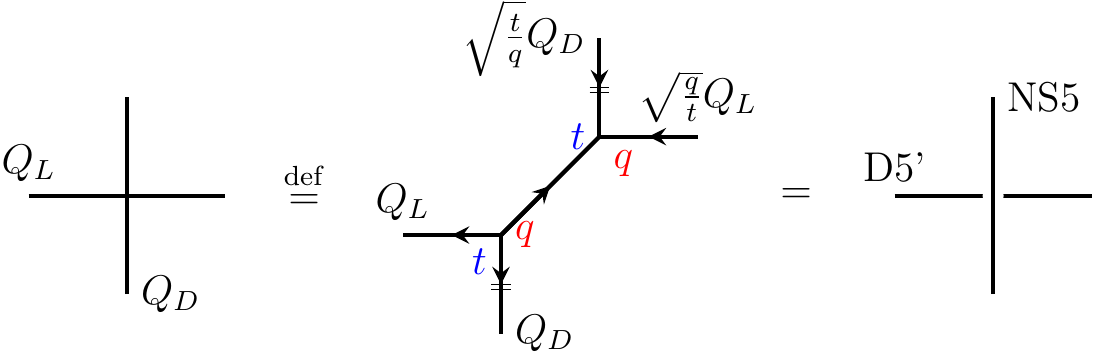}\\
    a)
  \end{tabular}\\[.5cm]
  \begin{tabular}{c}
    \includegraphics{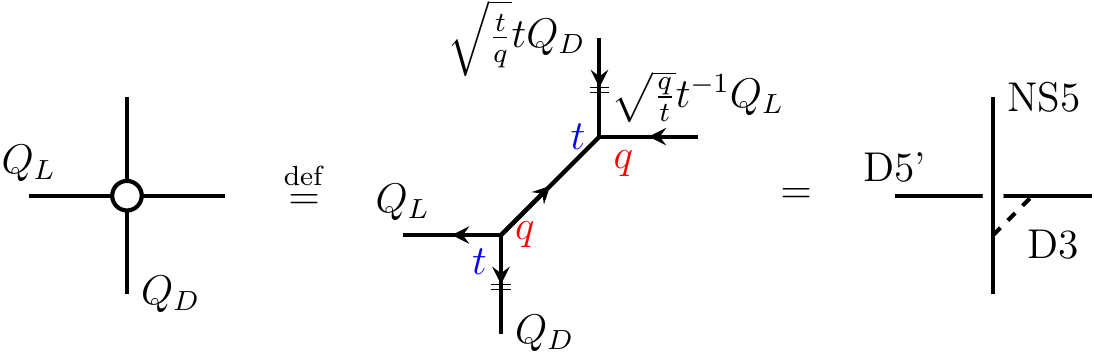}\\
    b)
  \end{tabular}
  \caption{Higgsing of the resolved conifold geometry leading to two
    different types of crossings. Notice the particular values of the
    K\"ahler parameters on the legs. a) An ``empty'' crossing, i.e.\
    without D3 branes stretched between the NS5 and D5'. b) a ``full''
    crossing, i.e.\ with one D3 brane stretched between the NS5 and
    D5'.}
  \label{fig:105}
\end{figure}

The crucial point for our Higgsing construction is that for quantized
$Q$ the function~(\ref{eq:22}) actually vanishes for a large subset of
``boundary conditions'' (external Young diagrams). Namely for the
situation pictured in Fig.~\ref{fig:105}, the \emph{lengths} of the
diagrams on the vertical leg before and after the crossing are
constrained as follows:
\begin{align}
  \label{eq:45}
  \text{for
  }\qquad &\includegraphics[valign=c]{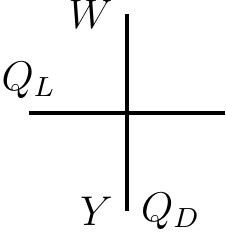} \qquad
  l(W) \leq l(Y),\\
    \text{for
  }\qquad &\includegraphics[valign=c]{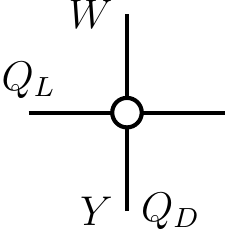} \qquad
  l(W) \leq l(Y)+1.\label{eq:46}
\end{align}
These constraints are valid irrespective of the diagrams propagating
on the \emph{horizontal} leg. For example if the resolved conifold
fragment~(\ref{eq:45}) sits in the lowest part of the diagram, then
$Y=\varnothing$ since it corresponds to an external leg, and therefore
$W$ is constrained to be empty. The block~(\ref{eq:46}) in the same
situation would constraint the diagram $W$ to have just one column,
i.e.\ $W=[k]$, $k\in \mathbb{Z}_{\geq 0}$. The integers $k$ in this
construction will correspond to the summation variables in the $3d$
vortex series.

We will denote the ``Higgsed'' CY manifold (i.e.\ the CY with discrete
choice of K\"ahler parameters) corresponding to the $3d$ theory
$\mathcal{T}$ by $\mathcal{Y}$ and that corresponding to
$\mathcal{T}'$ by $\mathcal{Y}'$. Of course, $\mathcal{Y}'$ is the
fiber-base dual (the mirror image along the diagonal) of
$\mathcal{Y}$. Below we give some details of the topological string
computations for $\mathcal{Y}$ and $\mathcal{Y}'$.

\subsubsection*{CY $\mathcal{Y}$}
\label{sec:abelian-theory}

The toric diagram for the CY  $\mathcal{Y}$  in the case $N=4$ looks as follows:
\begin{multline}
  \label{eq:5}
  Z^{\mathcal{Y}}_{\mathrm{top}}(\vec{\mu}, \vec{\tau}, q, t) \quad =
  \quad \includegraphics[valign=c]{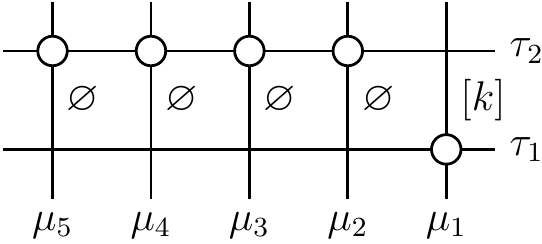} \quad
  =\\
  = Z^{\mathcal{Y}}_{\mathrm{top, }1\mathrm{-loop}} (\vec{\mu},
  \xi, q, t) Z_{\mathrm{vortex}}^{\mathcal{Y}}(\vec{\mu},
  \xi, q, t).
\end{multline}
where $\tau_1 = e^{2\pi b \xi}$, $\tau_2 = t^{\frac{1-N}{2}}$. Here we
have explicitly indicated the Young diagrams propagating on the
intermediate vertical legs. These diagrams are constrained by the
rules, ~(\ref{eq:45}), (\ref{eq:46}), so that $[k]$ is the single
column Young diagram. It is this variable over which the summation in
the vortex series is performed. We normalize our partition function so
that $Z_{\mathrm{vortex}}^{\mathcal{Y}}(\vec{\mu}, \xi, q, t)$ is a
series in $e^{2\pi b \xi}$ which starts with identity.  The partition
function can then be computed explicitly e.g.\ using the resolved
conifold formula from Eq.~(\ref{eq:22})\footnote{There is, however, a
  more compact and convenient operator product technique   
\cite{Awata:2016riz,Mironov:2016yue,Awata:2011ce}, which we don't present here not to overcomplicate the
  presentation with technicalities.} and the result coincides with the
series vortex series~(\ref{eq:39}).

The relative prefactor $Z^{\mathcal{Y}}_{\mathrm{top,
  }1\mathrm{-loop}} (\vec{\mu}, \xi, q, t)$ is easy to
calculate --- it is what remains of the partition function when
$\frac{\tau_1}{\tau_2}$ goes to zero.  This limit corresponds to an
infinitely large K\"ahler parameter between the two horizontal legs
in~(\ref{eq:5}), so that the diagram splits into a product of two
horizontal strip partition functions. Indeed, in this limit only $k=0$
contributes and we have:
\begin{equation}
  \label{eq:7}
  Z^{\mathcal{Y}}_{\mathrm{top, }1\mathrm{-loop}}
(\vec{\mu}, \xi, q, t) \quad = \quad \includegraphics[valign=c]{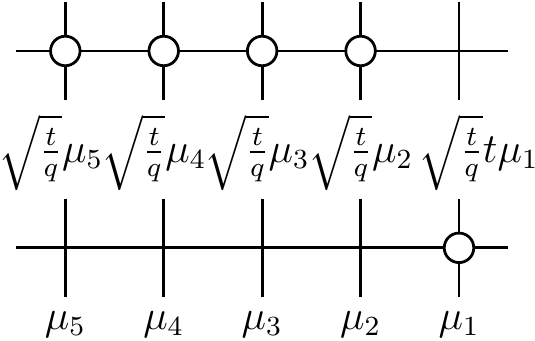}.
\end{equation}
Here we have written out the spectral parameter of the vertical legs
explicitly. The well-known formula for the refined strip
partition function gives  \cite{Taki:2007dh}:
\begin{equation}
  \label{eq:9}
  Z^{\mathcal{Y}}_{\mathrm{top, }1\mathrm{-loop}}
  (\vec{\mu}, \xi, q, t) = \prod_{i=2}^{N+1} \prod_{j=i+1}^{N+1} \frac{\left( \frac{q}{t} \frac{\mu_i}{\mu_j};q \right)_{\infty}}{\left( t \frac{\mu_i}{\mu_j};
      q \right)_{\infty}} \prod_{k=2}^{N+1}  \frac{\left( q \frac{\mu_1}{\mu_k};q \right)_{\infty}}{\left( t \frac{\mu_1}{\mu_k};
      q \right)_{\infty}}.
\end{equation}

\subsubsection*{CY $\mathcal{Y}'$}
\label{sec:spec-d-abelian-quiver}
The toric diagram for the spectral dual CY
$\mathcal{Y}'$ is simply the mirror image along the diagonal of that
of $\mathcal{Y}$~\eqref{eq:5}, so that:
\begin{equation}
  \label{eq:6}
  Z^{\mathcal{Y}'}_{\mathrm{top, }1\mathrm{-loop}}
  (\vec{\mu}, \vec{\tau}, q, t) Z^{\mathcal{Y}'}_{\mathrm{vortex}}
  (\vec{\mu}, \vec{\tau}, q, t) =  \includegraphics[valign=c]{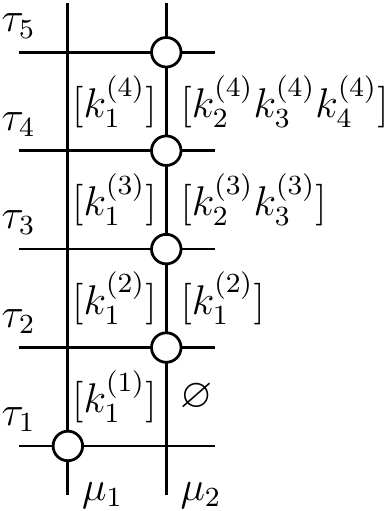}
\end{equation}
Here we have again used the rules~(\ref{eq:45}), (\ref{eq:46}) to
constraint the Young diagrams on the vertical legs. The integers
$k_i^{(a)}$ are precisely the integers in the $3d$ vortex sum and
$Z^{\mathcal{Y}'}_{\mathrm{vortex}}$ can be checked to
reproduce~(\ref{eq:33}).

The computation of the one-loop factor is similar to
sec.~\ref{sec:abelian-theory}: the toric diagram in the limit
$\frac{\tau_a}{\tau_{a+1}} \to 0$ splits into $(N+1)$ horizontal
strips. Using the result for the strips we obtain
\begin{equation}
  \label{eq:10}
  Z^{\mathcal{Y}'}_{\mathrm{top, }1\mathrm{-loop}}
  (\vec{\mu}, \vec{\tau}, q, t) =  \frac{\left( q
      t^{\frac{N-1}{2}}e^{2\pi \beta \xi};q \right)_{\infty}}{\left(
      t^{\frac{N+1}{2}}e^{2\pi \beta \xi} ;
      q \right)_{\infty}}.
\end{equation}

The fiber-base duality of the topological string partition function
(after Higgsing) yields the following equality:
\begin{equation}
  Z^{\mathcal{Y}}_{1 \mathrm{-loop},
    \mathrm{top}}Z^{\mathcal{Y}}_{\mathrm{vort}, \mathrm{top}}=
  Z^{\mathcal{Y}'}_{1 \mathrm{-loop}, \mathrm{top}}Z^{\mathcal{Y}'}_{\mathrm{vort}, \mathrm{top}}\label{eq:11}
\end{equation}
A simple brute force check of Eq.~(\ref{eq:11}) to lower orders in the
K\"ahler parameters is given in Appendix~\ref{sec:an-explicit-check}.

\subsection*{Match of field theory with $Z_{\mathrm{top}}$}

Finally we relate our gauge theory results for the holomorphic blocks with the results
of the Higgsing prescription. We find that:
\begin{equation}
  {\mathcal B}^{\alpha_0}_{\mathcal{T}}=\mathcal{E}_{\mathcal{T}}
  Z^{\alpha_0}_{\mathrm{cl}, \mathcal{T}}Z^{\alpha_0}_{1 \mathrm{-loop}, \mathcal{T}}
  Z^{\alpha_0}_{\mathrm{vort}, \mathcal{T}}= G_{\mathcal{T}} Z^{\mathcal{Y}}_{1 \mathrm{-loop, top}}Z^{\mathcal{Y}}_{\mathrm{vort}, \mathrm{top}}
\end{equation}
and 
\begin{equation}
  {\mathcal B}^{\alpha_0}_{\mathcal{T}'}=\mathcal{E}_{\mathcal{T}'}
  Z^{\alpha_0}_{\mathrm{cl}, \mathcal{T}'} Z^{\alpha_0}_{1
    \mathrm{-loop}, \mathcal{T}'} Z^{\alpha_0}_{\mathrm{vort},
    \mathcal{T}'}= G_{\mathcal{T}'} Z^{\mathcal{Y}'}_{1 \mathrm{-loop}, \mathrm{top}}Z^{\mathcal{Y}'}_{\mathrm{vort}, top}\,,
\end{equation}
where $G_{\mathcal{T}}, G_{\mathcal{T}'}$ denote fiber-base invariant prefactors.

Since we checked   the fiber-base duality of the refined string
Eq.~(\ref{eq:11}))we are left to check that:
\begin{equation}
  \label{eq:19}
  \frac{G_{\mathcal{T}}}{ G_{\mathcal{T}'}}=1,
\end{equation}
or
\begin{equation}
\frac{\mathcal{E}_{\mathcal{T}}  Z^{\alpha_0}_{\mathrm{cl},
    \mathcal{T}}Z^{\alpha_0}_{1 \mathrm{-loop}, \mathcal{T}}
}{\mathcal{E}_{\mathcal{T}'} Z^{\alpha_0}_{cl,
    \mathcal{T}'}Z^{\alpha_0}_{1 \mathrm{-loop}, \mathcal{T}'} }
\frac{Z^{\mathcal{Y}'}_{1 \mathrm{-loop},
    \mathrm{top}}}{Z^{\mathcal{Y}}_{1 \mathrm{-loop}, \mathrm{top}}}=
1.\label{eq:1}
\end{equation}
In fact Eqs.~(\ref{eq:1}),(\ref{eq:19}) can be relaxed slightly: the r.h.s.\ can be a
$q$-periodic function, e.g.\ a combination of $\theta_q$-functions
which also becomes an identity when glued into the
$S^3_b$ partition function. Notice that the topological string
partition function lacks the classical (i.e.\ power function) part, so
the relation~(\ref{eq:1}) is essentially the requirement that the
\emph{classical} part of the field theory holomorphic block be
fiber-base duality invariant on its own.

We evaluate~(\ref{eq:1}) in two steps. We combine
Eqs.~(\ref{eq:31}), (\ref{eq:14}), (\ref{eq:9}) to get
\begin{multline}
  \label{eq:13}
  \frac{ Z^{\alpha_0}_{\mathrm{cl}, \mathcal{T}}Z^{\alpha_0}_{1
      \mathrm{-loop}, \mathcal{T}}}{Z^{\mathcal{Y}}_{1 \mathrm{-loop},
      \mathrm{top}}} = \frac{\mathcal{F}_{\mathcal{T}}
    \mathcal{I}^{\alpha_0}_{0,\mathcal{T}}(\vec{\mu}, \xi,q,t)
  }{Z^{\mathcal{Y}}_{1 \mathrm{-loop}, \mathrm{top}}} =\\
  = \frac{1}{(t)_{\infty}^N} e^{-2\pi i b M_1 \xi - \pi b \beta (N+1)
    M_1} \prod_{k=2}^{N+1} \frac{\theta_q \left( t \frac{\mu_k}{\mu_1}
    \right)}{\theta_q \left( \frac{\mu_k}{\mu_1} \right)}
  \prod_{i>j}^{N+1} \frac{\theta_q \left( \frac{q}{t}
      \frac{\mu_i}{\mu_j} \right)}{\theta_q \left( t
      \frac{\mu_i}{\mu_j}
    \right)} \sim\\
  \sim \frac{e^{-2\pi i b M_1 \xi + \pi b \beta (N+1) M_1 - 2\pi b
      \beta \sum_{k=1}^{N+1} M_k +2\pi b (1-2\beta)
      \sum_{j=1}^{N+1}M_j (N+2 - 2j) }}{(t)_{\infty}^N}.
\end{multline}
where the last equality is up to $q$-periodic function of
$\mu_k$. Similarly we combine
Eqs.~\eqref{eq:32}, (\ref{eq:18}), (\ref{eq:10}) and obtain
\begin{multline}
  \label{eq:29}
  \frac{ Z^{\alpha_0}_{\mathrm{cl}, \mathcal{T}'}Z^{\alpha_0}_{1
      \mathrm{-loop}, \mathcal{T}'}}{Z^{\mathcal{Y}'}_{1 \mathrm{-loop},
      \mathrm{top}}} = \frac{\mathcal{F}_{\mathcal{T}'}
    \mathcal{I}^{\alpha_0}_{0,\mathcal{T}'}(\vec{\mu},
    \xi,q,t) }{Z^{\mathcal{Y}'}_{1 \mathrm{-loop}, \mathrm{top}}}=\\
  = \frac{1}{(t)_{\infty}^N}e^{\pi i b^2 \beta^2 \frac{N(N^2-1)}{6} + \pi b\beta (N+1) M_1 + \pi b \beta \sum_{a=1}^{N+1} (2a-3-N)
    M_a - 2 \pi i \xi (M_1 - M_{N+1} - i b\beta N) } \frac{\theta_q
    \left( t^{\frac{N+1}{2}} e^{2\pi b \xi} \right)}{\theta_q
    \left( q t^{\frac{N-1}{2}} e^{2\pi b \xi} \right)} \sim\\
  \sim \frac{1}{(t)_{\infty}^N} e^{\pi i b^2 \beta^2 \frac{N(N^2-1)}{6} + \pi b\beta (N+1) M_1 + \pi b \beta \sum_{a=1}^{N+1} (2a-3-N)
    M_a - 2 \pi i \xi (M_1 - M_{N+1} - i b\beta N) + 2\pi b \xi (1-\beta)}\,.
\end{multline}
Dividing Eq.~(\ref{eq:13}) by Eq.~(\ref{eq:29}) we get
\begin{multline}
  \label{eq:30}
  \frac{ Z^{\alpha_0}_{\mathrm{cl}, \mathcal{T}}Z^{\alpha_0}_{1
      \mathrm{-loop}, \mathcal{T}}}{Z^{\mathcal{Y}}_{1 \mathrm{-loop},
      \mathrm{top}}} \frac{Z^{\mathcal{Y}'}_{1 \mathrm{-loop},
      \mathrm{top}}}{ Z^{\alpha_0}_{\mathrm{cl}, \mathcal{T}'}Z^{\alpha_0}_{1
      \mathrm{-loop}, \mathcal{T}'}} \sim\\
  \sim e^{-\pi i b^2 \beta^2 \frac{N(N^2-1)}{6}  + \pi b \beta \sum_{a=1}^{N+1} (N+1-2a)
    M_a - 2 \pi i \xi ( M_{N+1} + i b \beta N) - 2\pi b \xi (1-\beta)+2\pi b (1-2\beta)
      \sum_{j=1}^{N+1}M_j (N+2 - 2j) }\,.
\end{multline}

Thus, to get the invariance we need to have
\begin{multline}
  \label{eq:2}
  \frac{\mathcal{E}_{\mathcal{T}'}}{\mathcal{E}_{\mathcal{T}}} = \exp
  \Big[ -2 \pi i \xi M_{N+1} - \pi b \beta \sum_{k=1}^{N+1} (2k-N-1) M_k +\\+2\pi b (1-2\beta)
      \sum_{j=1}^{N+1}M_j (N+2 - 2j)
   + 2 \pi b (\beta (N+1) - 1 ) \xi - \frac{\pi i b^2 \beta^2}{6} N(N^2 - 1) \Big]\,.
\end{multline}
And indeed we obtained the  ratio of the contact terms as a
determined from the gauge theory partition function calculation in Eq.~\eqref{eq:35} in
the Appendix~\ref{sec:holom-blocks-calc}.

We have thus established the spectral duality for theories
$\mathcal{T}$ and $\mathcal{T}'$ using topological string
computation. It is remarkable that the field theory computation
matches the topological string not only qualitatively but with such a
\emph{quantitative} finesse.

\section*{Acknowledgements}
We are very grateful to Sergio Benvenuti, Noppadol Mekareeya, Simone Giacomelli and Matteo Sacchi for enlightening
discussions.  We are especially thankful to Alberto Zaffaroni for several illuminating comments during the various stages of this work. 
FA acknowledges Galileo Galilei Institute for
Theoretical Physics for hospitality during the workshop `Supersymmetric Quantum Field Theories in the Non-perturbative Regime'. 
FA, SP, and YZ are  supported by the ERC-STG grant 637844-HBQFTNCER and by the INFN.

\appendix

\section{More details on Nilpotent Higgsing}\label{Higgs}

We describe some additional details of the Higgsing process studied in Section \ref{nilpot_higg}.

Let us recall our notation for Theory $C$: Bifundamentals of type $S$ transform in the $(\Box,\overline{\Box})$ of $U(k)\times U(k+1)$. 
Bifundamentals of type $\widetilde{S}$ transform in the $(\Box,\overline{\Box})$ of $U(k+1)\times U(k)$. In matrix notation, the reps are
\be
\Box=\left[ \begin{array}{c} v_1 \\ \vdots \\ v_k \end{array} \right]    \qquad \overline{\Box}= \left[ \begin{array}{lll} v_1,& \ldots &, v_{k+1} \end{array}\right]
\ee
Covariant derivatives $DS^{(k,k+1)}$ and $D\widetilde{S}^{(k,k+1)}$, on a link $(k,k+1)$, with $U(k)$ connection on the left and $U(k+1)$ on the right, are defined as usual as
\bea
DS^{(k,k+1)} &=& d S - i\mathcal{A}_{k} S^{(k,k+1)} +  S^{(k,k+1)} i\mathcal{A}_{k+1},\\
D\widetilde{S}^{(k,k+1)} &=& d S +  \widetilde{S}^{(k,k+1)} i\mathcal{A}_{k} -  i\mathcal{A}_{k+1} \widetilde{S}^{(k,k+1)}.
\eea
The two objects $D_\mu S^{(k,k+1)}$ and $D_\mu \widetilde{S}^{(k,k+1)}$, are themselves bifundamentals. 
The covariant derivative for the adjoint scalars on a node $U(k)$ is
\be
D\Psi_k = d \Psi_k -i[\mathcal{A}_{k},\Psi_k].
\ee

\subsection*{Nilpotent vev and D-terms}
We discussed in the main text the role of D-terms in the solution of our nilpotent vev. 
Our notation for a D-term there was the following: For a gauge node $U(k)$, 
with bifundamentals on the left,  $L=(k-1,k)$, and on the right, $R=(k,k+1)$, 
we have
\bea
D^a&=& D^a\Big|_{hyper}  + \Tr_k \big( T^a [\Psi_k^\dagger,\Psi_k]\big)\\
D^a\Big|_{hyper} &=& \Tr_{k}\left(  T^a\left( S^R S^{R\dagger} -\widetilde{S}^{R\dagger}\widetilde{S}^R + \widetilde{S}^L \widetilde{S}^{L\dagger} - S^{L\dagger} S^L \right) \right)
\eea
Then, it is straightforward to compute on the nilpotent vev \eqref{nilp_vev_tail} the following matrix products 
\bea
 S^{(k-1,k)\dagger}S^{(k-1,k)}&=&{\rm diag}(0^{2}, 1^{k-2}),\\
 S^{(k-1,k)}S^{(k-1,k)\dagger}&=&{\rm diag}(0^{1}, 1^{k-2})= \widetilde{S}^{(k,k-1)\dagger}\widetilde{S}^{(k,k-1)}\\
\widetilde{S}^{(k-1,k)}\widetilde{S}^{(k-1,k)\dagger}&=&{\rm diag}(0,1^{k-2},0)
\eea

\subsection*{Gauge Multiplets Mass Matrix}
Given the Lagrangian of the theory, the mass matrix for spin-$1$ fields 
can be obtained from the covariant derivatives of the charged fields.  
We expect that the bifundamentals $S$ and $\widetilde{S}$, whose kinetic term is 
$$
\Tr_{k+1}\left[ (D_\mu S)^\dagger (D_\mu S)  + (D_\mu \widetilde{S}) (D_\mu \widetilde{S})^\dagger \right]
$$
will be responsible for mass terms between different gauge nodes. 
We quote the form of the mass matrix coming from the bifundamentals $S^{(k,k+1)}$, since it is instructive:
\be\label{mass1_S}
\sum_{k=1}^{N}
\left[\begin{array}{ll}  \scriptstyle \mathcal{A}^a_k,& \scriptstyle \mathcal{A}^a_{k+1} \end{array}\right] 
\Tr\otimes \left[ \begin{array}{lcl} 			\scriptstyle T^a_k S^{(k,k+1)} S^{(k,k+1)\dagger} T_k^b      & \  &	 \scriptstyle S^{(k,k+1)\dagger} T_k^a S^{(k,k+1) } T^b_{k+1} \\
							\scriptstyle  T^b_k S^{(k,k+1)} T^a_{k+1}S^{(k,k+1)\dagger} &   &    \scriptstyle T^a_{k+1} S^{(k,k+1)\dagger} S^{(k,k+1)} T_{k+1}^b  \end{array}\right] 
							\left[\begin{array}{l} \scriptstyle \mathcal{A}^b_k \\ \scriptstyle \mathcal{A}^b_{k+1} \end{array}\right]
\ee
In this formula $\mathcal{A}_{N+1}=0$ since the last node is a flavor node, i.e. it  is ungauged. Matrix elements are understood on the nilpotent vev. 
The contribution of $\widetilde{S}$-type bifundamentals is similar to \eqref{mass1_S}. 
Then, if we split the total mass matrix into the contributions of $S$, $\widetilde{S}$, and $\Phi$,
adjoint fields will not couple different gauge nodes. 

The total mass matrix has the following block structure,
\be\label{mass_sp1_appA}
\left(\begin{array}{c|c|c|c|c|c|c}
U(1)  & 0		 & 0 			 & 0  & 0 & 0 & \ldots  \\
\hline
0 	& U(2) 	&  \times &  0 & 0  & 0	  & \ldots \\
\hline
0     &  \times  &  U(3) & \times & 0 & 0  & \ldots \\
\hline
0 	& 0 		& \times &  U(4) & \times & 0 & \ldots  \\
\hline
0 	& 0 		&   0 	       & \times & \ldots   & \ldots &\ldots
\end{array}\right)
\ee 
with non zero crossed blocks. This structure resemble indeed that of the quiver: 
all but the first gauge node get two contributions, one from bifundamentals on the left, one from the right. 

After careful evaluation of \eqref{mass_sp1_appA} we were able to double-check the solution quoted in \eqref{U1quivergaugeF}. 
This same solution can then be understood in a simpler way by thinking about the action of broken gauge generators, 
along the lines of what we stated in Section \ref{nilpot_higg}.

\subsection*{A basis for massive chiral fields}

When discussing Theory C we described, within the set of UV fields,  an explicit basis for the massless fields on the nilpotent vev.  
This basis contained two subspaces: physical IR massless fields and goldstone bosons. 
In the physical sector we then had a further splitting: bifundamentals, and adjoints. This splitting is orthogonal by default. 
However, physical massless fields are not orthogonal to goldstone bosons. 
(This is OK, since both are in the kernel of the matrix, and it might happen that is just  
convenient, but not needed, that physical massless are taken to be orthogonal among themselves). 

In order to obtain a basis for massive chiral fields we can adopt the following strategy. 
\begin{itemize}
\item  We split the set of UV fields, call them $\mathcal{B}$, into the set of physical IR fields and its orthogonal, 
hereafter denoted by $\mathcal{K}^{\perp}$. (This is not $ker^\perp$). 
The only non trivial construction in $\mathcal{K}^{\perp}$ regards the adjoints, 
since as we mentioned, bifundamentals and adjoints are orthogonal by default. In practise we construct 
\be
\mathcal{B}=\{ v_1\ldots ,v_{\#_{ir} }\}\cup\{  v_{\#_{ir+1}},\ldots v_{\#_{uv} }\}
\ee
where the first set contains only physical massless fields in the IR. We check that 
$\{ v_1\ldots ,v_{\#_{ir} }\}\cup \{ {\rm \, goldstone\, bosons\,} \}$ is a set of independent fields.\footnote{
For example we show that there is no non trivial solution to $\sum_{i=1}^{\#_{ir}} x_iv_i+ \sum_{k=2}^N G_k=0$.}
Given $\mathcal{K}^{\perp}=\{  v_{\#_{ir+1}},\ldots v_{\#_{uv} }\}$, then we know that $\{{\rm goldostone\, bosons} \}\subset \mathcal{K}^\perp$. 
We do not impose orthogonality among the vectors in $\{  v_{\#_{ir+1}},\ldots v_{\#_{uv} }\}$. 
\item  For each goldstone boson, call it $G_k$, we impose the orthogonality condition 
\be
G_k v=0 \qquad v=\sum_{j\in \mathcal{K}^{\perp} } \alpha_j v_j
\ee
These linear equations fix a number of parameters equal to the number of goldstone bosons. 
The resulting free parameters provide a span for the massive fields, i.e. the actual $ker^\perp$. 
Vectors in this basis are not orthogonal among themselves, 
but they are automatically orthogonal to physical massless fields which is what we were looking for. 
Concluding we have splitted $\mathcal{B}$ in the form
\be
\mathcal{B}=\underbrace{\mathcal{K}\oplus G}_{ker} \oplus\, {ker^\perp}
\ee
\end{itemize}
Let us come back on the first part of this construction, i.e. a convenient basis for adjoint fields.
Note indeed that massless IR fields in the adjoint are not directly aligned with a basis of hermitian matrices for $U(N)$, 
so it is better to use an alternative basis. 
Consider the map $\iota_k: \mathbb{R}^{d\times d}\rightarrow \mathbb{R}^k$ with $k\leq d$ defined as
\be
\iota_k\cdot\left[ \begin{array}{ ccccc } a_{1,1} & a_{1,2} & \ldots & \ldots & a_{1,d} \\
						    a_{2,1} & a_{2,2}& a_{2,3} & \dots  & a_{2,d} \\
						    \vdots &\vdots &\vdots &\vdots & \vdots \end{array}\right] = \left[\begin{array}{c} a_{1,d-k+1} \\ a_{2,d-k+2} \\ \vdots \end{array}\right]
\ee
For example, if $k=d$, the map $\iota_d$ returns the diagonal of the matrix. 
We can find a basis for $\mathbb{R}^{d\times d}$ by considering for each $k\leq d$ an orthogonal basis of $\mathbb{R}^k$ of the form, 
\be
\left[\begin{array}{c} 1 \\ 1 \\ \vdots \\ 1 \end{array}\right]  , \left[ \begin{array}{c} +1 \\ -1 \\ 0 \\ \vdots\end{array} \right] ,\ldots 
\ee
Then, for each node $U(n)$ we construct a basis ${\Phi}_n$ of the adjoint rep recursively. 
Define ${\Phi}_{n-1}$ to be the basis of $U(n-1)$ built out of $\iota_k$ for $k\le n-1$. 
We can embed ${\Phi}_{n-1}$ in $\Phi_n$ in two ways
\be\label{embdPhinnplus1}
{\Phi}_n\supset \left( \begin{array}{cc} 0 & 0 \\ 0 & {\Phi}_{n-1} \end{array}\right) \qquad{\rm or}\qquad  
{\Phi}_n\supset \left( \begin{array}{cc} {\Phi}_{n-1} & 0 \\ 0 & 0 \end{array}\right)
\ee
The embedding on the right of \eqref{embdPhinnplus1} will be needed for the $U(N+1)$ flavor node. 
The other one is used on the gauge nodes of the tail. 
In order to find a complete orthogonal basis we only need elements 
parametrizing the remaining row and a column of $\Phi_n$. 
Finally we normalize. 
The basis $\oplus_{n=1}^{N+1}{\Phi}_n$ parametrize the 2$N$+3 fields 
$\{\Gamma_i,\psi_k, \Sigma_{\pm},\delta\}$, in a natural way. 
A basis orthogonal to these $2N$+3 fields is also simple to construct.

\subsection*{More general nilpotent deformations}
The nilpotent vev we studied, together with the Higgsing, can be generalized outside next-to-extremality. For example,
let us label the F-term deformation generated by the monopoles using a partition, 
i.e. the following set of integers: $\mathcal{I}=\{ n_1,\ldots , n_{N}\}$ with $n_i\ge 0$ and $\sum_{l=1}^{N} l n_l = N$. 
In \eqref{start_rec} we considered $\mathcal{I}=(0^{N-1},1)$, which naturally generalize to,
\be\label{start_rec_gen}
 \Tr_{N} \mathbb{S}^{(N,N+1)}  = \bigoplus_{l=1}^{N} \mathbb{J}^{\oplus n_l}_l \oplus \mathbb{J}_1
\ee
This equation is solved block by block in the same way as in \eqref{vevstep1}. Then
\bea
\begin{array}{cll}
\langle \widetilde{S}^{(N,N+1)}\rangle&= \bigoplus_{l=1}^{N} \mathbb{J}_l^{\oplus n_l}\oplus \mathbb{J}_1 &\qquad { \rm drop\ the\ last\ column}.\\[.2cm]
\langle{S}^{(N,N+1)}\rangle&= \bigoplus_{l=1}^{N} (\mathbb{J}_1\oplus\mathbb{I}_{l-1})^{\oplus n_l}\oplus \mathbb{J}_1 &\qquad { \rm drop\ the\ last\ row}.
\end{array}
\eea
and
\bea
\begin{array}{cll}
\langle \widetilde{S}^{(N-k,N+1-k)}\rangle&= \bigoplus_{l=1}^{N} (\mathbb{J}_1^{\oplus k}\oplus\mathbb{J}_{l-k})^{\oplus n_l}  & {\rm\qquad drop\ the\ first\ k\ column\ and\ k-1\ rows } \\[.2cm]
\langle {S}^{(N-k,N+1-k)}\rangle&= \bigoplus_{l=1}^N(\mathbb{J}_1^{\oplus k}\oplus \mathbb{J}_1\oplus\mathbb{I}_{k-2})^{\oplus n_l}   & {\rm\qquad drop\ the\ first\ k\ row\ and\ k-1\ columns} 
\end{array}\nn\\
\eea

\section{Bookkeeping Integrals}\label{App_Integrals}

In this Appendix we collect some useful results about hyperbolic integrals.

\subsection*{Double-sine function}
The double-sine function, $s_b$, appeared in the very first computation of \cite{Hama:2011ea}, 
as a building block for the localized partition function of $3d$  $\mathcal{N}=2$ theories on the squashed sphere ${S}^3_b$. 
It can be introduced with an infinite product representation, which is perhaps familiar to the physics literature,
\be
s_b= \prod_{m,n\ge 0} \frac{ mb+n/b+Q/2-ix}{mb+n/b+Q/2+ix},\qquad Q=b+b^{-1}.
\ee
It satisfies the following non trivial properties
\bea
\label{factosb}
s_b(x)s_b(-x)&=&1\\
s_b(\tfrac{ib}{2}-x)s_b(\tfrac{ib}{2}+x)&=&\frac{1}{2\cosh(\pi bx)}\\[.2cm]
s_b(x)&=&e^{+\tfrac{i\pi}{2} B_{22}[Q/2-ix] } ( e^{2\pi i b (Q/2-i x)},q)_{\infty} ( e^{2\pi i/b (Q/2-ix)},\tilde{q} )_{\infty} \\
&=&e^{-\tfrac{i\pi}{2} B_{22}[Q/2+ix] } ( e^{2\pi i b (Q/2+i x)},q)^{-1}_{\infty} ( e^{2\pi i/b (Q/2+ix)},\tilde{q} )^{-1}_{\infty}
\eea
where $q\equiv e^{2i\pi b^2}=e^{2i\pi b Q}$, $\tilde{q}\equiv e^{\frac{ 2i\pi}{ b^2}}=e^{2i\pi/ b Q}$. The Bernulli numbers relevant to the factorization formulas are 
\be
B_{22}(x)=(x-Q/2)^2-(b^2+b^{-2})/12.
\ee 
A slightly more compact notation distinguishes $s_b$ with argument $x\pm{iQ}/2$, i.e. 
\bea
\overline{s_b}\equiv s_b(x+\tfrac{iQ}{2})^{+1}, \qquad \underline{s_b}\equiv s_b(x-\tfrac{iQ}{2})^{-1}.
\eea
Then, we find
\bea
&&
\frac{ \overline{s_b}(x+\tfrac{mi}{b}+nib)}{ \overline{s_b}(x) }= 
\frac{  (-)^{nm+n+m}   q^{\frac{n(n+1)}{4} } \tilde{q}^{\frac{m(m+1)}{4} }   e^{+\pi n b x + \frac{ \pi mx}{b}  }    }{ 
 (qe^{ 2\pi b x },q)_{n}  (\tilde{q} e^{ \frac{ 2\pi x}{b}},\tilde{q})_{m}  }  \\[.2cm]
 &&
 \frac{  \underline{s_b}(x+inb+\tfrac{im}{b}) }{  \underline{s_b}(x)  } = 
 \frac{ (e^{2\pi b x},q)_n  (e^{\frac{2\pi x}{ b} },\tilde{q})_m }{  
 (-)^{nm+m+n}   q^{\frac{n(n+1)}{4} } \tilde{q}^{\frac{m(m+1)}{4} } e^{\pi b(x-iQ) n+\frac{\pi m (x-iQ)}{b}}  } 
\eea

\subsection*{Abelian integrals}
In section \ref{sec_TS2_full} we studied in details the commutative diagram for $T[SU(2)]$. The computations involving $T[SU(2)]$ reduce to abelian integrals of the form
\be
\int dx\, z^{x} \, \prod_{i=1}^{N_f} \frac{ s_b(x+M_i+u_i+\tfrac{iQ}{2} ) }{s_b(x+M_i-u_i-\tfrac{iQ}{2} )  }
\ee
where $z=e^{i\pi\lambda}$ and $u_i$ are arbitrary. 
For example, if we take $u_i=-\tfrac{iQ}{4}-\tfrac{m_A}{2}$, we find $Z_{\rm bif}$ as defined in the main text with a minus sign for the masses.

In order to compute such a generic abelian integral, we pick poles from the two $s_b$ functions at the numerators: 
Let us focus first on the computation involving the first set of poles, i.e $X_1=x+M_1+u_1=i nb +im\big/b$, belonging to the contour $\mathcal{C}_1$, since 
the computation on the other contours $\mathcal{C}_{i=2,..,N_f}$ will be very similar.  
When $X_1=x+M_1+u_1=i nb +im\big/b$ we find a series made out of
\bea
&&
\sum_{n,m\ge 0}z^{X_1-M_1-u_1} {\rm Res}[ \overline{s_b}(X_1)]\prod_{j\neq 1} \overline{s_b}(D_{j1}+X_1) \prod_j \underline{s_b}(C_{j1}+X_1)\qquad
\eea
which upon evaluation gives
\bea
&&
z^{-M_1-u_1} \prod_{j}\left[ {s_b}[D_{j1}+\tfrac{iQ}{2}]/{s_b}[C_{j1}-\tfrac{iQ}{2} ] \right] \times\nonumber\\
&&
\rule{2.5cm}{0pt}
\sum_{n\ge 0} 
\left[ \frac{ (e^{2\pi b C_{11} })_n  (e^{2\pi b C_{21}})_n }{ (q)_n (q e^{2\pi b D_{21}})_n }\right] \left[ z^{ib} e^{2\pi b \sum_{j} ( u_j+iQ/2) } \right]^n\times\nonumber\\
&&
\rule{2.5cm}{0pt}
\sum_{m\ge 0} 
 \left[ \frac{ (e^{ {2\pi}C_{11}/b })_m  (e^{{2\pi} C_{21}/b})_m }{ (\widetilde{q}\,)_m (\widetilde{q} e^{ {2\pi} D_{21}/b})_m }\right] \left[ z^{i/b} e^{{2\pi}/{b}\, \sum_{j} ( u_j+iQ/2) } \right]^m\nonumber\\
\eea
We defined the quantities,
\be
C_{ij}=(M_i-M_j)-(u_i+u_j)\qquad D_{ij}=(m_i-m_j)+(u_i-u_j)\ .
\ee
Here $u_i=-\mu_i$ if we want to compare with $\widetilde{Z}$:
\bea
 \widetilde{Z}^{(N_c,N_f)}_{\rm }[\vec{M},\vec{\mu}\,]&=&\prod_{i=1}^{N_c}\prod_{j=1}^{N_f} s_b\left(\tfrac{iQ}{2}-\mu_j\pm(x^{(N_c)}_i-M_j)\right)\,
\eea

\subsection*{BC tranformations and its real mass deformations}
The integral identities \eqref{mofor} and \eqref{aharec} have been derived in \cite{Benini:2017dud} 
starting from the master relation between multivariate integrals with $BC$ symmetry.
The transformation between $BC_n$ and $BC_m$ hyperbolic integrals has been proved by E.Rains in Corollary 4.2 of \cite{Rains:2006dfy}. 
For convenience, we repeat here below the main statement.  Consider the integral 
\bea\label{RainsI}
&&
\mathcal{I}^{(m)}_n[ \vec{\mu}; \omega_1,\omega_2]=\\
&&\nonumber
\frac{1}{(-4\omega_1\omega_2)^{n/2}n!} \int_{\mathcal{C}^n} \prod_{i=1}^n dx_i \frac{ \prod_{i=1}^n \prod_{r=1}^{2m+2n+4} \Gamma_h(\mu_r\pm x_i;\omega_1,\omega_2) }{ \prod_{i<j} \Gamma_h(\pm x_i\pm x_j;\omega_1,\omega_2) \prod_{i=1}^n \Gamma_h(\pm 2x_i,\omega_1,\omega_2) }
\eea
where the contour $\mathcal{C}$ can be closed on the `positive' poles of the form $\mu_r+i\omega_1+i\omega_2$, excluding the `negative' poles, or viceversa. Equivalently $\mathcal{C}$ is a Barnes contour which agrees with $\mathbb{R}$.
Then, 
\be\label{Rains}
\mathcal{I}^{(m)}_n[ \vec{\mu}; \omega_1,\omega_2]=\left( \prod_{s>r\geq 1}^{2m+2n+4} \Gamma_h[\mu_r+\mu_s;\omega_1,\omega_2]\right) \mathcal{I}^{(n)}_m[\tfrac{\omega_1+\omega_2}{2}-\vec{\mu};\omega_1,\omega_2]
\ee
with the constraint
\be
\sum_r \mu_r=(m+1)(\omega_1+\omega_2)\ .
\ee
%
Notice that \eqref{Rains} provides an evaluation formula when $m=0$. Gauge theory parameters $N_c$ and $N_f$ 
enter with the following dictionary: $N_c=n$ and $N_f=(m+n+2)$, thus $m=N_f-N_c-2$. 
The background parameter $b$, which measures the squashing of the three-sphere, 
enters through $\omega_1=ib$ and $\omega_2=i/b$, thus $\omega_1+\omega_2=iQ$. 
Finally $\Gamma_h(x)=s_b(\tfrac{iQ}{2}-x)$. 

\subsection*{Summary}
In the notation of \cite{Benini:2017dud}, the equality $\mathcal{Z}_{\mathcal{T}_{\mathfrak{M}}}=\mathcal{Z}_{\mathcal{T}_{\mathfrak{M}'}}$ is obtained from \eqref{Rains} by taking the limit
\be\label{TM_sara}
\mu_{i}=m_i+s,\qquad \mu_{\,i+N_f}=\widetilde{m}_i-s,\qquad s\rightarrow \infty,\qquad 1\leq i\leq N_f
\ee
Then \cite{Benini:2017dud} find other two results: 
\begin{itemize}
\item[$\star$] Derive the monopole duality [Section~$8$ of \cite{Benini:2017dud}], which we used in this paper,
\be
\, U(N_c)\oplus\,N_{flav.}\ {\rm and}\ W=\Mon^{+} \ \leftrightarrow\ N_f^2\oplus1\ {\rm singlets}\ M_{ij}\oplus\gamma\ {\rm and}\ W=\gamma\det M\qquad\nn
\ee
\item[$\bullet$] Recover Aharony duality,
\end{itemize}
The corresponding integral identities can be deduced from \eqref{Rains} as follows,
\bea
\label{monopole_sara_appendix}
\star&\qquad& t\rightarrow \infty,\qquad m_{N_f+1}=\tfrac{\zeta}{2}+ t,\qquad \widetilde{m}_{N_f+1}=\tfrac{\zeta}{2}- t,\rule{3cm}{0pt}\\
& & {\rm constraint\ becomes\ }\ \sum_{a=1}^{N_f}(m_a+\widetilde{m}_a) +\zeta = iQ(N_f-N_c)\\[.5cm]
\label{aharony_sara_appendix}
\bullet&\qquad& t\rightarrow\infty,\qquad
\begin{array}{ll}
m_{N_f+2}=\tfrac{\zeta-\lambda}{4}-t,\qquad&  \widetilde{m}_{N_f+2}=\tfrac{\zeta-\lambda}{4}+t,\\[.2cm]
m_{N_f+1}=\tfrac{\zeta+\lambda}{4}+t,\qquad&  \widetilde{m}_{N_f+1}=\tfrac{\zeta+\lambda}{4}-t,
\end{array}\\
& & {\rm constraint\ becomes\ }\ \sum_{a=1}^{N_f}(m_a+\widetilde{m}_a) +\zeta = iQ(N_f-N_c+1)
\eea
%
In the next paragraphs we present some important details on contact terms involved in these computation.

\subsection*{Details on \eqref{TM_sara}}
Starting from \eqref{Rains}, it is trivial to substitute the $m,n$ dependence with gauge theory parameters $N_c$ and $N_f$. 
The strategy of \cite{Benini:2017dud} is to rewrite the integrals as
\be
\int_{-\infty}^{\infty} d\sigma f(\sigma)=2\int_{-s}^{\infty} dx f(x+s)
\ee
and take the limit \eqref{TM_sara} on $s$ with the asymptotics expansion of the $s_b$, 
\be\label{sara_expa_asympt}
\lim_{x\rightarrow \pm \infty} s_b(x)= e^{\pm i \pi x^2/2}\ .
\ee
In terms of $N_f$, the vector $\vec{\mu}$ has $2N_f$ component, and $s$ enter with different signs, specified in  \eqref{TM_sara}. Since the integration variables are also shifted by $+s$,
it will happen that out of the combinations, $\mu_r\pm x_i$ and $\pm x_i \pm x_j$, which appear in \eqref{RainsI},
some are invariant and the others are shifted twice. For example
\bea
\mu_{i+N_f}+x_j\rightarrow \tilde{m}_{i}+x_j \qquad \mu_{i}-x_j\rightarrow m_i-x_j
\eea
Instead $\pm 2x_i\rightarrow \pm 2x_i \pm 2s$. 

Taking into account these shifts, and the asymptotics expansion \eqref{sara_expa_asympt}, 
we find two (different) prefactor in \eqref{Rains}, one for the l.h.s and one for r.h.s. These two prefactors depend on $m_i,\tilde{m}_j$, the integration variables, and $s$.
In particular there is a divergent part. However, upon imposing the constraint, the dependence on the integration variables drops, and the simplified prefactors cancels each other from r.h.s. to l.h.s. 
It follows that 
\bea
\mathcal{Z}_{\mathcal{T}_{\mathfrak{M}}}=\mathcal{Z}_{\mathcal{T}_{\mathfrak{M}'}}
\eea
where
\bea
\mathcal{Z}_{\mathcal{T}_{\mathfrak{M} } } = \frac{1}{N_c!} \int \prod_{i=1}^{N_c} dx_i \ \frac{  \prod_{j=1}^{N_c} \prod_{a=1}^{N_f} s_b( \tfrac{iQ}{2} +x_j-m_a) s_b( \tfrac{iQ}{2} -x_j-\tilde{m}_b ) }{  \prod_{i<j}^{N_c} s_b(\tfrac{iQ}{2} \pm (x_i-x_j) ) }
\eea
and 
\bea
&&
\mathcal{Z}_{\mathcal{T}_{\mathfrak{M}' } } =
\frac{1}{(N_f-N_c-2)!} \prod_{a,b=1}^{N_f} s_b(\tfrac{iQ}{2} -(m_a+\tilde{m}_b))\times \rule{3.5cm}{0pt}\nn\\
&&
\rule{2cm}{0pt}
\int \prod_{i=1}^{N_f-N_c-2} dx_i \ \frac{  \prod_{j=1}^{N_f-N_c-2} \prod_{a=1}^{N_f} s_b( x_j+m_a) s_b( -x_j+\tilde{m}_b ) }{  \prod_{i<j}^{N_f-N_c-2} s_b(\tfrac{iQ}{2} \pm (x_i-x_j) ) }
\eea

\subsection*{Details on \eqref{monopole_sara_appendix}}
On the electric side, i.e. $\mathcal{Z}_{\mathcal{T}_{\mathfrak{M} } }$ with $N_f+1$ flavors, 
we consider the following manipulations on the integrand: split the product over $j=1,\ldots N_f+1$ into $1\leq j\leq N_f$ and the last one, and take the limit \eqref{monopole_sara_appendix}, 
\bea
s_b\left( \tfrac{iQ}{2} +x_i -m_{N_f+1}\right) s_b\left( \tfrac{iQ}{2} -x_i -\widetilde{m}_{N_f+1}\right) \prod_{j=1}^{N_f} s_b\left(\tfrac{iQ}{2} + x_i-m_j\right) s_b\left(\tfrac{iQ}{2} - x_i-\widetilde{m}_j\right)\nn\\
\rightarrow e^{i\pi  t(iQ-\zeta) } e^{i\pi (\zeta-iQ) x_i}  \prod_{j=1}^{N_f} s_b\left(\tfrac{iQ}{2} + x_i-m_j\right) s_b\left(\tfrac{iQ}{2} - x_i-\widetilde{m}_j\right)\nonumber
\eea
The total prefactor will be $\prod_{i=1}^{N_c} e^{i\pi  t(iQ-\zeta) } e^{i\pi (\zeta-iQ) x_i}$. 
On the magnetic side, i.e. $\mathcal{Z}_{\mathcal{T}_{\mathfrak{M}' } }$, the same kind of manipulations lead to
\bea
\prod_{i=1}^{N_f-N_c-1}s_b\left(m_{N_f+1}+x_i\right) s_b\left( \widetilde{m}_{N_f+1}-x_i\right) \prod_{j=1}^{N_f} s_b\left(m_j+x_i\right) s_b\left(\widetilde{m}_j-x_i\right)\nonumber\\
\prod_{i=1}^{N_f-N_c-1} e^{i\pi  \zeta (t+x_i)}  \prod_{j=1}^{N_f} s_b\left(m_j+x_i\right) s_b\left(\widetilde{m}_j-x_i\right)
\eea
Additionally, on the magnetic side the prefactors produce extra terms,
\bea
\frac{ \prod_{a,b=1}^{N_f+1}s_b\left(\tfrac{iQ}{2} -m_a-\widetilde{m}_b\right)}{  \prod_{a,b=1}^{N_f}s_b\left(\tfrac{iQ}{2} -m_a-\widetilde{m}_b\right) }\rightarrow 
s_b(\tfrac{iQ}{2}-\eta)\prod_{a=1}^{N_f} e^{\tfrac{i \pi}{2}(m_a-\widetilde{m}_a)(m_a+\widetilde{m}_a+\zeta-iQ) }e^{i\pi t(iQ-\zeta-m_a-\widetilde{m}_a)}\nonumber\\
\eea
Comparing the divergences on the electric and magnetic side, we find that they are equal and cancel out due to the constraint, i.e. 
\be
e^{i\pi N_c t(iQ-\zeta) } = e^{i\pi t\sum_{a} (iQ-\zeta-m_a-\widetilde{m}_a)} e^{i\pi  (N_f-N_c-1) t \zeta }
\ee
We can finally re-introduce the notation 
\be
m_i=\mu_i-M_i\qquad \widetilde{m}_i=\mu_i+M_i
\ee
and get the relation used in the main text
\bea
&&
\int dx^{(N_c)} e^{+i\pi  (\zeta-iQ) \sum_{a=1}^{N_c} x_a }Z_{\rm vec}^{(N_c)}  \widetilde{Z}^{(N_c,N_f)}[-\vec{M},\vec{\mu}]= \nonumber\\
&&
\rule{.5cm}{0pt}
e^{-i \pi\sum_{a=1}^{N_f} M_a(2\mu_a +\zeta-iQ) } 
 s_b(\tfrac{iQ}{2}-\eta)  \prod_{a,b=1}^{N_f}s_b\left(\tfrac{iQ}{2} -\mu_a-{\mu}_b+M_a-M_b\right) \times \nonumber\\
&&
\rule{1cm}{0pt} 
 \int dx^{(N_f-N_c-1)} e^{+i\pi \zeta\sum_{a=1}^{N_f} x_a } Z_{\rm vec}^{(N_f-N_c-1)}  \widetilde{Z}^{(N_f-N_c-1,N_f)}[\vec{M},\tfrac{iQ}{2}-\vec{\mu}] \qquad
\eea
where
\bea
 \widetilde{Z}^{(N_c,N_f)}_{\rm }[\vec{M},\vec{\mu}\,]&=&\prod_{i=1}^{N_c}\prod_{j=1}^{N_f} s_b\left(\tfrac{iQ}{2}-\mu_j\pm(x^{(N_c)}_i-M_j)\right)\,
\eea
\subsection*{Details on \eqref{aharony_sara_appendix}}
On the electric side, i.e. $\mathcal{Z}_{\mathcal{T}_{\mathfrak{M} } }$ with $N_f+2$ flavors, we single out the two extra flavors and take the limit, thus producing
\bea
\prod_{a=1}^2
s_b\left( \tfrac{iQ}{2} +x_i -m_{N_f+a}\right) s_b\left( \tfrac{iQ}{2} -x_i -\widetilde{m}_{N_f+a}\right)
\prod_{j=1}^{N_f} s_b\left(\tfrac{iQ}{2} + x_i-m_j\right) s_b\left(\tfrac{iQ}{2} - x_i-\widetilde{m}_j\right)\nn\\
\rightarrow e^{i\pi  t(2iQ-\zeta) } e^{i\pi \lambda x_i}  \prod_{j=1}^{N_f} s_b\left(\tfrac{iQ}{2} + x_i-m_j\right) s_b\left(\tfrac{iQ}{2} - x_i-\widetilde{m}_j\right)\nonumber\\
\eea
Then, the total prefactor is $\prod_{i=1}^{N_c} e^{i\pi  t(2iQ-\zeta) } e^{i\pi \lambda x_i}$. 
Similarly on the magnetic side, 
\bea
\prod_{i=1}^{N_f-N_c}\prod_{a=1}^2 s_b\left(m_{N_f+a}+x_i\right) s_b\left( \widetilde{m}_{N_f+a}-x_i\right) \prod_{j=1}^{N_f} s_b\left(m_j+x_i\right) s_b\left(\widetilde{m}_j-x_i\right)\nonumber\\
\prod_{i=1}^{N_f-N_c} e^{i\pi (\zeta t+\lambda x_i)}  \prod_{j=1}^{N_f} s_b\left(m_j+x_i\right) s_b\left(\widetilde{m}_j-x_i\right)
\eea
and the extra prefactors,
\bea
\frac{ \prod_{a,b=1}^{N_f+2}s_b\left(\tfrac{iQ}{2} -m_a-\widetilde{m}_b\right)}{  \prod_{a,b=1}^{N_f}s_b\left(\tfrac{iQ}{2} -m_a-\widetilde{m}_b\right) }\rightarrow  e^{2i\pi t(  iQ -\zeta) } s_b\left( \tfrac{iQ-\zeta\pm\lambda}{2} \right) 
\prod_{a=1}^{N_f}e^{\frac{i\pi}{2} \lambda (m_a-\widetilde{m}_a)+ i\pi (2iQ-\zeta-2m_a-2\widetilde{m}_a)t}\nonumber\\
\eea
By using the constraint, the divercences cancel each other, i.e
\be
 e^{i\pi N_c t(2iQ-\zeta) } =e^{i\pi (N_f-N_c) \zeta t} e^{2i\pi t(  iQ -\zeta) } e^{i\pi \sum_a (2iQ-\zeta-2m_a-2\widetilde{m}_a)t}
 \ee
We can finally re-introduce the notation 
\be
m_i=\mu_i-M_i\qquad \widetilde{m}_i=\mu_i+M_i
\ee
The result is 
\bea
&&
\int dx^{(N_c)} e^{i\pi \lambda \sum_{a=1}^{} x_i^{(N_c)} } Z_{\rm vec}^{(N_c)} \widetilde{Z}^{(N_c,N_f)}[-\vec{M},\vec{\mu}]=\nonumber\\
&&
\rule{0.5cm}{0pt}
s_b\left( \tfrac{iQ-\zeta\pm\lambda}{2} \right) \prod_{a,b=1}^{N_f}s_b\left(\tfrac{iQ}{2} -\mu_a-\mu_b+M_a-M_b\right)\nonumber\\
&&
\rule{.8cm}{0pt}
e^{-i\pi \lambda \sum_{a=1}^{N_f } M_a } \int dx^{(N_f-N_c)} e^{i\pi \lambda \sum_{a=1} x_i^{(N_f-N_c)}} Z_{\rm vec}^{(N_f-N_c)} \widetilde{Z}^{(N_f-N_c,N_f)}[\vec{M},\tfrac{iQ}{2}-\vec{\mu}] \qquad\nn\\
\label{Aharony_offshell}
\eea

\section{Holomorphic Blocks calculations for $\mathcal{T}$ and $\mathcal{T}'$}
\label{sec:holom-blocks-calc}
In this section we evaluate the holomorphic blocks for $\mathcal{T}$ and $\mathcal{T}'$ over the reference
contours.

We first list the integrals obtained via factorisation of
the $S^3_b$ partion function, which is a consequence of the factorisation property of
the double sine (\ref{factosb}).

For theory $\mathcal{T}$ we have
\begin{equation}
\label{upst}
\Upsilon_{\mathcal{T}}=\mathcal{E}_{\mathcal{T}}\mathcal{F}_{\mathcal{T}}\left[ z^{(1)} \right]^{- 
  \frac{i \xi}{b} - \beta \frac{1+N}{2} } 
\frac{\left(t \frac{\mu_i}{z}\right)_\infty}{\left(\frac{\mu_i}{z}\right)_\infty}
\end{equation}
where 
\begin{equation}
\mathcal{F}_{\mathcal{T}}=\frac{(q)_{\infty} }{(t)_\infty^N} \prod_{i>j}^{N+1}
\frac{\left( \frac{q}{t}\frac{\mu_i}{\mu_j}\right)_\infty}{\left(t
    \frac{\mu_i}{\mu_j}\right)_\infty} \label{eq:31}
\end{equation}
is the contribution of the flipping fields and
$\mathcal{E}_{\mathcal{T}}$ is a contact term.
We have also introduced the exponentiated variables $\mu_i=e^{2\pi b  M_i}$ and $q=e^{2\pi i b^2}$, $t = q^{\beta}$.

On the dual side, for theory $\mathcal{T}'$ we have:
\begin{equation}
\label{upstp}
\Upsilon_{\mathcal{T}'}=\mathcal{E}_{\mathcal{T}'}\mathcal{F}_{\mathcal{T}'} 
 \prod_{a=1}^N \prod_{i=1}^a \left[
    z^{(a)}_i\right]^{-i \frac{M_a - M_{a+1}}{b}-\beta}
  B_{\mathrm{tail}} \prod_{i=1}^{N} \frac{\left( \frac{te^{2\pi
          b \xi}}{z^{(N)}_i}\right)_{\infty}}{\left( \frac{e^{2\pi b
          \xi}}{z^{(N)}_i}\right)_{\infty}}
  \frac{\left(\frac{t^{({1+N})/{2}} }{z^{(N)}_i} \right)_{\infty} }
  {\left(\frac{t^{{(1-N)}/{2}} }{z^{(N)}_i}\right)_{\infty}}
 \end{equation} 
with 
\begin{equation}
  B_{\textrm{tail}}=  \frac{{(q)_{\infty}^{\frac{N(N+1)}{2} }}}{ (t)_{\infty}^{\frac{N(N+1)}{2}} } \prod_{a=1}^{N} \prod_{i=1}^a\prod_{k=1\&k\neq i}^a\frac{ \left(\frac{z^{(a)}_k}{z^{(a)}_i}\right)_{\infty} }{  \left(t \frac{z^{(a)}_k}{z^{(a)}_i}\right)_{\infty} } 
  \prod_{a=1}^{N-1}
  \prod_{i=1}^a
  \prod_{j=1}^{a+1} \frac{
    \left(t \frac{z^{(a+1)}_j}{z^{(a)}_i}\right)_{\infty} }{  \left(\frac{z^{(a+1)}_j}{z^{(a)}_i}\right)_{\infty} }
\end{equation}  
 and
\begin{equation}
  \mathcal{F}_{\mathcal{T}'}= \frac{\left(q t^{-\frac{N+1}{2}}
        e^{-2\pi b \xi} ;q \right)_{\infty} } {\left(
        t^{\frac{1+N}{2}} e^{-2\pi b
          \xi}  ;q \right)_{\infty}
    }\prod_{l=1}^{N-1}\frac{1}{(t^{l+1})_\infty}\,.\label{eq:32}
\end{equation}

The factorization procedure gives rather complicated expressions for
the contact terms $\mathcal{E}_{\mathcal{T}}$ and
$\mathcal{E}_{\mathcal{T}'}$. However, all we need to check the
spectral duality is the \emph{ratio} of the contact
terms which is comparatively easy to write down:
\begin{equation}
\frac{\mathcal{E}_{\mathcal{T}'}}{\mathcal{E}_{\mathcal{T}}}= R\,
  e^{2 \pi b (  -i \xi/b M_{N+1}  +  (\beta (N+1) -1 ) \xi - \frac{1}{2} \beta \sum_{k=1}^{N+1} (2k-N-1) M_k + (2\beta-1)\sum_{k=1}^{N+1} (2k-N-2)M_k )}\label{eq:35}
\end{equation}
where $R=R(N,b,\beta)$ is a prefactor which we henceforth discard.

We would like to evaluate the block integrals, i.e.\ integrals of
$\Upsilon_{\mathcal{T}}$ and $\Upsilon_{\mathcal{T}'}$ on the
reference contours, which for both theories we denote by
$\Gamma_{\alpha_0}$.

Let's start with the theory $\mathcal{T}$. We focus on the part of
$\Upsilon_{\mathcal{T}}$ which does depend on the integration
variables and take as the reference contour $\Gamma_{\alpha_0}$ as the
contour $\mathcal{C}_{1,0,\ldots,0}$ encircling the poles at $z^{(1)}
= \mu_1 q^k$, $k \in \mathbb{Z}_{ \geq 1}$. We find:
\begin{eqnarray}
  \label{eq:38}
  {\mathcal I}^{\alpha_0}_{\mathcal{T}} (\vec{\mu}, \vec{\tau}, q, t) &=& 
  \oint_{\mathcal{C}_{1,0,\ldots,0}} \frac{dz^{(1)}}{z^{(1)}} \left( z^{(1)} \right)^{-\frac{i \xi}{b}
    -\beta \frac{N+1}{2} }   \prod_{i=1}^{N+1} \frac{\left(t \frac{\mu_i}{z^{(1)}} ;q \right)_{\infty}}{\left(\frac{\mu_i}{z^{(1)}} ;q \right)_{\infty}}=\nonumber\\
    &&=\mathcal{I}^{\alpha_0}_{0,\mathcal{T}}(\vec{\mu}, \vec{\tau},q,t)   Z_{\mathrm{vort},\mathcal{T}}^{\alpha_0}(\vec{\mu}, \vec{\tau}, q, t)\,,
\end{eqnarray}
where factored out the contribution of the  \emph{first} pole at $z^{(1)} =\mu_1$: 
\begin{multline}
  \label{eq:14}
  \mathcal{I}^{\alpha_0}_{0,\mathcal{T}}(\vec{\mu},
  \xi,q,t) = \Res_{z^{(1)} = \mu_1} \left( z^{(1)} \right)^{-\frac{i
      \xi}{b} -\beta \frac{N+1}{2} -1} \prod_{i=1}^{N+1} \frac{\left(t
      \frac{\mu_i}{z^{(1)}} ;q
    \right)_{\infty}}{\left(\frac{\mu_i}{z^{(1)}} ;q
    \right)_{\infty}} =\\
  =\frac{(t;q)_{\infty}}{(q;q)_{\infty}} e^{-2\pi i b  M_1 \xi - \pi b \beta (N+1)  M_1}
   \prod_{i=2}^{N+1}
  \frac{\left(t \frac{\mu_i}{\mu_1} ;q
    \right)_{\infty}}{\left(\frac{\mu_i}{\mu_1} ;q \right)_{\infty}}\,,
\end{multline}
and the vortex series
\begin{equation}
  \label{eq:39}
  Z^{\alpha_0}_{
    \mathrm{vort}, \mathcal{T}}(\vec{\mu}, \xi, q,t) = \sum_{k
    \geq 0} \left(t^{\frac{N+1}{2}}  e^{2\pi b\xi}\right)^k \prod_{i=1}^{N+1} \frac{\left( \frac{q \mu_1}{t \mu_i};q
    \right)_k}{\left( q \frac{\mu_1}{\mu_i};q \right)_k}.
\end{equation}

Taking into account also the contribution of the flipping fields $\mathcal{F}_{\mathcal{T}}$ we have:
\be
\label{colt}
  Z^{\alpha_0}_{\mathrm{cl}, \mathcal{T}}Z^{\alpha_0}_{1 \mathrm{-loop}, \mathcal{T}}\equiv
\mathcal{F}_{\mathcal{T}} \mathcal{I}^{\alpha_0}_{0, \mathcal{T}}(\vec{\mu}, \xi, q,t)\,.
\ee

For the dual theory we argue that the reference contour
$\Gamma_{\alpha_0}$ is described iteratively as a sequence of contours
$\{ \mathcal{C}_{1,0}, \mathcal{C}_{1,1}, \mathcal{C}_{1,2}, \ldots,
\mathcal{C}_{1,N-1} \}$:
\begin{multline}
  \label{eq:16}
{\mathcal I}^{\alpha_0}_{\mathcal{T'}}=
  \oint_{\mathcal{C}_{1,0}} dz^{(1)}_1 \oint_{\mathcal{C}_{1,1}}
  d^2z^{(2)} \oint_{\mathcal{C}_{1,2}} d^3z^{(3)} \cdots
  \oint_{\mathcal{C}_{1,N-1}} d^Nz^{(N)} \prod_{a=1}^N
  \prod_{i=1}^a\left( z^{(a)}_i \right)^{-i\frac{M_a -
      M_{a+1}}{b} - \beta - 1} \times \\
  \times \left\{ \prod_{a=2}^N \prod_{i\neq j}^a \frac{\left(
        \frac{z^{(a)}_i}{z^{(a)}_j};q \right)_{\infty}}{\left( t
        \frac{z^{(a)}_i}{z^{(a)}_j};q \right)_{\infty}} \right\}
  \left\{ \prod_{a=1}^{N-1} \prod_{i=1}^a \prod_{j=1}^{a+1}
    \frac{\left( t \frac{z^{(a+1)}_j}{z^{(a)}_i};q
      \right)_{\infty}}{\left( \frac{z^{(a+1)}_j}{z^{(a)}_i};q
      \right)_{\infty}}\right\} \left\{ \prod_{i=1}^N \frac{\left(
        t \frac{e^{2\pi b \xi}}{z^{(N)}_i};q \right)_{\infty}}{\left(
        \frac{e^{2\pi b \xi}}{z^{(N)}_i};q \right)_{\infty}} \frac{\left(
        \frac{t^{\frac{1+N}{2}}}{z^{(N)}_i};q \right)_{\infty}}{\left(
        \frac{t^{\frac{1-N}{2}}}{z^{(N)}_i};q \right)_{\infty}} \right\}.
\end{multline}
The pole structure is complicated: at each ``level'' $a$ of
integration ($a$ running from $1$ to $N$) corresponding to the gauge
group $U(a)$ the poles split into two groups encoded by to two Young
diagrams $^{(a)}Y^{(1)}$ and $^{(a)}Y^{(2)}$. The first group consists
of one variable $z^{(a)}_1 = e^{2\pi b \xi} q^{^{(a)}Y^{(1)}_1}$,
while the second one has $(a-1)$ variables $z^{(a)}_i =
t^{\frac{1-N}{2}} q^{^{(a)}Y^{(2)}_{i-1}} t^{a-i}$,
$i=2,\ldots,a$. Similarly to the previous section we have the
decomposition:
\begin{equation}
  \label{eq:17}
  {\mathcal I}^{\alpha_0}_{\mathcal{T}} (\vec{\mu}, \xi, q, t)=
  \mathcal{I}^{\alpha_0}_{0, \mathcal{T}'} (\vec{\mu}, \xi,  q,t)   Z_{\mathrm{vort},\mathcal{T}'}^{\alpha_0}(\vec{\mu}, \xi, q, t)\,,
\end{equation}
with
\begin{multline}
  \label{eq:18}
 \mathcal{I}^{\alpha_0}_{0, \mathcal{T}'}(\vec{\mu}, \xi, q,t) =
  \Res_{
    \begin{smallmatrix}
      z^{(a)}_1 =
      e^{2\pi b \xi}\\
      z^{(a)}_i = t^{\frac{1-N}{2}+a-i},\,\, i=2,\ldots, a
    \end{smallmatrix}} \{ \text{integrand}\}=\\
  = e^{\pi i b^2 \beta^2 \frac{N(N^2-1)}{6} + \pi b\beta (N+1) M_1 + \pi b \beta \sum_{a=1}^{N+1} (2a-3-N)
    M_a - 2 \pi i \xi (M_1 - M_{N+1} - i b\beta N) } \times\\
  \times
  \prod_{i=1}^N \frac{(t^i;q)_{\infty}}{(t;q)_{\infty}}
  \frac{\left( t^{\frac{N+1}{2}} e^{-2\pi b \xi} ;q
    \right)_{\infty}}{\left(t^{\frac{1-N}{2}} e^{-2\pi b \xi};q \right)_{\infty}}.
\end{multline}
One can also write down the vortex series explicitly as a sum over a
set of Young diagrams $^{(a)}Y^1$ and $^{(a)}Y^2$, but it is probably
easier and definitely more compact to notice that the
integral~(\ref{eq:16}) can be obtained from that of $T[SU(N+1)]$
theory given in eq. (2.16) of \cite{Zenkevich:2017ylb}. Indeed, if we set the $T[SU(N+1)]$ mass parameters
$\tau^{T[SU(N+1)]}_1 = e^{2\pi b \xi}$, $\tau^{T[SU(N+1)]}_a = t^{a-2}
t^{\frac{1-N}{2}}$ for $a = 2,\ldots, (N+1)$ we obtain precisely the
integral representation~(\ref{eq:16}), and even the contour of
integration (the sequence $\mathcal{C}_{1,0}$, $\mathcal{C}_{1,1}$
etc.) is neatly matched. We thus have
\begin{multline}
  \label{eq:33}
  Z_{\mathrm{vort},\mathcal{T}'}^{\alpha_0}(\vec{\mu},\xi, q,t)=\\ = \sum_{ k_i^a }
  \prod_{a=1}^N \left[\left( t \frac{\mu_a}{\mu_{a+1}}
    \right)^{\sum_{i=1}^a k^{(a)}_i} \prod_{i\neq j}^a \frac{\left(t
        \frac{\tau_i}{\tau_j};q\right)_{k^{(a)}_i - k^{(a)}_j}}{\left(
        \frac{\tau_i}{\tau_j};q\right)_{k^{(a)}_i - k^{(a)}_j}}
    \prod_{i=1}^a \prod_{j=1}^{a+1} \frac{\left( \frac{q}{t}
        \frac{\tau_i}{\tau_j}; q \right)_{k_i^{(a)}-k_j^{(a+1)}}}{\left(
        q \frac{\tau_i}{\tau_j}; q \right)_{k_i^{(a)}-k_j^{(a+1)}}}
  \right]_{
    \begin{smallmatrix}
      \tau_1 = e^{2\pi b \xi}, \\
      \tau_i = t^{i-2} t^{\frac{1-N}{2}}, \, i \geq 2
    \end{smallmatrix}
}
\end{multline}
where $k_i^{(a)}$ satisfy
\begin{equation}
  \label{eq:34}
      \begin{array}{ccccc}
        k^{(1)}_1 \geq & k^{(2)}_1 \geq & k^{(3)}_1 \geq & \cdots \geq &
    k^{(N)}_1 \geq 0\\
    & k^{(2)}_2 \geq & k^{(3)}_2 \geq &\cdots \geq & k^{(N)}_2 \geq
 0\\
    &  & k^{(3)}_3 \geq &\cdots \geq & k^{(N)}_3 \geq
 0\\
 & & &\ddots  & \vdots\\
    &  &  & & k^{(N)}_N \geq 0
 \end{array}
\end{equation}
The integers $k_i^{(a)}$ are of course just another way of writing the
sequence of Young diagrams $^{(a)}Y^1$ and $^{(a)}Y^2$. And as before we define:
\be
\label{coltp}
  Z^{\alpha_0}_{\mathrm{cl}, \mathcal{T}'}Z^{\alpha_0}_{1 \mathrm{-loop}, \mathcal{T}'}=
\mathcal{F}_{\mathcal{T}'} \mathcal{I}^{\alpha_0}_{0, \mathcal{T}}(\vec{\mu}, \xi, q,t)\,.
\ee

\section{Fiber-base invariance and Higgsing}
\label{sec:an-explicit-check}
In this Appendix we check the equality of 
topological string partitions, for fiber-base dual pairs, after Higgsing, and for the first nontrivial case of $N=2$. We write down the first terms of the vortex series and
then expand the partition functions in a double expansion in masses
and FI parameters to check the equality~(\ref{eq:11}) for the first
few orders.

For the  topological string on CY $\mathcal{Y}$)
we have following Eqs.~(\ref{eq:9}) and~(\ref{eq:39})
\begin{multline}
  \label{eq:8}
  Z^{\mathcal{Y}}_{1 \mathrm{-loop},
    \mathrm{top}}Z^{\mathcal{Y}}_{\mathrm{vort}, \mathrm{top}} =
  \frac{\left( \frac{q}{t} \frac{\mu_2}{\mu_3};q
    \right)_{\infty}}{\left( t \frac{\mu_2}{\mu_3};q \right)_{\infty}}
  \frac{\left( q \frac{\mu_1}{\mu_2};q
    \right)_{\infty}}{\left( t \frac{\mu_1}{\mu_2};q \right)_{\infty}}
   \frac{\left( q \frac{\mu_1}{\mu_3};q
    \right)_{\infty}}{\left( t \frac{\mu_1}{\mu_3};q \right)_{\infty}}
  \times\\
  \times \left[ 1 + t^N \frac{\tau_1}{\tau_2} \frac{\left( 1 -\frac{q}{t}\right)
      \left( 1 -\frac{q}{t} \frac{\mu_1}{\mu_2}\right) \left( 1
        -\frac{q}{t} \frac{\mu_1}{\mu_3}\right)}{( 1 -q )
      \left( 1 - q \frac{\mu_1}{\mu_2}\right) \left( 1
        - q \frac{\mu_1}{\mu_3}\right)} + \mathcal{O}\left( \left(
      \frac{\tau_1}{\tau_2}\right)^2 \right)\right].
\end{multline}
On the dual side we have (see Eqs.~(\ref{eq:10}) and~(\ref{eq:33}))
\begin{multline}
  \label{eq:12}
    Z^{\mathcal{Y}'}_{1 \mathrm{-loop},
    \mathrm{top}} Z^{\mathcal{Y}'}_{\mathrm{vort}, \mathrm{top}} =
  \frac{\left( q \frac{\tau_1}{\tau_2} ;q
    \right)_{\infty}}{\left( t \frac{\tau_1}{\tau_2} ;q
    \right)_{\infty}} \Biggl[ 1 + t \frac{\mu_1}{\mu_2} \frac{\left( 1
        - \frac{q}{t} \right) \left( 1
        - \frac{q}{t} \frac{\tau_1}{\tau_2} \right)}{\left( 1
        - q \right) \left( 1
        - q \frac{\tau_1}{\tau_2} \right)} + t \frac{\mu_2}{\mu_3} \frac{\left( 1
        - \frac{q}{t^2} \right) }{\left( 1
        - q \right) } +\\
    +t^2 \frac{\mu_1}{\mu_3} \frac{\left( 1 -
        \frac{q}{t} \right) \left( 1 - \frac{q}{t^2} \right) \left( 1
        - \frac{\tau_1}{t \tau_2} \right)}{(1-q)^2 \left( 1 -
        \frac{\tau_1}{\tau_2} \right)} + \mathcal{O}\left( \left( \frac{\mu_a}{\mu_{a+1}} \right)^2\right) \Biggr]
\end{multline}
Expanding Eqs.~(\ref{eq:8}) and~(\ref{eq:12}) to first order in
$\frac{\tau_1}{\tau_2}$, $\frac{\mu_1}{\mu_2}$ and
$\frac{\mu_2}{\mu_3}$ we get
\begin{multline}
      Z^{\mathcal{Y}}_{1 \mathrm{-loop},
    \mathrm{top}} Z^{\mathcal{Y}}_{\mathrm{vort}, \mathrm{top}} = 1 + \frac{\mu _1 (t+1) \tau _1 \left(q^2 t-q^2+q-t^2\right) (q-t)^2}{\mu _3 (q-1)^3
    t^2 \tau _2}+\\
  +\frac{\mu _1 \tau _1 \left(q^2 t-q^2+q-t^2\right) (q-t)}{\mu _2
   (q-1)^2 t \tau _2}+\frac{\mu _2 \tau _1 \left(q-t^2\right) (q-t)}{\mu _3
   (q-1)^2 t \tau _2}+\\
 +\frac{\mu _2 \left(q-t^2\right)}{\mu _3 (q-1) t}+\frac{\mu
   _1 (t+1) (q-t)^2}{\mu _3 (q-1)^2 t}+\frac{\mu _1 (q-t)}{\mu _2
   (q-1)}+\frac{\tau _1 (q-t)}{(q-1) \tau _2} + \mathcal{O}\left(
   \left( \frac{\mu_a}{\mu_{a+1}} \right)^2, \left(
     \frac{\tau_1}{\tau_2} \right)^2\right) = \\
 = Z^{\mathcal{Y}'}_{1 \mathrm{-loop},
    \mathrm{top}} Z^{\mathcal{Y}'}_{\mathrm{vort}, \mathrm{top}}.
\end{multline}
The equality~(\ref{eq:11}) can be checked to higher orders quite
easily on a computer and turns out to be valid.



\begin{thebibliography}{99}
%
%
%
\bibitem{Pestun:2016zxk}
  V.~Pestun {\it et al.},
  J.\ Phys.\ A {\bf 50} (2017) no.44,  440301
  doi:10.1088/1751-8121/aa63c1
  [arXiv:1608.02952 [hep-th]].
%
%
%
%
%
%
%
%
%
%
%


\bibitem{Aharony:2013dha}
  O.~Aharony, S.~S.~Razamat, N.~Seiberg and B.~Willett,
  JHEP {\bf 1307} (2013) 149
  doi:10.1007/JHEP07(2013)149
  [arXiv:1305.3924 [hep-th]].


\bibitem{Benvenuti:2018bav}
  S.~Benvenuti,
  arXiv:1809.03925 [hep-th].

\bibitem{Amariti:2018wht}
  A.~Amariti and L.~Cassia,
  arXiv:1809.03796 [hep-th].


\bibitem{Gaiotto:2008ak} 
  D.~Gaiotto and E.~Witten,
  ``S-Duality of Boundary Conditions In N=4 Super Yang-Mills Theory,''
  Adv.\ Theor.\ Math.\ Phys.\  {\bf 13}, no. 3, 721 (2009)
  doi:10.4310/ATMP.2009.v13.n3.a5
  [arXiv:0807.3720 [hep-th]].

\bibitem{Aharony:1997gp}
  O.~Aharony,
  Phys.\ Lett.\ B {\bf 404} (1997) 71
  doi:10.1016/S0370-2693(97)00530-3
  [hep-th/9703215].

\bibitem{Spiridonov:2009za}
  V.~P.~Spiridonov and G.~S.~Vartanov,
  Commun.\ Math.\ Phys.\  {\bf 304} (2011) 797
  doi:10.1007/s00220-011-1218-9
  [arXiv:0910.5944 [hep-th]].


\bibitem{Rains:2006dfy} 
  E.~M.~Rains,
  ``Limits of elliptic hypergeometric integrals,''
  Ramanujan J.\  {\bf 18}, no. 3, 257 (2007)
  doi:10.1007/s11139-007-9055-3
  [math/0607093 [math.CA]].

\bibitem{Fokko}
  F. ~J. ~van de Bult 
  ``Hyperbolic hypergeometric functions ,''
  Ph.D. thesis, University of Amsterdam.
  http://math.caltech.edu/vdbult/Thesis.pdf

\bibitem{Bullimore:2014awa} 
  M.~Bullimore, H.~C.~Kim and P.~Koroteev,
  ``Defects and Quantum Seiberg-Witten Geometry,''
  JHEP {\bf 1505}, 095 (2015)
  doi:10.1007/JHEP05(2015)095
  [arXiv:1412.6081 [hep-th]].

  
\bibitem{Benvenuti:2017kud} 
  S.~Benvenuti and S.~Giacomelli,
  ``Abelianization and sequential confinement in $2+1$ dimensions,''
  JHEP {\bf 1710}, 173 (2017)
  doi:10.1007/JHEP10(2017)173
  [arXiv:1706.04949 [hep-th]].
 
  
\bibitem{Giacomelli:2017vgk} 
  S.~Giacomelli and N.~Mekareeya,
  ``Mirror theories of 3d $ \mathcal{N} $ = 2 SQCD,''
  JHEP {\bf 1803}, 126 (2018)
  doi:10.1007/JHEP03(2018)126
  [arXiv:1711.11525 [hep-th]].
  
  

\bibitem{Benini:2017dud} 
  F.~Benini, S.~Benvenuti and S.~Pasquetti,
  ``SUSY monopole potentials in 2+1 dimensions,''
  JHEP {\bf 1708}, 086 (2017)
  doi:10.1007/JHEP08(2017)086
  [arXiv:1703.08460 [hep-th]].

\bibitem{Hanany:2004ea}
  A.~Hanany and D.~Tong,
  JHEP {\bf 0404} (2004) 066
  doi:10.1088/1126-6708/2004/04/066
  [hep-th/0403158].


\bibitem{Gaiotto:2012xa}
  D.~Gaiotto, L.~Rastelli and S.~S.~Razamat,
  JHEP {\bf 1301} (2013) 022
  doi:10.1007/JHEP01(2013)022
  [arXiv:1207.3577 [hep-th]].

\bibitem{Mironov:2009qt}
  A.~Mironov and A.~Morozov,
  Phys.\ Lett.\ B {\bf 680} (2009) 188
  doi:10.1016/j.physletb.2009.08.061
  [arXiv:0908.2190 [hep-th]].
\bibitem{Kozcaz:2010af}
  C.~Kozcaz, S.~Pasquetti and N.~Wyllard,
  JHEP {\bf 1008} (2010) 042
  doi:10.1007/JHEP08(2010)042
  [arXiv:1004.2025 [hep-th]].
\cite{Dimofte:2010tz}
\bibitem{Dimofte:2010tz}
  T.~Dimofte, S.~Gukov and L.~Hollands,
  Lett.\ Math.\ Phys.\  {\bf 98} (2011) 225
  doi:10.1007/s11005-011-0531-8
  [arXiv:1006.0977 [hep-th]].
\bibitem{Dorey:2011pa}
  N.~Dorey, S.~Lee and T.~J.~Hollowood,
  JHEP {\bf 1110} (2011) 077
  doi:10.1007/JHEP10(2011)077
  [arXiv:1103.5726 [hep-th]].



\bibitem{Nieri:2013vba}
  F.~Nieri, S.~Pasquetti, F.~Passerini and A.~Torrielli,
  JHEP {\bf 1412} (2014) 040
  doi:10.1007/JHEP12(2014)040
  [arXiv:1312.1294 [hep-th]].


\bibitem{Gaiotto:2014ina}
  D.~Gaiotto and H.~C.~Kim,
  JHEP {\bf 1610} (2016) 012
  doi:10.1007/JHEP10(2016)012
  [arXiv:1412.2781 [hep-th]].


\bibitem{Nieri:2018pev}
  F.~Nieri, Y.~Pan and M.~Zabzine,
  arXiv:1809.00736 [hep-th].


\bibitem{Zenkevich:2017ylb}
  A.~Nedelin, S.~Pasquetti and Y.~Zenkevich,
  arXiv:1712.08140 [hep-th].



\bibitem{Tong:2000ky}
  D.~Tong,
  JHEP {\bf 0007} (2000) 019
  doi:10.1088/1126-6708/2000/07/019
  [hep-th/0005186].
  
\bibitem{Jafferis:2010un}
  D.~L.~Jafferis,
  JHEP {\bf 1205} (2012) 159
  doi:10.1007/JHEP05(2012)159
  [arXiv:1012.3210 [hep-th]].
 
  
\bibitem{Cabrera:2016vvv}
  S.~Cabrera and A.~Hanany,
  JHEP {\bf 1611} (2016) 175
  doi:10.1007/JHEP11(2016)175
  [arXiv:1609.07798 [hep-th]].

  
  
\bibitem{Borokhov:2002cg}
  V.~Borokhov, A.~Kapustin and X.~k.~Wu,
  JHEP {\bf 0212} (2002) 044
  doi:10.1088/1126-6708/2002/12/044
  [hep-th/0207074].
 


\bibitem{Benini:2011cma}
  F.~Benini, C.~Closset and S.~Cremonesi,
  JHEP {\bf 1109} (2011) 005
  doi:10.1007/JHEP09(2011)005
  [arXiv:1105.2299 [hep-th]].
  
  







  
  
\bibitem{Benvenuti:2017lle} 
  S.~Benvenuti and S.~Giacomelli,
  ``Supersymmetric gauge theories with decoupled operators and chiral ring stability,''
  Phys.\ Rev.\ Lett.\  {\bf 119}, no. 25, 251601 (2017)
  doi:10.1103/PhysRevLett.119.251601
  [arXiv:1706.02225 [hep-th]].
  
  
  
  


\bibitem{Hanany:1996ie}
  A.~Hanany and E.~Witten,
  Nucl.\ Phys.\ B {\bf 492} (1997) 152
  doi:10.1016/S0550-3213(97)00157-0, 10.1016/S0550-3213(97)80030-2
  [hep-th/9611230].

  
    

\bibitem{Gaiotto:2013bwa}
  D.~Gaiotto and P.~Koroteev,
  JHEP {\bf 1305} (2013) 126
  doi:10.1007/JHEP05(2013)126
  [arXiv:1304.0779 [hep-th]].
    


  
  
\bibitem{Agarwal:2014rua} 
  P.~Agarwal, I.~Bah, K.~Maruyoshi and J.~Song,
  ``Quiver tails and $ \mathcal{N}=1 $ SCFTs from M5-branes,''
  JHEP {\bf 1503}, 049 (2015)
  doi:10.1007/JHEP03(2015)049
  [arXiv:1409.1908 [hep-th]].
  

  
\bibitem{Kapustin:1999ha}
  A.~Kapustin and M.~J.~Strassler,
  JHEP {\bf 9904} (1999) 021
  doi:10.1088/1126-6708/1999/04/021
  [hep-th/9902033].

  
  
\bibitem{Benvenuti:2016wet} 
  S.~Benvenuti and S.~Pasquetti,
  ``3d $ \mathcal{N} $ = 2 mirror symmetry, pq-webs and monopole superpotentials,''
  JHEP {\bf 1608}, 136 (2016)
  doi:10.1007/JHEP08(2016)136
  [arXiv:1605.02675 [hep-th]].
  
\bibitem{Hama:2011ea} 
  N.~Hama, K.~Hosomichi and S.~Lee,
  ``SUSY Gauge Theories on Squashed Three-Spheres,''
  JHEP {\bf 1105}, 014 (2011)
  doi:10.1007/JHEP05(2011)014
  [arXiv:1102.4716 [hep-th]].
  
\bibitem{Pasquetti:2011fj} 
  S.~Pasquetti,
  ``Factorisation of N = 2 Theories on the Squashed 3-Sphere,''
  JHEP {\bf 1204}, 120 (2012)
  doi:10.1007/JHEP04(2012)120
  [arXiv:1111.6905 [hep-th]].
  
\bibitem{Beem:2012mb}
  C.~Beem, T.~Dimofte and S.~Pasquetti,
  JHEP {\bf 1412} (2014) 177
  doi:10.1007/JHEP12(2014)177
  [arXiv:1211.1986 [hep-th]].
  
\bibitem{Pasquetti:2016dyl}
  S.~Pasquetti,
  J.\ Phys.\ A {\bf 50} (2017) no.44,  443016
  doi:10.1088/1751-8121/aa60fe
  [arXiv:1608.02968 [hep-th]].


\bibitem{Gasper}
  G.Gasper and M.Rahman,
  ``Basic Hypergeometric Series,''
   Encyclopedia of Mathematics and its Applications, Cambridge,
  ISBN-10: 0521833574



\bibitem{Nieri:2015yia} 
  F.~Nieri and S.~Pasquetti,
  ``Factorisation and holomorphic blocks in 4d,''
  JHEP {\bf 1511}, 155 (2015)
  doi:10.1007/JHEP11(2015)155
  [arXiv:1507.00261 [hep-th]].
  
  
\bibitem{Giveon:1998sr}
  A.~Giveon and D.~Kutasov,
  Rev.\ Mod.\ Phys.\  {\bf 71} (1999) 983
  doi:10.1103/RevModPhys.71.983
  [hep-th/9802067].

\bibitem{Iqbal:2007ii}
  A.~Iqbal, C.~Kozcaz and C.~Vafa,
  JHEP {\bf 0910} (2009) 069
  doi:10.1088/1126-6708/2009/10/069
  [hep-th/0701156].


\bibitem{Awata:2011ce}
  H.~Awata, B.~Feigin and J.~Shiraishi,
  JHEP {\bf 1203} (2012) 041
  doi:10.1007/JHEP03(2012)041
  [arXiv:1112.6074 [hep-th]].
\bibitem{Mironov:2016yue}
  A.~Mironov, A.~Morozov and Y.~Zenkevich,
  Phys.\ Lett.\ B {\bf 762} (2016) 196
  doi:10.1016/j.physletb.2016.09.033
  [arXiv:1603.05467 [hep-th]].

\bibitem{Awata:2016riz}
  H.~Awata, H.~Kanno, T.~Matsumoto, A.~Mironov, A.~Morozov, A.~Morozov, Y.~Ohkubo and Y.~Zenkevich,
  JHEP {\bf 1607} (2016) 103
  doi:10.1007/JHEP07(2016)103
  [arXiv:1604.08366 [hep-th]].
\bibitem{Taki:2007dh}
  M.~Taki,
  JHEP {\bf 0803} (2008) 048
  doi:10.1088/1126-6708/2008/03/048
  [arXiv:0710.1776 [hep-th]].

\end{thebibliography}
\end{document}